\documentclass{article}
\usepackage{amsmath}
\usepackage{amssymb}
\usepackage{bm}
\begin{document}
\title{$N$ Identical Particles
Under Quantum Confinement: A Many-Body Dimensional Perturbation
Theory Approach II,\\ The Lowest-Order Wave Function I}
\author{M. Dunn, D. K. Watson\\
University of Oklahoma \\ Department of Physics and Astronomy \\
Norman, OK 73019 \and J. G. Loeser \\
Oregon State University \\ Department of Chemistry \\ Corvallis,
OR 97331}
\date{\today}
\maketitle

\begin{abstract}
In this paper we continue our development of a dimensional
perturbation theory (DPT) treatment of $N$ identical particles
under quantum confinement. DPT is a beyond-mean-field method which
is applicable to both weakly and strongly-interacting systems and
can be used to connect both limits. In a previous paper we
developed the formalism for low-order energies and excitation
frequencies. This formalism has been applied to atoms,
Bose-Einstein condensates and quantum dots. One major advantage of
the method is that $N$ appears as a parameter in the analytical
expressions for the energy and so results for $N$ up to a few
thousand are easy to obtain. Other properties however, are also of
interest, for example the density profile in the case of a BEC,
and larger $N$ results are desirable as well. The latter case
requires us to go to higher orders in DPT. These calculations
require as input zeroth-order wave functions and this paper, along
with a subsequent paper, address this issue.
\end{abstract}

\section{Introduction}
Systems that involve $N$ identical interacting particles under
quantum confinement appear throughout many areas of physics,
including chemical, condensed matter, and atomic
physics\cite{setting}. In a previous paper\cite{FGpaper}, we
applied the methods of dimensional perturbation theory
(DPT)\cite{copen92}, a powerful set of tools that uses symmetry to
yield simple results, to such $N$-body systems. We presented a
detailed discussion of the dimensional continuation of the
$N$-particle Schr\"odinger equation, the $D\to\infty$ equilibrium
($D^0$) structure, and the ($D^{-1}$) energy.  We used the FG
matrix method to derive general, analytical expressions for the
$N$-body normal-mode vibrational frequencies, and we gave specific
analytical results for three confined $N$-body quantum systems:
the $N$-electron atom, the $N$-electron quantum dot, and the
$N$-atom inhomogeneous Bose-Einstein condensate with a repulsive
hardcore potential.

In a subsequent paper\cite{energy} we further pursued our analysis
of the $N$-atom inhomogeneous Bose-Einstein condensate (BEC)with a
repulsive hardcore potential and optimized our low-order DPT
energy by fitting to low-$N$ DMC data\cite{blume} and then
extrapolated the result out to large $N$. Our low-order energies
are accurate out to $N \approx 10,000$ and larger $N$ could in
principle be obtained by going to higher orders in the
perturbation expansion\cite{matrix_method}. Ten thousand atoms may
not seem particularly impressive given the fact that the
Gross-Pitaevskii (GP) equation can yield higher-$N$ results, but
the advantage of DPT is that it is not a
weak-interatomic-interaction approximation. At lowest orders it
includes beyond-mean-field effects. Thus DPT is effective for
large-$a$ systems that have been created in the lab by exploiting
Feshbach resonances\cite{wieman}, systems for which the GP and
modified Gross-Pitaevskii (MGP) equations fail. This we
demonstrated for a BEC which had an $a$ one hundred times that of
the natural scattering length of $^{87}$Rb.

In another paper\cite{loeser}, low-order $N$-body dimensional
perturbation methods have been applied to the $N$-electron atom.
In this instrumental paper, Loeser obtained low-order, analytical
expressions for the ground-state energy of neutral atoms. For
$Z=1$ to $127$, the numerical results compare well to Hartree-Fock
energies with a correlation correction.

All of the previous work has focussed on low-order calculations of
energies and thus far little attention has been given to wave
functions, even at lowest order. Yet the lowest-order wave
function yields important information about the system. The normal
mode coordinates tell us the nature of the excitations of the
system and the lowest-order wave function gives us expectation
values at low orders. For macroscopic quantum-confined systems,
such as the BEC, the wave function is uncloaked in an explicit way
since the density profile may be viewed in a direct fashion. Also,
calculating energies and wave functions to higher orders in $1/D$
requires as input the lowest-order wave function.

We briefly outlined the derivation of the lowest-order wave
function in a previous letter\cite{PRL} and in this paper we
expand considerably upon that discussion. In Section~\ref{sec:SE}
we discuss the $S$-wave, Jacobian-weighted Hamiltonian and wave
function. In Section~\ref{sec:infD} we dimensionally scale the
system to regularize the large dimension limit for which we then
solve. In Section~\ref{sec:firstorder} we begin the normal mode
analysis which will result in the energy through first order in
$1/D$ and the lowest-order wave function. In
Section~\ref{sec:symm} we discuss the symmetry of the $N$-body
quantum-confinement problem. This symmetry simplifies the problem
to an extraordinary degree, making an exact solution of this
$N$-body problem, with $N(N-1)/2$ interparticle interactions,
possible at large $D$\,. In Section~\ref{sec:symnorm} we exploit
this symmetry further and discuss the use of symmetry coordinates
in the calculation of the large-dimension, zeroth-order normal
modes of the system. In Section~\ref{sec:DetS} we introduce a
particular approach to determining a suitable set of symmetry
coordinates. The method uses what we term primitive irreducible
coordinates, from which the symmetry coordinates are derived. This
is illustrated in detail in a simple example involving the
coordinates of a particular species of the $N$-body
quantum-confinement problem (symmetry coordinates transforming
under the same irreducible representation are said to belong to
the same species). In Section~\ref{sec:FreqNorModN} we apply the
general theory discussed in earlier sections to derive in detail
the frequencies and normal-mode coordinates of this species.
Section~\ref{sec:SumConc} is a summary and conclusion section. The
frequencies and normal-mode coordinates of the remaining, and more
complicated species of the system are derived in a subsequent
paper\cite{paperII}.

\section{The ${\mathbf{D}}$-dimensional ${\mathbf{N}}$-body Schr\"odinger
Equation}\label{sec:SE}  For an $N$-body system of particles
confined by a spherically symmetric potential and interacting via
a common two-body potential $g_{ij}$, the Schr\"odinger equation
in $D$-dimensional Cartesian coordinates is
\begin{equation}
\label{generalH} H \Psi = \left[ \sum\limits_{i=1}^{N} h_{i} +
\sum_{i=1}^{N-1}\sum\limits_{j=i+1}^{N} g_{ij} \right] \Psi = E
\Psi \,,
\end{equation}
where
\begin{equation} \label{eq:hi}
h_{i}=-\frac{\hbar^2}{2
m_{i}}\sum\limits_{\nu=1}^{D}\frac{\partial^2}{\partial
x_{i\nu}^2} +
V_{\mathtt{conf}}\left(\sqrt{\sum\nolimits_{\nu=1}^{D}x_{i\nu}^2}\right)
\,,
\end{equation}
\begin{equation}
\mbox{and} \;\;\;
g_{ij}=V_{\mathtt{int}}\left(\sqrt{\sum\nolimits_{\nu=1}^{D}\left(x_{i\nu}-x_{j\nu}
\right)^2}\right)
\end{equation}
are the single-particle Hamiltonian and the two-body interaction
potential, respectively. The operator $H$ is the $D$-dimensional
Hamiltonian, and $x_{i\nu}$ is the $\nu^{th}$ Cartesian component
of the $i^{th}$ particle. The term $V_{\mathtt{conf}}$ is the
confining potential. For the $N$-electron atom,
$V_{\mathtt{conf}}$ is the Coulomb attraction between the
electrons and the nucleus, while for the $N$-electron quantum dot
and $N$-atom hard-sphere problem, a model appropriate for trapped
gaseous BECs, we model the confinement as a harmonic trapping
potential. The two-body interaction potential $V_{\mathtt{int}}$
is Coulombic in the first two systems and a hard sphere in the
third.

\subsection{Transformation of the Laplacian}
Restricting our attention to spherically symmetric ($L=0$) states
of the $N$-body system, we transform from Cartesian to internal
coordinates. A convenient internal coordinate system for confined
systems is
\begin{equation}\label{eq:int_coords}
r_i=\sqrt{\sum_{\nu=1}^{D} x_{i\nu}^2} \;\;\; (1 \le i \le N)
\;\;\; \mbox{and} \;\;\;
\gamma_{ij}=cos(\theta_{ij})=\left(\sum_{\nu=1}^{D}
x_{i\nu}x_{j\nu}\right) / r_i r_j \;\;\; (1 \le i < j \le N),
\end{equation}
which are the $D$-dimensional scalar radii $r_i$ of the $N$
particles from the center of the confining potential and the
cosines $\gamma_{ij}$ of the $N(N-1)/2$ angles between the radial
vectors.

Now for a function $\Psi$ dependent on only two functions $r(x)$
and $\gamma(x)$ one can write
\begin{equation}
\frac{d^2 \Psi(r(x),\gamma(x))}{ d x^2} = \frac{d^2 r}{d x^2}
\frac{d \Psi}{d r} + \frac{d^2 \gamma}{d x^2} \frac{d \Psi}{d
\gamma} + \left( \frac{d r}{d x} \right)^2 \frac{d^2 \Psi}{d r^2}
+ \left( \frac{d \gamma}{d x} \right)^2 \frac{d^2 \Psi}{d
\gamma^2} + 2 \frac{d \gamma}{d x} \frac{d r}{d x} \frac{d^2
\Psi}{d r d \gamma}.
\end{equation}
Generalizing this, when operating on the state
$\Psi(r_i(x_{i\nu}),\gamma_{i1}(x_{i\nu})\ldots
\gamma_{ik}(x_{i\nu}) \ldots  \gamma_{iN}(x_{i\nu}))$ (where
$k\not= i$ and $\nu=1,\ldots,D$)), $\nabla_i^2$ can be written in
terms of the internal coordinates of Eq.~(\ref{eq:int_coords}) as
\begin{eqnarray}
\nabla_i^2\Psi\equiv\sum\limits_{\nu=1}^{D}\frac{\partial^2}{\partial
x_{i\nu}^2}\Psi  & = & \sum\limits_{\nu=1}^{D}\left(
\frac{\partial^2 r_i}{{\partial x_{i\nu}}^2}\right)
\frac{\partial}{\partial r_i}\Psi +
\sum\limits_{\nu=1}^{D}\sum\limits_{j\not= i} \left(
\frac{\partial^2 \gamma_{ij}}{{\partial x_{i\nu}}^2} \right)
\frac{\partial}{\partial \gamma_{ij}}\Psi +\nonumber\\
&&\sum\limits_{\nu=1}^{D} \left( \frac{\partial r_i}{\partial
x_{i\nu}} \right)^2 \frac{\partial^2}{{\partial r_i}^2}\Psi +
\sum\limits_{\nu=1}^{D}\sum\limits_{j\not= i}\sum\limits_{k\not=
i} \left( \frac{\partial \gamma_{ij}}{\partial x_{i\nu}} \right)
\left ( \frac{\partial \gamma_{ik}} {\partial x_{i\nu}} \right)
\frac{\partial^2}{\partial \gamma_{ij} \partial  \gamma_{ik}}\Psi + \\
&&2 \sum\limits_{\nu=1}^{D}\sum\limits_{j\not=i} \left(
\frac{\partial r_i}{\partial x_{i\nu}} \right) \left(
\frac{\partial\gamma_{ij}}{\partial x_{i\nu}} \right)
\frac{\partial^2}{\partial r_i \partial
\gamma_{ij}}\Psi\,.\nonumber
\end{eqnarray}

The relevant derivatives of the internal coordinates are
\begin{eqnarray}
\frac{\partial r_i}{\partial x_{i\nu}} = \frac{x_{i\nu}}{r_i} &&
\frac{\partial \gamma_{ij}}{\partial x_{i\nu}} =
\frac{1}{r_i}\left( \frac{x_{j\nu}}{r_j} -
\frac{x_{i\nu}}{r_i}\gamma_{ij} \right) \\
\frac{\partial^2 r_i}{{\partial x_{i\nu}}^2} = \frac{1}{r_i}\left(
1 - \frac{x_{i\nu}^2}{r_i^2} \right) && \frac{\partial^2
\gamma_{ij}}{{\partial x_{i\nu}}^2} = \frac{1}{r_i^2} \left( 3
\frac{x_{i\nu}^2}{r_i^2}\gamma_{ij} - 2 \frac{x_{i\nu}}{r_i}
\frac{x_{j\nu}}{r_j}-\gamma_{ij}\right),
\end{eqnarray}
which lead to the effective $S$-wave Laplacian in internal
coordinates:
\begin{eqnarray}
\label{eq:laplacian} \sum\limits_{i}\nabla_i^2 \Psi & = &
\sum\limits_{i}\frac{D-1}{r_i}\frac{\partial}{\partial
r_i}\Psi-\sum\limits_{i}\frac{D-1}{r_i^2}\sum\limits_{j\not=
i}\gamma_{ij}\frac{\partial}
{\partial \gamma_{ij}}\Psi +\nonumber\\
&&\sum\limits_{i}\frac{\partial^2}{{\partial r_i}^2}\Psi
+\sum\limits_{i}\sum\limits_{j\not= i}\sum\limits_{k\not=
i}\frac{\gamma_{jk}-\gamma_{ij}\gamma_{ik}}
{r_i^2}\frac{\partial^2}{\partial \gamma_{ij} \partial
\gamma_{ik}}\Psi.
\end{eqnarray}

\subsection{The Jacobian-Weighted Hamiltonian}
\label{sub:simtransf} All of the $N$-body DPT work up till now has
utilized a similarity transformation so that the kinetic energy
operator is transformed to a sum of terms of two kinds, namely,
derivative terms {\em and} a repulsive centrifugal-like term,
which when attractive interparticle potentials are present,
stabilizes the system against collapse in the large-$D$ limit. The
particular transformation which has been used in almost all the
work to date, is chosen so that first-order derivatives resulting
from the Laplacian (Eq.~(\ref{eq:laplacian})) are removed from the
transformed Hamiltonian. When this is done the zeroth and first
orders of the dimensional ($1/D$) expansion of the Hamiltonian
become exactly soluble for any value of $N$. In the $D\to\infty$
limit, the second derivative terms drop out, resulting in a static
problem at zeroth order, while first order corrections correspond
to simple harmonic normal-mode oscillations about the
infinite-dimensional structure.

Writing the similarity transformation of the wave function $\Psi$
and operators $\widehat{O}$ as
\begin{equation}
\label{eq:simtransf}
\Phi = \chi^{-1} \Psi, \;\; \mbox{and} \;\;
\widetilde{O}=\chi^{-1} \widehat{O} \chi,
\end{equation}
the transforming function is chosen to be
\begin{equation}\label{eq:chi1}
\chi = (r_1 r_2 \ldots r_N)^{-(D-1)/2} \Gamma^{- (D-1)/4}.
\end{equation}
where $\Gamma$ is the Gramian determinant, the determinant of the
matrix whose elements are $\gamma_{ij}$ (see
Appendix~\ref{app:gram}).

Carrying out the transformation of the Hamiltonian and wave
function of Eq.~(\ref{generalH}) via Eqs.~(\ref{eq:simtransf}) and
(\ref{eq:chi1}), the Schr\"odinger equation becomes\cite{loeser}:
\begin{eqnarray}
&&{\displaystyle ( _{(iii)}T+V) \,{_{(iii)}}\Phi = E \,{_{(iii)}}\Phi} \label{eq:SE1} \\
&&{\displaystyle
_{(iii)}T=\sum\limits_{i=1}^{N}\left(-\frac{\hbar^2}{2
m_i}\frac{\partial^2}{{\partial r_i}^2}-
\sum\limits_{j\not=i}\sum\limits_{k\not=i}\frac{\hbar^2(\gamma_{jk}-\gamma_{ij}
\gamma_{ik})}{2 m_i
r_i^2}\frac{\partial^2}{\partial\gamma_{ij}\partial\gamma_{ik}}
+\frac{\hbar^2(D- 1)(D-2N-1)}{8 m_i
r_i^2}\frac{\Gamma^{(i)}}{\Gamma} \right)} \label{eq:SE_Tf}\\
&&{\displaystyle V=\sum\limits_{i=1}^{N}V_{\mathtt{conf}}(r_i)+
\sum\limits_{i=1}^{N-1}\sum\limits_{j=i+1}^{N}
V_{\mathtt{int}}(r_{ij}).} \label{eq:V}
\end{eqnarray}
The Gramian matrix whose determinant is $\Gamma^{(i)}$ is the
$i^{th}$ principal minor formed by deleting from $\Gamma$ the row
and column corresponding to the $i^{th}$ particle. The quantity
$r_{ij}=\sqrt{r_{i}^2+r_{j}^2-2r_{i}r_{j}\gamma_{ij}}$ is the
interparticle separation.  The similarity transformed Hamiltonian
for the energy eigenstate $_{(iii)}\Phi$ is $_{(iii)}H=\chi^{-1} H
\chi$, where $ _{(iii)}H=( _{(iii)}T+V)$. As we can see from
Eq.~(\ref{eq:SE_Tf}) all first-order derivatives have been
eliminated from the Hamiltonian.

While the similarity transformation of Eqs.~(\ref{eq:simtransf})
and (\ref{eq:chi1}) is convenient in that all first-order
derivative terms are eliminated, it complicates the interpretation
of the normal-mode structure of the large-$D$ wave function. Only
when the weight function, $W$, for the matrix elements is equal to
unity, i.e.\
\begin{equation}
W=J\chi^2=1 \label{eq:Weight}
\end{equation}
or
\begin{equation} \chi=J^{-1/2},
\label{eq:chiEqJ}
\end{equation}
where $J$ is the Jacobian of the transformation to internal
coordinates, will a first derivative of an internal coordinate be
the conjugate momentum to that coordinate. The transforming
function of Eq.~(\ref{eq:chi1}) does not satisfy
Eq.~(\ref{eq:chiEqJ}) since the Jacobian $J$ is\cite{avery}
\begin{equation}
J = (r_1 r_2 \ldots r_N)^{D-1} \Gamma^{(D-N-1)/2}.
\end{equation}
Furthermore, the development of higher-order DPT expansions
involves the matrix elements of coordinates and their derivatives
between the zeroth-order normal-mode functions. These matrix
elements are much easier to calculate when the weight function in
the integrals is unity, and this only occurs when
Eq~(\ref{eq:chiEqJ}) is satisfied.

In Ref.~\cite{avery}, Avery {\sl et al.} considered the problem of
performing the similarity transformation of
Eq.~(\ref{eq:simtransf}) with a more general transforming function
$\chi$, which has adjustable parameters $\alpha$ and $\beta$, and
is of the form:
\begin{equation}\label{eq:chiG}
\chi = (r_1 r_2 \ldots r_N)^{-\alpha} \Gamma^{- \beta/2}.
\end{equation}
One of the cases considered by Avery {\sl et al.}\cite{avery} has
$\alpha=(D-1)/2$ and $\beta=(D-N-1)/2$ so that
\begin{equation}
\chi = (r_1 r_2 \ldots r_N)^{-(D-1)/2} \Gamma^{-(D-N-1)/4} \,.
\label{eq:chi}
\end{equation}
The transforming function $\chi$ of Eq.~(\ref{eq:chi}) does
satisfy Eqs.~(\ref{eq:Weight}) and (\ref{eq:chiEqJ}).

Carrying out the transformation of the Hamiltonian and wave
function of Eq.~(\ref{generalH}) via Eqs.~(\ref{eq:simtransf}) and
(\ref{eq:chi}), the Schr\"odinger equation becomes\cite{avery}:
\begin{equation}
 (_{(i)}T+V)\, {_{(i)}}\Phi = E \,{_{(i)}}\Phi
\label{eq:SE}
\end{equation}
where
\begin{eqnarray}
_{(i)}T&=& {\displaystyle \hbar^2
\sum\limits_{i=1}^{N}\Biggl[-\frac{1}{2
m_i}\frac{\partial^2}{{\partial r_i}^2}- \frac{1}{2 m_i r_i^2}
\Bigg(
\sum\limits_{j\not=i}\sum\limits_{k\not=i}(\gamma_{jk}-\gamma_{ij}
\gamma_{ik})
\frac{\partial^2}{\partial\gamma_{ij}\partial\gamma_{ik}} -
N\sum\limits_{j\not=i} \gamma_{ij}
\frac{\partial}{\partial\gamma_{ij}} \Bigg)} \nonumber \\
&& {\displaystyle +\frac{N(N-2)+(D-N-1)^2 \left(
\frac{\Gamma^{(i)}}{\Gamma} \right) }{8 m_i r_i^2} \Biggr] }
\nonumber \\
&=& {\displaystyle \hbar^2 \sum\limits_{i=1}^{N}\Biggl[-\frac{1}{2
m_i}\frac{\partial^2}{{\partial r_i}^2}- \frac{1}{2 m_i r_i^2}
\sum\limits_{j\not=i}\sum\limits_{k\not=i}
\frac{\partial}{\partial\gamma_{ij}}(\gamma_{jk}-\gamma_{ij}
\gamma_{ik})
\frac{\partial}{\partial\gamma_{ik}}} \nonumber \\
&& {\displaystyle +\frac{N(N-2)+(D-N-1)^2 \left(
\frac{\Gamma^{(i)}}{\Gamma} \right) }{8 m_i r_i^2} \Biggr] }
 \label{eq:SE_T}
\end{eqnarray}
The latter expression for $_{(i)}T$ is explicitly self-adjoint
since the weight function, $W$, for the matrix elements is equal
to unity, i.e.\ Eq.~(\ref{eq:Weight}) is satisfied. The similarity
transformed Hamiltonian for the energy eigenstate $_{(i)}\Phi$ is
$ _{(i)}H=( _{(i)}T+V)$.

\section{Infinite-${\mathbf{D}}$ analysis: Leading order
energy}\label{sec:infD}

To begin the perturbation analysis we regularize the
large-dimension limit of the Schr\"odinger equation by defining
dimensionally scaled variables:
\begin{equation} \label{eq:kappascale}
\bar{r}_i = r_i/\kappa(D) \;\;\; , \;\;\; \bar{E} = \kappa(D) E
\;\;\; \mbox{and} \;\;\; \bar{H} = \kappa(D) \,\, {_{(i)}H}
\end{equation}
with dimension-dependent scale factor $\kappa(D)$.  From
Eq.~(\ref{eq:SE_T}) one can see that the kinetic energy T scales
in the same way as $1/r^2$, so the scaled version of
Eq.~(\ref{eq:SE}) becomes
\begin{equation} \label{eq:scale1}
\bar{H} \Phi =
\left(\frac{1}{\kappa(D)}\bar{T}+\bar{V}_{\mathtt{eff}}
\right)\Phi = \bar{E} \Phi,
\end{equation}
where barred quantities simply indicate that the variables are now
in scaled units. Because of the quadratic $D$ dependence in the
centrifugal-like term in T of Eq.~(\ref{eq:SE_T}), we conclude
that the scale factor $\kappa(D)$ must also be quadratic in $D$,
otherwise the $D\to\infty$ limit of the Hamiltonian would not be
finite. The precise form of $\kappa(D)$ is chosen so that the
result of the scaling is as simple as possible and depends on the
system in question. In previous work\cite{FGpaper} we have chosen
$\kappa(D)=(D-1)(D-2N-1)/(4Z)$ for the $S$-wave, $N$-electron
atom; $\kappa(D)=\Omega \, l_{\mathtt{ho}}$ for the $N$-electron
quantum dot where $\Omega=(D-1)(D-2N-1)/4$ and the
dimensionally-scaled harmonic oscillator length and trap frequency
respectively are
${l}_{\mathtt{ho}}=\sqrt{\frac{\hbar}{m^*\bar{\omega}_{\mathtt{ho}}}}$
and $\bar{\omega}^2_{\mathtt{ho}}=\Omega^3
{\omega}^2_{\mathtt{ho}}$\,; and $\kappa(D)=D^2
\bar{a}_{\mathtt{ho}}$ for the BEC where
$\bar{a}_{\mathtt{ho}}=\sqrt{\frac{\hbar}{m
\bar{\omega}_{\mathtt{ho}}}}$ and
${\bar{\omega}_{\mathtt{ho}}}=D^3{\omega_{\mathtt{ho}}}$\,. The
factor of $\kappa(D)$ in the denominator of Eq.~(\ref{eq:scale1})
acts as an effective mass that increases with $D$, causing the
derivative terms to become suppressed while leaving behind a
centrifugal-like term in an effective potential,
\begin{equation}
\label{veff}
\bar{V}_{\mathtt{eff}}(\bar{r},\gamma;\delta=0)=\sum\limits_{i=1}^{N}\left(\frac{\hbar^2}{8
m_i
\bar{r}_i^2}\frac{\Gamma^{(i)}}{\Gamma}+\bar{V}_{\mathtt{conf}}(\bar{r},\gamma;\delta=0)\right)+\sum\limits_{i=1}^{N-1}\sum\limits_{j=i+1}^{N}
\bar{V}_{\mathtt{int}}(\bar{r},\gamma;\delta=0)\,,
\end{equation}
where $\delta=1/D$, in which the particles become frozen at large
$D$. In the $D\to\infty$ ($\delta \to 0$) limit, the excited
states have collapsed onto the ground state, which is found at the
minimum of $V_{\mathtt{eff}}$.

We assume a totally symmetric minimum characterized by the
equality of all radii and angle cosines of the particles when
$D\to\infty$, i.e.\
\begin{equation}
\bar{r}_{i}=\bar{r}_{\infty} \;\; (1 \le i \le N), \;\;\;\;
\gamma_{ij}=\overline{\gamma}_\infty \;\; (1 \le i < j \le N).
\end{equation}
Since each particle radius and angle cosine is equivalent, we can
take derivatives with respect to an arbitrary $\bar{r}_{i}$ and
$\gamma_{ij}$ in the minimization procedure. Then evaluating all
$\bar{r}_{i}$ and $\gamma_{ij}$ at the infinite-$D$ radius and
angle cosine, $\bar{r}_{\infty}$ and $\overline{\gamma}_\infty$,
respectively, we find that $\bar{r}_{\infty}$ and
$\overline{\gamma}_\infty$ satisfy
\begin{eqnarray}
\label{minimum1} \left[ \frac{\partial
\bar{V}_{\mathtt{eff}}(\bar{r},\gamma;\delta)}{\partial
\bar{r}_{i}} \right]_{\delta=0}&=&0
\\ \label{minimum2} \left[ \frac{\partial \bar{V}_{\mathtt{eff}}(\bar{r},\gamma;\delta)}{\partial
\gamma_{ij}}\right]_{\delta=0}&=&0,
\end{eqnarray}
where the $\delta=0$ subscript means to evaluate all $\bar{r}_{i}$
at $\bar{r}_{\infty}$ and all $\gamma_{ij}$ at
$\overline{\gamma}_\infty$. In scaled units the zeroth-order
($D\to\infty$) approximation for the energy becomes
\begin{equation}
\label{zeroth}
\bar{E}_{\infty}=\bar{V}_{\mathtt{eff}}(\bar{r}_{\infty},\overline{\gamma}_\infty;0).
\end{equation}
In this leading order approximation, the centrifugal-like term
that appears in $\bar{V}_{\mathtt{eff}}$, even for the ground
state, is a zero-point energy contribution required by the minimum
uncertainty principle\cite{chat}.

\section{Normal-mode analysis and the ${\mathbf{1/D}}$ first-order
quantum energy correction}\label{sec:firstorder}

At zeroth-order, the particles can be viewed as frozen in a
completely symmetric, high-$D$ configuration or simplex, which is
somewhat analogous to the Lewis structure in atomic physics
terminology. Likewise, the first-order $1/D$ correction can be
viewed as small oscillations of this structure, analogous to
Langmuir oscillations. Solving Eqs. (\ref{minimum1}) and
(\ref{minimum2}) for $\bar{r}_{\infty}$ and
$\overline{\gamma}_\infty$ gives the infinite-$D$ structure and
zeroth-order energy and provides the starting point for the $1/D$
expansion. To obtain the $1/D$ quantum correction to the energy
for large but finite values of $D$, we expand about the minimum of
the $D\to\infty$ effective potential. We first define a position
vector, consisting of all $N(N+1)/2$ internal coordinates:
\begin{equation}\label{eq:ytranspose}
{\bar{\bm{y}}} = \left( \begin{array}{c} \bar{\bm{r}} \\
\bm{\gamma} \end{array} \right) \,,
\end{equation}
where
\begin{equation}
\bar{\bm{r}} = \left(
\begin{array}{c}
\bar{r}_1 \\
\bar{r}_2 \\
\vdots \\
\bar{r}_N
\end{array}
\right)
\end{equation}
and
\begin{equation}
\bm{\gamma} = \left(
\begin{array}{c}
\gamma_{12} \\ \cline{1-1}
\gamma_{13} \\
\gamma_{23} \\ \cline{1-1}
\gamma_{14} \\
\gamma_{24} \\
\gamma_{34} \\ \cline{1-1}
\gamma_{15} \\
\gamma_{25} \\
\vdots \\
\gamma_{N-2,N} \\
\gamma_{N-1,N} \end{array} \right) \,.
\end{equation}
We then make the following substitutions for all
radii and angle cosines:
\begin{eqnarray}
\label{eq:taylor1}
&&\bar{r}_{i} = \bar{r}_{\infty}+\delta^{1/2}\bar{r}'_{i}\\
&&\gamma_{ij} =
\overline{\gamma}_{\infty}+\delta^{1/2}\overline{\gamma}'_{ij}
\label{eq:taylor2},
\end{eqnarray}
where $\delta=1/D$ is the expansion parameter. Now in all
practical situations $\bar{V}_{\mathtt{eff}}(\bar{y}; \delta)$ is
a function of $\bar{y}$ and $\delta$ and so we may then obtain a
power series in $\delta^{1/2}$ of the effective potential about
the $D\to\infty$ symmetric minimum as
\begin{eqnarray}
\label{eq:Taylor1} \bar{V}_{\mathtt{eff}}(\bar{y}; \delta) &=&
\left[ \bar{V}_{\mathtt{eff}} \right]_{\delta^{1/2}=0} + \nonumber
\\
&&+ \delta^{1/2} \left\{\left[ \left. \frac{\partial
\bar{V}_{\mathtt{eff}}}{\partial \delta} \right|_{\bar{y}_{\mu}}
\right]_{\delta^{1/2}=0} \left[ \frac{\partial \delta}{\partial
\delta^{1/2}} \right]_{\delta^{1/2}=0} + \sum\limits_{\mu=1}^{P}
\left[ \left. \frac{\partial \bar{V}_{\mathtt{eff}}}{\partial
\bar{y}_{\mu}} \right|_{\delta^{1/2}} \right]_{\delta^{1/2}=0}
\left[ \frac{\partial \bar{y}_{\mu}}{\partial \delta^{1/2}}
\right]_{\delta^{1/2}=0} \right\} + \nonumber \\
&& + \frac{1}{2} \delta \left\{ \left[ \left. \frac{\partial
\bar{V}_{\mathtt{eff}}}{\partial \delta} \right|_{\bar{y}_{\mu}}
\right]_{\delta^{1/2}=0} \left[ \frac{\partial^2 \delta}{(\partial
\delta^{1/2})^2 } \right]_{\delta^{1/2}=0} + \left[ \left.
\frac{\partial^2 \bar{V}_{\mathtt{eff}}}{\partial \delta^2}
\right|_{\bar{y}_{\mu}} \right]_{\delta^{1/2}=0} \left[
\frac{\partial \delta}{\partial \delta^{1/2} }
\right]_{\delta^{1/2}=0}^2 + \right. \nonumber \\
&& \left. + \sum\limits_{\mu=1}^{P} \left( 2\left[ \left.
\frac{\partial}{\partial \delta} \right|_{\bar{y}_{\mu}}
\left(\left. \frac{\partial \bar{V}_{\mathtt{eff}}}{\partial
\bar{y}_{\mu}} \right|_{\delta^{1/2}} \right)
\right]_{\delta^{1/2}=0} \left[ \frac{\partial
\bar{y}_{\mu}}{\partial \delta^{1/2}} \right]_{\delta^{1/2}=0}
\left[ \frac{\partial \delta}{\partial \delta^{1/2}}
\right]_{\delta^{1/2}=0} \right. \right. + \nonumber \\
&& \left. + \left[ \left. \frac{\partial
\bar{V}_{\mathtt{eff}}}{\partial \bar{y}_{\mu}}
\right|_{\delta^{1/2}} \right]_{\delta^{1/2}=0} \left[
\frac{\partial^2 \bar{y}_{\mu}}{(\partial \delta^{1/2})^2}
\right]_{\delta^{1/2}=0}
 \vphantom{2\left[ \left.
\frac{\partial}{\partial \delta} \right|_{\bar{y}_{\mu}}
\left(\left. \frac{\partial \bar{V}_{\mathtt{eff}}}{\partial
\bar{y}_{\mu}} \right|_{\delta^{1/2}} \right)
\right]_{\delta^{1/2}=0}} \right) + \nonumber \\
&& \left. + \sum\limits_{\mu=1}^{P} \sum\limits_{\nu=1}^{P} \left[
\frac{\partial \bar{y}_{\mu}}{\partial \delta^{1/2}}
\right]_{\delta^{1/2}=0} \left[\frac{\partial^2
\bar{V}_{\mathtt{eff}}}{\partial \bar{y}_{\mu}
\partial \bar{y}_{\nu}}\right]_{\delta^{1/2}=0} \left[ \frac{\partial \bar{y}_{\nu}}{\partial \delta^{1/2}}
\right]_{\delta^{1/2}=0} \right\} + O\left(\delta^{3/2}\right),
\end{eqnarray}
where
\begin{equation}
P \equiv N(N+1)/2
\end{equation}
is the number of internal coordinates. The $O((\delta^{1/2})^0)$
term in the power series (Eq.~\ref{Taylor}) is simply the
zeroth-order energy (Eq.~\ref{zeroth}). The $O((\delta^{1/2})^1)$
term is zero since
\begin{equation}
\left[ \frac{\partial \delta}{\partial \delta^{1/2}}
\right]_{\delta^{1/2}=0} = 0 \end{equation}
and we are expanding about the minimum of the effective potential
so
\begin{equation}
\left[ \left. \frac{\partial \bar{V}_{\mathtt{eff}}}{\partial
\bar{y}_{\mu}} \right|_{\delta^{1/2}} \right]_{\delta^{1/2}=0} = 0
\,,
\end{equation}
i.e.\ the system is said to be in equilibrium since the forces
acting on the system vanish [Eqs. (\ref{minimum1}) and
(\ref{minimum2})].

Defining a displacement vector consisting of the internal
displacement coordinates [primed in Eqs. (\ref{eq:taylor1}) and
(\ref{eq:taylor2})]
\begin{equation}\label{eq:ytransposeP}
{\bar{\bm{y}}'} = \left( \begin{array}{c} \bar{\mathbf{r}}' \\
\overline{\bm{\gamma}}' \end{array} \right) \,,
\end{equation}
where
\begin{equation}\label{eq:bfrp}
\bar{\bm{r}}' = \left(
\begin{array}{c}
\bar{r}'_1 \\
\bar{r}'_2 \\
\vdots \\
\bar{r}'_N
\end{array}
\right)
\end{equation}
and
\begin{equation}\label{eq:bfgammap}
\overline{\bm{\gamma}}' = \left(
\begin{array}{c}
\overline{\gamma}'_{12} \\ \cline{1-1}
\overline{\gamma}'_{13} \\
\overline{\gamma}'_{23} \\ \cline{1-1}
\overline{\gamma}'_{14} \\
\overline{\gamma}'_{24} \\
\overline{\gamma}'_{34} \\ \cline{1-1}
\overline{\gamma}'_{15} \\
\overline{\gamma}'_{25} \\
\vdots \\
\overline{\gamma}'_{N-2,N} \\
\overline{\gamma}'_{N-1,N} \end{array} \right) \,,
\end{equation}
we then have
\begin{equation}
\left[ \frac{\partial \bar{y}_{\mu}}{\partial \delta^{1/2}}
\right]_{\delta^{1/2}=0} = \bar{y}'_{\mu} \end{equation}
and also defining
\begin{equation}
v_o = \left[ \left. \frac{\partial
\bar{V}_{\mathtt{eff}}}{\partial \delta} \right|_{\bar{y}_{\mu}}
\right]_{\delta^{1/2}=0}\,,
\end{equation}
Eq.~(\ref{eq:Taylor1}) becomes
\begin{equation}
\label{Taylor}
\bar{V}_{\mathtt{eff}}({\bar{\bm{y}}'};\delta)=\left[
\bar{V}_{\mathtt{eff}} \right]_{\delta^{1/2}=0} + \frac{1}{2} \,
\delta \left\{ \sum\limits_{\mu=1}^{P} \sum\limits_{\nu=1}^{P}
\bar{y}'_{\mu} \left[\frac{\partial^2
\bar{V}_{\mathtt{eff}}}{\partial \bar{y}_{\mu}
\partial \bar{y}_{\nu}}\right]_{\delta^{1/2}=0} \bar{y}'_{\nu} + v_o \right\} +
O\left(\delta^{3/2}\right),
\end{equation}
The first term of the $O((\delta^{1/2})^2)$ term defines the
elements of the Hessian matrix\cite{strang} ${\bf F}$ of
Eq.~(\ref{Gham}) below. The derivative terms in the kinetic energy
are taken into account by a similar series expansion, beginning
with a first-order term that is bilinear in ${\partial/\partial
\bar{y}'}$, i.e.\,
\begin{equation}\label{eq:T}
{\mathcal T}=-\frac{1}{2} \delta \sum\limits_{\mu=1}^{P}
\sum\limits_{\nu=1}^{P} {G}_{\mu\nu}
\partial_{\bar{y}'_{\mu}}
\partial_{\bar{y}'_{\nu}} + O\left(\delta^{3/2}\right),
\end{equation}
where ${\mathcal T}$ is the derivative portion of the kinetic
energy $T$ (see Eq.~(\ref{eq:SE_T})).  Thus, obtaining the
first-order energy correction is reduced to a harmonic problem,
which is solved by obtaining the normal modes of the system.

We use the Wilson FG matrix method\cite{dcw} to obtain the
normal-mode vibrations and, thereby, the first-order energy
correction. It follows from Eqs. (\ref{Taylor}) and (\ref{eq:T})
that $\bm{G}$ and ${\bf F}$, both constant matrices, are defined
in the first-order $\delta=1/D$ Hamiltonian as follows:
\begin{equation}\label{Gham}
\widehat{H}_1=-\frac{1}{2} {\partial_{\bar{y}'}}^{T} {\bm G}
{\partial_{\bar{y}'}} + \frac{1}{2} \bar{\bm{y}}^{\prime T} {\bm
F} {{\bar{\bm{y}}'}} + v_o \,.
\end{equation}
After the Schr\"odinger equation (\ref{eq:SE}) has been
dimensionally scaled, the second-order derivative terms are of
order $\delta$, and by comparing these terms with the first part
of $\widehat{H}_1$, the elements of the kinetic-energy matrix
$\bm{G}$ are easily determined.  The elements of the Hessian
matrix\cite{strang}, ${F}_{\mu\nu}=\left[\frac{\partial^2
\bar{V}_{\mathtt{eff}}}{\partial \bar{y}_{\mu}
\partial \bar{y}_{\nu}}\right]_{\delta^{1/2}=0}$, on the other hand, require a bit more
effort to obtain\cite{FGpaper}.

We include a derivation of the FG matrix method in Appendix
\ref{app:wilson}, but we state here the main results of the
method. The $b^{\rm th}$ normal mode coordinate may be written as
(Eq.~(\ref{eq:qy}))
\begin{equation} \label{eq:qyt}
[{\bm q'}]_b = {\bm{b}}^T {\bar{\bm{y}}'} \,,
\end{equation}
where the coefficient vector ${\bm{b}}$ satisfies the eigenvalue
equation (Eq.~(\ref{eq:FGi}))
\begin{equation} \label{eq:FGit}
{\bf F} \, \bm{G} \, {\bm{b}} = \lambda_b \, {\bm{b}}
\end{equation}
with the resultant secular equation (Eq.~(\ref{eq:detFG}))
\begin{equation} \label{eq:character}
\det({\bf F}\bm{G}-\lambda{\bf I})=0.
\end{equation}
The coefficient vector also satisfies the normalization condition
(Eq.~(\ref{eq:normi}))
\begin{equation} \label{eq:normit}
{\bm{b}}^T \bm{G} \, {\bm{b}} = 1.
\end{equation}
As can be seen from Eq.~(\ref{eq:appH1}) in Appendix
\ref{app:wilson} the frequencies are given by
\begin{equation}\label{eq:omega_b}
\lambda_b=\bar{\omega}_b^2,
\end{equation}
while the wave function is a product of $P = N(N+1)/2$ harmonic
oscillator wave functions
\begin{equation}
\Phi_0({\bar{\bm{y}}'}) = \prod_{b=1}^{P} h_{n_b}\left(
\bar{\omega}^{1/2}_b [{\bm q'}]_b \right) \,, \label{eq:Phi_0}
\end{equation}
where $h_{n_b}\left( \bar{\omega}^{1/2}_b [{\bm q'}]_b \right)$ is
a one-dimensional harmonic-oscillator wave function of frequency
$\bar{\omega}_b$\,, and $n_{b}$ is the oscillator quantum number,
$0 \leq n_{b} < \infty$, which counts the number of quanta in each
normal mode.

Depending on the number of particles, the number of roots
$\lambda$ of Eq.~(\ref{eq:character}) -- there are $P \equiv
N(N+1)/2$ roots -- is potentially huge. However, due to the $S_N$
symmetry (see Ref.~\cite{hamermesh} and App.~\ref{app:Char}) of
the problem (see Sect.~\ref{sec:symm}), there is a remarkable
simplification. Equation~(\ref{eq:character}) has only five
distinct roots, $\lambda_{\mu}$, where $\mu$ runs over ${\bf
0}^-$, ${\bf 0}^+$, ${\bf 1}^-$, ${\bf 1}^+$, and ${\bf 2}$,
regardless of the number of particles in the system (see
Refs.~\cite{FGpaper, loeser} and Sect.~\ref{subsec:eigreduct}).
Thus the energy through first-order (see Eq.~(\ref{eq:E1})) can be
written in terms of the five distinct normal-mode vibrational
frequencies, which are related to the roots $\lambda_{\mu}$ of
$\bm{FG}$ by
\begin{equation}\label{eq:omega_p}
\lambda_{\mu}=\bar{\omega}_{\mu}^2 \,.
\end{equation}

The energy through first-order in $\delta=1/D$ is then\cite{FGpaper}
\begin{equation}
\overline{E} = \overline{E}_{\infty} + \delta \Biggl[
\sum_{\renewcommand{\arraystretch}{0}
\begin{array}[t]{r@{}l@{}c@{}l@{}l} \scriptstyle \mu = \{
  & \scriptstyle \bm{0}^\pm,\hspace{0.5ex}
  & \scriptstyle \bm{1}^\pm & , & \\
  & & \scriptstyle \bm{2} & & \scriptstyle  \}
            \end{array}
            \renewcommand{\arraystretch}{1} }
\hspace{-0.50em} \sum_{\mathsf{n}_{\mu}=0}^\infty
({\mathsf{n}}_{\mu}+\frac{1}{2}) d_{\mu,\mathsf{n}_{\mu}}
\bar{\omega}_{\mu} \, + \, v_o \Biggr] \,, \label{eq:E1}
\end{equation}
where the $\mathsf{n}_{\mu}$ are the vibrational quantum numbers
of the normal modes of the same frequency $\bar{\omega}_{\mu}$ (as
such, $\mathsf{n}_{\mu}$ counts the number of nodes in a given
normal mode). The quantity $d_{\mu,\mathsf{n}_{\mu}}$ is the
occupancy of the manifold of normal modes with vibrational quantum
number $\mathsf{n}_{\mu}$ and normal mode frequency
$\bar{\omega}_{\mu}$, i.e.\ it is the number of normal modes with
the same frequency $\bar{\omega}_{\mu}$ and the same number of
quanta $\mathsf{n}_{\mu}$.  The total occupancy of the normal
modes with frequency $\bar{\omega}_{\mu}$ is equal to the
multiplicity of the root $\lambda_{\mu}$, i.e.\
\begin{equation}
d_{\mu} = \sum_{\mathsf{n}_{\mu}=0}^\infty
            d_{\mu,\mathsf{n}_{\mu}} \,,
\end{equation}
where $d_{\mu}$ is the multiplicity of the $\mu^{th}$ root. The
multiplicities of the five roots are\cite{FGpaper}
\begin{eqnarray}
d_{{\bf 0}^+} &=& 1 \,,\nonumber\\
d_{{\bf 0}^-} &=& 1 \,,\nonumber\\
d_{{\bf 1}^+} &=& N-1 \,,\\
d_{{\bf 1}^-} &=& N-1 \,,\nonumber\\
d_{{\bf 2}} &=& N(N-3)/2 \,.\nonumber
\end{eqnarray}
Note that although the equation in Ref.~\cite{loeser} for the
energy through Langmuir order is the same as Eq.~(\ref{eq:E1}), it
is expressed a little differently (See Ref. \cite{different}).

\section{The Symmetry of the Large-${\mathbf{D}}$, ${\mathbf{N}}$-body Quantum-Confinement Problem} \label{sec:symm}

Such a high degree of degeneracy of the large-$D$, $N$-body
quantum-confinement problem indicates a high degree of symmetry.
In this section we study this in more detail.

\subsection{The Indical structure of ${\mathbf{F}}$,
${\bm{G}}$, and ${\bm{FG}}$ matrices}

The ${\bf F}$, $\bm{G}$, and $\bm{FG}$ matrices, which we
generically denote by ${\bf Q}$, are $P \times P$ matrices with
the same indical structure as $\bar{y} \bar{y}^T$:
\begin{equation}\label{eq:yTy}
\bar{y}\bar{y}^T=\left(
\begin{array}{ccccccccc}
\bar{r}'_1\bar{r}'_1     &\bar{r}'_1\bar{r}'_2  &\cdots                &\bar{r}'_1\bar{r}'_N     &\vline&\bar{r}'_1\overline{\gamma}'_{12}&\bar{r}'_1\overline{\gamma}'_{13}&\cdots&\bar{r}'_1\overline{\gamma}'_{N-1 N}\\
\bar{r}'_2\bar{r}'_1     &\bar{r}'_2\bar{r}'_2  &\cdots                &                       &\vline&\bar{r}'_2\overline{\gamma}'_{12}&\bar{r}'_2\overline{\gamma}'_{13}&\cdots&\bar{r}'_2\overline{\gamma}'_{N-1 N} \\
\vdots                 &\vdots              &\ddots                &    \vdots             &\vline&\vdots              &\vdots              & \ddots          &\vdots \\
\bar{r}'_N\bar{r}'_1     &\cdots              &                      &\bar{r}'_N\bar{r}'_N     &\vline&\bar{r}'_N\overline{\gamma}'_{12}& &\cdots&\bar{r}'_N\overline{\gamma}'_{N-1 N} \\
\hline
\overline{\gamma}'_{12}\bar{r}'_1   &\overline{\gamma}'_{12}\bar{r}'_2& \cdots               &\overline{\gamma}'_{12}\bar{r}'_N   &\vline&\overline{\gamma}'_{12}\overline{\gamma}'_{12}&\overline{\gamma}'_{12}\overline{\gamma}'_{13}&\cdots&\overline{\gamma}'_{12}\overline{\gamma}'_{N-1 N}\\
\overline{\gamma}'_{13}\bar{r}'_1   &\overline{\gamma}'_{13}\bar{r}'_2& \cdots               &\overline{\gamma}'_{12}\bar{r}'_1   &\vline&\overline{\gamma}'_{13}\overline{\gamma}'_{12}&\overline{\gamma}'_{13}\overline{\gamma}'_{13}&\cdots&\overline{\gamma}'_{13}\overline{\gamma}'_{N-1 N}\\
\vdots                 &\vdots              & \ddots               &\vdots                 &\vline&\vdots                &\vdots                &\ddots&\vdots \\
\overline{\gamma}'_{N-1 N}\bar{r}'_1& & \cdots&
\overline{\gamma}'_{N-1 N}\bar{r}'_N &\vline
&\overline{\gamma}'_{N-1
N}\overline{\gamma}'_{12}&\overline{\gamma}'_{N-1
N}\overline{\gamma}'_{13} &\cdots&\overline{\gamma}'_{N-1
N}\overline{\gamma}'_{N-1 N}
\end{array}
\right),
\end{equation}
where $\bar{y}$ is defined by Eq.~(\ref{eq:ytranspose}). The
indical structure of this matrix suggests a convenient shorthand
for referencing the elements of the ${\bf Q}$ matrices. The upper
left block of Eq.~(\ref{eq:yTy}) is an $(N \times N)$ matrix with
elements associated with $(\bar{r}'_i,\bar{r}'_j)$; hence we use
the subscript $(i,j)$ to refer to these elements. The upper right
block is an $(N \times N(N-1)/2)$ matrix with elements associated
with $(\bar{r}'_i, \overline{\gamma}'_{jk})$; hence, we use the
subscript $(i,jk)$ to refer to these elements. The lower left
block is an $(N(N-1)/2 \times N)$ matrix with elements associated
with $(\overline{\gamma}'_{ij},\bar{r}'_k)$; hence, we use the
subscript $(ij,k)$ to refer to these elements. Finally, the lower
right block is an $(N(N-1)/2 \times N(N-1)/2)$ matrix with
elements associated with
$(\overline{\gamma}'_{ij},\overline{\gamma}'_{kl})$; hence, we use
the subscript $(ij,kl)$ to refer to these elements.

\subsection{The ${\mathbf{S_N}}$ symmetry of the ${\bm{Q}}$ matrices}
As the number of particles $N$ increases, diagonalizing the $P
\times P$ $\bm{FG}$ matrix (where $P \equiv N(N+1)/2$) becomes,
prima facie, a daunting task. However, one of the advantages of
dimensional perturbation theory is the simplifications that occur
in the large-dimension limit. In particular, since we are dealing
with identical particles in a totally symmetric configuration (the
Lewis structure) in which all the particles are equivalent, the
${\bf Q}$ matrices display a high degree of symmetry with many
identical elements.  Specifically,
\begin{equation}\label{eq:GFsyma}
\begin{array}{llllll}
Q_{i,i}&=&Q_{i',i'} &\equiv& Q_a   & \\
Q_{i,j}&=&Q_{i',j'} &\equiv& Q_b   & (i \neq j) \;\; \mbox{and} \;\; (i' \neq j') \\
Q_{ij,i}&=&Q_{i'j',i'} &\equiv& Q_c & (i \neq j) \;\; \mbox{and} \;\; (i' \neq j')\\
Q_{jk,i}&=&Q_{j'k',i'} &\equiv& Q_d & (i \neq j \neq k) \;\; \mbox{and} \;\; (i' \neq j' \neq k')\\
Q_{i,ij}&=&Q_{i',i'j'} &\equiv& Q_e & (i \neq j) \;\; \mbox{and} \;\; (i' \neq j') \\
Q_{i,jk}&=&Q_{i',j'k'} &\equiv& Q_f & (i \neq j \neq k) \;\; \mbox{and} \;\; (i' \neq j' \neq k')\\
Q_{ij,ij}&=& Q_{i'j',i'j'} &\equiv& Q_g & (i \neq j) \;\; \mbox{and} \;\; (i' \neq j')\\
Q_{ij,jk}&=& Q_{i'j',j'k'} &\equiv& Q_h & (i \neq j \neq k) \;\; \mbox{and} \;\; (i' \neq j' \neq k')\\
Q_{ij,kl}&=&Q_{i'j',k'l'} &\equiv& Q_{\iota} & (i \neq j \neq k
\neq l) \;\; \mbox{and} \;\; (i' \neq j' \neq k' \neq l').
\end{array}
\end{equation}
Note the indices in the relationships above run over all particles
$(1,2,\ldots,N)$ with the exceptions noted in the far right
column.  For example, $Q_{i,j}=Q_{i',j'}\equiv Q_b$, where $(i
\neq j)$ and $(i' \neq j')$, means that all off-diagonal elements
of the upper left block (the pure radial block) of ${\bf Q}$ are
equal to the same constant $Q_b$. Similarly,
$Q_{ij,kl}=Q_{i'j',k'l'}\equiv Q_{\iota}$, where $(i \neq j \neq k
\neq l)$ and $(i' \neq j' \neq k' \neq l')$, means that any
elements of ${\bf Q}$ in the lower right block (the pure angular
block) that do not have a repeated index are all equal to the same
constant $Q_{\iota}$.

Equations~(\ref{eq:GFsyma}) imply that the $\bm{Q}$ matrices,
$\bm{F}$, $\bm{G}$, and $\bm{FG}$, are invariant under interchange
of the particle labels. Thus Eqs.~(\ref{eq:GFsyma}) show that the
system is invariant under the group of all the permutations of $N$
objects, where the objects are the particle label indices. This is
the symmetric group $S_N$ (see Appendix~\ref{app:Char}). As we
shall see, this $S_N$ invariance greatly simplifies the problem
which would otherwise be intractable for large $N$; allowing us to
solve for the $N(N+1)/2$ normal-mode coordinates, and hence the
Langmuir-order wave function. The $S_N$ symmetry is also behind
the remarkable reduction in the number of distinct frequencies we
noted at the end of Sec.~\ref{sec:firstorder}, from a possible
$N(N+1)/2$ to only five distinct frequencies.

Parenthetically, we should remark here that $\bm{G}$ and ${\bf F}$
are also symmetric matrices ($\bm{G}^T=\bm{G}$ and ${\bf F}^T={\bf
F}$); however, while $\bm{FG}$ does display the high degree of
symmetry of Eqs.~(\ref{eq:GFsyma}), it is not a symmetric matrix.

\subsection{Q matrices in terms of simple submatrices}
\label{subsec:Qsubm} The $S_N$ symmetry of the ${\bf Q}$ matrices
($\bm{F}$, $\bm{G}$, and $\bm{FG}$) described in
Eqs.~(\ref{eq:GFsyma}) allows us to write these matrices in terms
of six simple submatrices. We first define the number of
$\overline{\gamma}'_{ij}$ coordinates to be
\begin{equation}\label{eq:M}
M \equiv N(N-1)/2,
\end{equation}
and let ${\bf I}_N$ be an $N \times N$ identity matrix, ${\bf
I}_M$ an $M \times M$ identity matrix, ${\bf J}_N$ an $N \times N$
matrix of ones and ${\bf J}_M$ an $M \times M$ matrix of ones.
Further, we let ${\bf R}$ be an $N \times M$ matrix such that
${R}_{i,jk}=\delta_{ij}+\delta_{ik}$, ${\bf J}_{NM}$ be an $N
\times M$ matrix of ones, and ${\bf J}^T_{NM}={\bf J}_{MN}$.

We then write the ${\bf Q}$ matrices as
\begin{equation}\label{eq:Q}
{\bf Q}=\left(\begin{array}{cc} {\bf Q}_{\bar{\bm{r}}'
\bar{\bm{r}}'} & {\bf Q}_{\bar{\bm{r}}' \overline{\bm{\gamma}}'}
\\ {\bf Q}_{\overline{\bm{\gamma}}' \bar{\bm{r}}'} & {\bf
Q}_{\overline{\bm{\gamma}}' \overline{\bm{\gamma}}'}
\end{array}\right),
\end{equation}
where the block ${\bf Q}_{\bar{\bm{r}}' \bar{\bm{r}}'}$ has
dimension $(N \times N)$, block ${\bf Q}_{\bar{\bm{r}}'
\overline{\bm{\gamma}}'}$ has dimension $(N \times M)$, block
${\bf Q}_{\overline{\bm{\gamma}}' \bar{\bm{r}}'}$ has dimension
$(M \times N)$, and block ${\bf Q}_{\overline{\bm{\gamma}}'
\overline{\bm{\gamma}}'}$ has dimension $(M \times M)$.  Now, as
we show in Appendix~B of Ref.~\cite{FGpaper},
Eqs.~(\ref{eq:GFsyma}) allows us to write the following:
\begin{eqnarray}
{\bf Q}_{\bar{\bm{r}}'
\bar{\bm{r}}'} & = & (Q_a-Q_b) {\bf I}_N + Q_b {\bf J}_N \label{eq:Qrr}\\
{\bf Q}_{\bar{\bm{r}}'
\overline{\bm{\gamma}}'} & = & (Q_e-Q_f) {\bf R} + Q_f {\bf J}_{NM} \label{eq:Qrg} \\
{\bf Q}_{\overline{\bm{\gamma}}' \bar{\bm{r}}'}
& = & (Q_c-Q_d) {\bf R}^T + Q_d {\bf J}_{NM}^T \label{eq:Qgr} \\
{\bf Q}_{\overline{\bm{\gamma}}' \overline{\bm{\gamma}}'} & = &
(Q_g-2Q_h+Q_{\iota}) {\bf I}_M + (Q_h-Q_{\iota}) {\bf R}^T {\bf R}
+ Q_{\iota} {\bf J}_M \,. \label{eq:Qgg}
\end{eqnarray}
In particular, letting ${\bf Q}=\bm{FG}$, the matrix that must be
diagonalized, Eq.~(\ref{eq:Q}) becomes
\begin{equation} \label{GFsub}
\bm{FG}= \left(
\begin{array}{cc}
\tilde{a} {\bf I}_N + \tilde{b} {\bf J}_N & \tilde{e} {\bf R} + \tilde{f} {\bf J}_{NM} \\
\tilde{c} {\bf R}^T + \tilde{d} {\bf J}_{MN} & \tilde{g} {\bf I}_M
+ \tilde{h} {\bf R}^T {\bf R} + \tilde{\iota} {\bf J}_M
\end{array}\right) \,,
\end{equation}
where we have used the following abbreviations:
\begin{eqnarray}  \label{GFsym}
\tilde{a} & \equiv & ({FG})_{a} - ({FG})_{b} = ({F}_{a} - {F}_{b}) {G}_{a}   \nonumber \\
\tilde{b} & \equiv & ({FG})_{b} = {F}_{b}{G}_{a}   \nonumber\\
\tilde{c} & \equiv & ({FG})_{c} - ({FG})_{d} = ({F}_{e} - {F}_{f}) {G}_{a}   \nonumber \\
\tilde{d} & \equiv & ({FG})_{d} = {F}_{f}{G}_{a}   \nonumber\\
\tilde{e} & \equiv & ({FG})_{e} - ({FG})_{f} = ({F}_{e} - {F}_{f}) ({G}_{g}+(N-4){G}_{h}) \\
\tilde{f} & \equiv & ({FG})_{f} = 2{F}_{e}{G}_{h} +  {F}_{f}({G}_{g} + 2(N-3){G}_{h})    \nonumber \\
\tilde{g} & \equiv & ({FG})_{g} - 2({FG})_{h} + ({FG})_{\iota}
= ({F}_{g} - 2{F}_{h} + {F}_{\iota}) ({G}_{g} - 2{G}_{h}) \nonumber \\
\tilde{h} & \equiv & ({FG})_{h} - ({FG})_{\iota} =
{F}_{g}{G}_{h}+{F}_{h}({G}_{g}+(N-6){G}_{h})
- {F}_{\iota}({G}_{g} + (N-5){G}_{h}) \nonumber\\
\tilde{\iota} & \equiv & ({FG})_{\iota} =
4{F}_{h}{G}_{h}+{F}_{\iota}({G}_{g}+2(N-4){G}_{h}) \,. \nonumber
\end{eqnarray}
The right-hand sides of Eq.~(\ref{GFsym}), the $\bm{FG}$ matrices
expressed in terms of the ${\bf F}$ and $\bm{G}$ matrix elements,
may be derived using the graph-theoretic techniques discussed in
Appendix~B of Ref.~\cite{FGpaper}.

We also require the $\bm{G}$ matrix for the normalization
condition (Eq.~(\ref{eq:normit})). It has a simpler structure than
the $\bm{FG}$ matrix,
\begin{equation} \label{eq:Gsub}
\bm{G} = \left( \begin{array}{cc}
\tilde{a}' {\bf I}_N & \bm{0} \\
\bm{0} & \tilde{g}' {\bf I}_M + \tilde{h}' {\bf R}^T {\bf R}
\end{array} \right) \,,
\end{equation}
where
\begin{eqnarray}  \label{eq:Gsym}
\tilde{a}' & \equiv & ({G})_{a} \nonumber \\
\tilde{g}' & \equiv & ({G})_{g} - 2({G})_{h} \\
\tilde{h}' & \equiv & ({G})_{h} \nonumber
\end{eqnarray}
and
\begin{eqnarray} \label{eq:Goneorzero}
({G})_{b} & = & 0 \nonumber \\
({G})_{c} & = & 0 \nonumber \\
({G})_{d} & = & 0  \\
({G})_{e} & = & 0 \nonumber \\
({G})_{f} & = & 0 \nonumber \\
({G})_{\iota} & = & 0 \,. \nonumber
\end{eqnarray}
The quantities $({G})_{a}$, $({G})_{g}$ and $({G})_{h}$ depend on
the choice of $\kappa(D)$ (See Ref.~\cite{FGpaper}).

\section{Symmetry and Normal Coordinates} \label{sec:symnorm}
The $\bm{FG}$ matrix is a $N(N+1)/2 \times N(N+1)/2$ dimensional
matrix (there being $N(N+1)/2$ internal coordinates), and so
Eqs.~(\ref{eq:FGit}) and (\ref{eq:character}) could have up to
$N(N+1)/2$ distinct frequencies. However, as noted above, there
are only five distinct frequencies. The $S_N$ symmetry is
responsible for the remarkable reduction from $N(N+1)/2$ possible
distinct frequencies to five actual distinct frequencies. As we
shall also see, the $S_N$ symmetry greatly simplifies the
determination of the normal coordinates and hence the solution of
the large-$D$ problem.

\subsection{Symmetrized coordinates} \label{subsec:symCoor}
To understand why there are only five distinct frequencies and to
start the process of solving for the normal coordinates, we first
look at the $S_N$ transformation properties of the internal
coordinates.

We first note that the internal-coordinate displacement vectors
$\bar{\bm{r}}'$ and $\overline{\bm{\gamma}}'$ of
Eqs.~(\ref{eq:bfrp}) and (\ref{eq:bfgammap}) are basis functions
which transform under matrix representations of $S_N$ and each
span the corresponding carrier spaces. These representations of
$S_N$ however, are not irreducible representations  of $S_N$
(Appendix~\ref{app:Char}).

We will show that the reducible representation under which
$\bar{\bm{r}}'$ transforms is reducible to one $1$-dimensional
irreducible representation labelled by the partition $[N]$ (the
partition denotes a corresponding Young diagram ( = Young pattern
= Young shape) of an irreducible representation (see
Appendix~\ref{app:Char})) and one $(N-1)$-dimensional irreducible
representation labelled by the partition $[N-1, \hspace{1ex} 1]$.
We will also show that the reducible representation under which
$\overline{\bm{\gamma}}'$ transforms is reducible to one
$1$-dimensional irreducible representation labelled by the
partition $[N]$, one $(N-1)$-dimensional irreducible
representation labelled by the partition $[N-1, \hspace{1ex} 1]$
and one $N(N-3)/2$-dimensional irreducible representation labelled
by the partition $[N-2, \hspace{1ex} 2]$.

\subsubsection{The Characters of the Reducible Representations of
${\mathbf{S_N}}$ under which ${\mathbf{\overline{r}'}}$ and
${\mathbf{\overline{\bm{\gamma}}'}}$ Transform.}
\label{subsubsec:charSNrpgp} Let us denote the cycle structure of
a permutation (see Appendix~\ref{app:Char}) by the symbol
$(1^{\nu_1},2^{\nu_2},3^{\nu_3},\ldots,N^{\nu_N})$, where the
notation $j^{\nu_j}$ means a cycle of length $j$ and $\nu_j$
equals the number of cycles of length $j$ in that
permutation\cite{hamermesh}. Crucially the characters,
$\chi^{(1^{\nu_1},2^{\nu_2},\ldots,N^{\nu_N})}$, for all elements
of a matrix representation of $S_N$ corresponding to permutations
with the same cycle structure
$(1^{\nu_1},2^{\nu_2},3^{\nu_3},\ldots,N^{\nu_N})$, are identical
(Appendix~\ref{app:Char}).

We will use the theory of group characters to decompose
$\bar{\bm{r}}'$ and $\overline{\bm{\gamma}}'$ into basis functions
which transform under irreducible representations of $S_N$. The
theory of group characters is briefly discussed in
Appendix~\ref{app:Char}. To determine the characters,
$\chi_{\bar{\bm{r}}'}^{(1^{\nu_1},2^{\nu_2},\ldots,N^{\nu_N})}$,
of the representation under which $\bar{\bm{r}}'$ transforms we
need to calculate how many of the elements $\bar{\bm{r}}'$
transform into themselves under a permutation in the class
$(1^{\nu_1},2^{\nu_2},3^{\nu_3},\ldots,N^{\nu_N})$. As any number
occurring in a cycle of length greater than one is transformed
into another number, only those elements of $\bar{\bm{r}}'$,
$r_i$, for which the index $i$ occurs in a cycle of length one
transform into themselves. As there are $\nu_1$ cycles of length
one, there will also be $\nu_1$ elements of $\bar{\bm{r}}'$ which
transform into themselves. Hence the character of the
representation under which $\bar{\bm{r}}'$ transforms under a
permutation in the class
$(1^{\nu_1},2^{\nu_2},3^{\nu_3},\ldots,N^{\nu_N})$ is given by
\begin{equation}\label{eq:chi_r}
\chi_{\bar{\bm{r}}'}^{(1^{\nu_1},2^{\nu_2},\ldots,N^{\nu_N})} =
\nu_1 \,.
\end{equation}

The derivation of the characters,
$\chi_{\overline{\bm{\gamma}}'}^{(1^{\nu_1},2^{\nu_2},\ldots,N^{\nu_N})}$,
under which the $\overline{\bm{\gamma}}'$ transforms is similar.
For a given permutation in the class
$(1^{\nu_1},2^{\nu_2},3^{\nu_3},\ldots,N^{\nu_N})$ all pairs of
indices occurring in  cycle of length one correspond to an element
of $\overline{\bm{\gamma}}'$, $\overline{\gamma}'_{ij}$, for which
the indices $i$ or $j$ transform into themselves. There will be
$\nu_1(\nu_1-1)/2$ such pairs. Also every pair of indices
occurring in a cycle of length two correspond to an element of
$\overline{\bm{\gamma}}'$, $\overline{\gamma}'_{ij}$, for which
the indices $i$ and $j$ transform into themselves. Thus the
element transforms into itself. There are $\nu_2$ such pairs. Thus
the
$\chi_{\overline{\bm{\gamma}}'}^{(1^{\nu_1},2^{\nu_2},\ldots,N^{\nu_N})}$,
under which the $\overline{\bm{\gamma}}'$ transforms is the sum of
these two numbers, i.e.\
\begin{equation}\label{eq:chi_gamma}
\chi_{\overline{\bm{\gamma}}'}^{(1^{\nu_1},2^{\nu_2},\ldots,N^{\nu_N})}
= \frac{\nu_1(\nu_1-1)}{2}+\nu_2 \,.
\end{equation}

\subsubsection{The Characters of the Irreducible Representations of
${\mathbf{S_N}}$.} \label{subsubsec:charSN} Now let's turn to the
characters of the irreducible representations. It can be
shown\cite{hamermesh} that for an element of an irreducible matrix
representation which corresponds to a particular permutation with
cycle structure
$(1^{\nu_1},2^{\nu_2},3^{\nu_3},\ldots,N^{\nu_N})$, the characters
of the $[N]$, $[N-1, \hspace{1ex} 1]$ and $[N-2, \hspace{1ex} 2]$
representations are:
\begin{eqnarray}\label{eq:chi_S_N}
\chi_{[N]}^{(1^{\nu_1},2^{\nu_2},\ldots,N^{\nu_N})}& = & 1\nonumber\\
\chi_{[N-1, \hspace{1ex} 1]}^{(1^{\nu_1},2^{\nu_2},\ldots,N^{\nu_N})}& = &\nu_1-1\\
\chi_{[N-2, \hspace{1ex}
2]}^{(1^{\nu_1},2^{\nu_2},\ldots,N^{\nu_N})}& =
&\frac{(\nu_1-1)(\nu_1-2)}{2}+\nu_2-1 \,. \nonumber
\end{eqnarray}

Armed with these characters, we can now see which irreducible
representations of $S_N$ $\bar{\bm{r}}'$ and
$\overline{\bm{\gamma}}'$ span. In Appendix~\ref{app:Char} we note
that the characters of the representations under which
$\bar{\bm{r}}'$ and $\overline{\bm{\gamma}}'$ transform can be
written as
\begin{equation}\label{eq:chi_a_alphat}
\chi(R) = \sum_{\alpha} a_\alpha \chi_\alpha(R) \,,
\end{equation}
where $R$ denotes the element of the group,  $\chi(R)$ is the
character of the representation under which $\bar{\bm{r}}'$ or
$\overline{\bm{\gamma}}'$ transform, $\alpha$ labels the
irreducible representation of $S_N$\,, and $\chi_\alpha(R)$ is the
character of the irreducible representation denoted by $\alpha$.
The decomposition of Eq.~(\ref{eq:chi_a_alphat}) is unique where
the coefficient $a_\alpha$ is the number of times the irreducible
representation labelled by $\alpha$ appears in the decomposition
of the representation under which $\bar{\bm{r}}'$ or
$\overline{\bm{\gamma}}'$ transform.

Thus the $\bar{\bm{r}}'$ vector must span a $[N]$ and a $[N-1,
\hspace{1ex} 1]$ irreducible representation as from
Eqs.~(\ref{eq:chi_r}) and (\ref{eq:chi_S_N})
\begin{equation}
\chi_{\bar{\bm{r}}'}^{(1^{\nu_1},2^{\nu_2},\ldots,N^{\nu_N})} = 1
\times \chi_{[N]}^{(1^{\nu_1},2^{\nu_2},\ldots,N^{\nu_N})} + 1
\times \chi_{[N-1, \hspace{1ex}
1]}^{(1^{\nu_1},2^{\nu_2},\ldots,N^{\nu_N})} \,,
\end{equation}
i.e.\
\begin{eqnarray}
a_{[N]} & = & 1 \\
a_{[N-1, \hspace{1ex} 1]} & = & 1 \,.
\end{eqnarray}
The $\overline{\bm{\gamma}}'$ vector spans a $[N]$, a $[N-1,
\hspace{1ex} 1]$ and a $[N-2, \hspace{1ex} 2]$ irreducible
representation as from Eqs.~(\ref{eq:chi_gamma}) and
(\ref{eq:chi_S_N})
\begin{equation}
\chi_{\overline{\bm{\gamma}}'}^{(1^{\nu_1},2^{\nu_2},\ldots,N^{\nu_N})}
= 1 \times \chi_{[N]}^{(1^{\nu_1},2^{\nu_2},\ldots,N^{\nu_N})} + 1
\times \chi_{[N-1, \hspace{1ex}
1]}^{(1^{\nu_1},2^{\nu_2},\ldots,N^{\nu_N})} + 1 \times
\chi_{[N-2, \hspace{1ex}
2]}^{(1^{\nu_1},2^{\nu_2},\ldots,N^{\nu_N})} \,,
\end{equation}
i.e.\
\begin{eqnarray}
a_{[N]} & = & 1 \\
a_{[N-1, \hspace{1ex} 1]} & = & 1 \\
a_{[N-2, \hspace{1ex} 2]} & = & 1 \,.
\end{eqnarray}

Regarding the dimensionalities of the $[N]$, $[N-1, \hspace{1ex}
1]$ and $[N-2, \hspace{1ex} 2]$ irreducible representations there
are many rules to determine these. The best known is that the
dimensionality of an irreducible representation of $S_N$ is equal
to the number of standard Young tableaux that the Young diagram
for the irreducible representation allows. Perhaps the simplest
rule however\cite{grouptheory}, is to fill the boxes of the Young diagram
with numbers determined as follows. To the number of boxes in the
row to the right of a box add the number of boxes in the column
below the same box and then add one for the box itself. Once all
of the boxes of the Young tableau have been filled with a number
in this fashion, multiply all of the numbers in the boxes
together. If we denote this product by $N_T$, then the
dimensionality, $d$, of the representation is
\begin{equation}
d = \frac{N!}{N_T}
\end{equation}
from which we deduce that
\begin{equation}
\renewcommand{\arraystretch}{1.5}
\begin{array}{rcccl}
d_{[N]} & = & \displaystyle \frac{N!}{N!} & = & 1 \\
d_{[N-1, \hspace{1ex} 1]} & = & \displaystyle \frac{N!}{[N!/(N-1)]} & = & N-1 \\
d_{[N-2, \hspace{1ex} 2]} & = & \displaystyle
\frac{N!}{[2N!/\left( N(N-3) \right) ]} & = & \displaystyle
\frac{N(N-3)}{2} \,.
\end{array}
\renewcommand{\arraystretch}{1}
\end{equation}
We note that
\begin{equation}
d_{[N]} + d_{[N-1, \hspace{1ex} 1]} = N \,,
\end{equation}
giving correctly the dimension of the $\bar{\bm{r}}'$ vector and
that
\begin{equation}
d_{[N]} + d_{[N-1, \hspace{1ex} 1]} + d_{[N-2, \hspace{1ex} 2]} =
\frac{N(N-1)}{2} \,,
\end{equation}
giving correctly the dimension of the $\overline{\bm{\gamma}}'$
vector.

\subsubsection{The Reduction of ${\mathbf{D_{\bar{\bm{r}}'}(R)}}$ by ${\mathbf{W_{\bar{\bm{r}}'}}}$\,.}
Let $D_{\bar{\bm{r}}'}(R)$ be the matrix representation of $S_N$
under which the $\bar{\bm{r}}'$ vector transforms. The
considerations of Secs.~\ref{subsubsec:charSNrpgp} and
\ref{subsubsec:charSN} then imply that $D_{\bar{\bm{r}}'}(R)$ may
be block diagonalized under a similarity transformation to
\begin{equation} \label{eq:DSrpeq}
D_{S_{\bar{\bm{r}}'}}(R) = \left(
\begin{array}{c|c} D_{S_{\bar{\bm{r}}'}}^{[N]}(R) & \bm{0} \\ \hline \bm{0} &
D_{S_{\bar{\bm{r}}'}}^{[N-1, \hspace{1ex} 1]}(R) \,,
\end{array} \right) \,,
\end{equation}
where $D_{S_{\bar{\bm{r}}'}}^{[N]}(R)$ is the one-dimensional
irreducible matrix representation $[N]$ of $S_N$ and
$D_{S_{\bar{\bm{r}}'}}^{[N-1, \hspace{1ex} 1]}(R)$ is an $(N-1)
\times (N-1)$-dimensional irreducible matrix representation $[N-1,
\hspace{1ex} 1]$ of $S_N$.

If $W_{\bar{\bm{r}}'}$ is a matrix which effects the reduction of
$D_{\bar{\bm{r}}'}(R)$ to Eq.~(\ref{eq:DSrpeq}), then
$W_{\bar{\bm{r}}'}$ satisfies the equation
\begin{equation}
W_{\bar{\bm{r}}'} \, D_{\bar{\bm{r}}'}(R)\, W_{\bar{\bm{r}}'}^{-1}
\,\, [W_{\bar{\bm{r}}'} \bar{\bm{r}}']= D_{S_{\bar{\bm{r}}'}}(R)
\,\, [W_{\bar{\bm{r}}'} \bar{\bm{r}}'] \,.
\end{equation}
Thus we have that the symmetry coordinates,
$\bm{S}_{\bar{\bm{r}}'}$, of the $\bar{\bm{r}}'$ sector which
transform under irreducible representations of the group $S_N$ are
given by
\begin{equation}\label{eq:swr}
{\bm{S}}_{\bar{\bm{r}}'} = W_{\bar{\bm{r}}'} \bar{\bm{r}}'
\end{equation}
and so writing
\begin{equation}\label{eq:wr}
W_{\bar{\bm{r}}'} = \left( \begin{array}{l} W_{\bar{\bm{r}}'}^{[N]} \\
W_{\bar{\bm{r}}'}^{[N-1, \hspace{1ex} 1]} \end{array} \right) \,,
\end{equation}
where $W_{\bar{\bm{r}}'}^{[N]}$ is a $1 \times N$ dimensional
matrix and $W_{\bar{\bm{r}}'}^{[N-1, \hspace{1ex} 1]}$ is an
$(N-1) \times N$ dimensional matrix, then
\begin{equation}\label{eq:sr}
\bm{S}_{\bar{\bm{r}}'} = \left( \begin{array}{l} {\bm{S}}_{\bar{\bm{r}}'}^{[N]} \\
{\bm{S}}_{\bar{\bm{r}}'}^{[N-1, \hspace{1ex} 1]} \end{array} \right) = \left( \begin{array}{l} W_{\bar{\bm{r}}'}^{[N]} \bar{\bm{r}}' \\
W_{\bar{\bm{r}}'}^{[N-1, \hspace{1ex} 1]} \bar{\bm{r}}'
\end{array} \right) \,.
\end{equation}
The symmetry coordinate column vector
${\bm{S}_{\bar{\bm{r}}'}^{[N]}}$ is a one-element vector and
transforms under $D_{S_{\bar{\bm{r}}'}}^{[N]}(R)$, the
one-dimensional irreducible matrix representation $[N]$ of $S_N$.
The symmetry coordinate column vector
${\bm{S}}_{\bar{\bm{r}}'}^{[N-1, \hspace{1ex} 1]}$ is a
$(N-1)$-element vector and transforms under
$D_{S_{\bar{\bm{r}}'}}^{[N-1, \hspace{1ex} 1]}(R)$, the $(N-1)
\times (N-1)$-dimensional irreducible matrix representation $[N-1,
\hspace{1ex} 1]$ of $S_N$.

\subsubsection{Reduction of ${\mathbf{D_{\overline{\bm{\gamma}}'}(R)}}$ by
${\mathbf{W_{\overline{\bm{\gamma}}'}}}$}\,. Likewise, if
$W_{\overline{\bm{\gamma}}'}$ is a matrix which effects the
reduction of the representation, $D_{\overline{\bm{\gamma}}'}(R)$,
under which the $\overline{\bm{\gamma}}'$ vector transforms then
\begin{equation}
W_{\overline{\bm{\gamma}}'} \, D_{\overline{\bm{\gamma}}'}(R)\,
W_{\overline{\bm{\gamma}}'}^{-1} \,\, [W_{\overline{\bm{\gamma}}'}
\overline{\bm{\gamma}}']= D_{S_{\overline{\bm{\gamma}}'}}(R) \,\,
[W_{\overline{\bm{\gamma}}'} \overline{\bm{\gamma}}'] \,,
\end{equation}
where
\begin{equation}
D_{S_{\overline{\bm{\gamma}}'}}(R) = \left(
\begin{array}{c|c|c} D_{S_{\overline{\bm{\gamma}}'}}^{[N]}(R) & \bm{0} & \bm{0} \\ \hline \bm{0} &
D_{S_{\overline{\bm{\gamma}}'}}^{[N-1, \hspace{1ex} 1]}(R) &
\bm{0}
\\ \hline \bm{0} & \bm{0} & D_{S_{\overline{\bm{\gamma}}'}}^{[N-2,
\hspace{1ex} 2]}(R)
\end{array} \right) \,,
\end{equation}
$D_{S_{\overline{\bm{\gamma}}'}}^{[N]}(R)$ is the one-dimensional
irreducible matrix representation $[N]$ of $S_N$,
$D_{S_{\overline{\bm{\gamma}}'}}^{[N-1, \hspace{1ex} 1]}(R)$ is
the $(N-1) \times (N-1)$-dimensional irreducible matrix
representation $[N-1, \hspace{1ex} 1]$ of $S_N$\,, and
$D_{S_{\overline{\bm{\gamma}}'}}^{[N-2, \hspace{1ex} 2]}(R)$ is
the $\{N(N-3)/2\} \times \{N(N-3)/2\}$-dimensional irreducible
matrix representation $[N-2, \hspace{1ex} 2]$ of $S_N$.

Note that while $D_{\bar{\bm{r}}'}^{\alpha}(R)$ and
$D_{\overline{\bm{\gamma}}'}^{\alpha}(R)$ both belong to the
$\alpha$ irreducible representation of $S_N$, where $\alpha$ is
the partition $[N]$ or $[N-1, \hspace{1ex} 1]$, they are not
necessarily the same matrices. Generally they are related by an
equivalence transformation
\begin{equation}\label{eq:DralphaDgammaalpha}
D_{\bar{\bm{r}}'}^{\alpha}(R) = t \,
D_{\overline{\bm{\gamma}}'}^{\alpha}(R) \, t^{-1} \,,
\end{equation}
where $t$ is the transformation matrix.

Thus we have that the symmetry coordinates,
$\bm{S}_{\overline{\bm{\gamma}}'}$, of the
$\overline{\bm{\gamma}}'$ sector which transform under irreducible
representations of the group $S_N$ are given by
\begin{equation}\label{eq:swgamma}
{\bm{S}}_{\overline{\bm{\gamma}}'} = W_{\overline{\bm{\gamma}}'}
\, \overline{\bm{\gamma}}'
\end{equation}
and so writing
\begin{equation}\label{eq:wgamma}
W_{\overline{\bm{\gamma}}'} = \left( \begin{array}{l} W_{\overline{\bm{\gamma}}'}^{[N]} \\
W_{\overline{\bm{\gamma}}'}^{[N-1, \hspace{1ex} 1]} \\
W_{\overline{\bm{\gamma}}'}^{[N-2, \hspace{1ex} 2]}
\end{array} \right) \,,
\end{equation}
where $W_{\overline{\bm{\gamma}}'}^{[N]}$ is a $1 \times N(N-1)/2$
dimensional matrix, $W_{\overline{\bm{\gamma}}'}^{[N-1,
\hspace{1ex} 1]}$ is an $(N-1) \times N(N-1)/2$ dimensional matrix
and $W_{\overline{\bm{\gamma}}'}^{[N-2, \hspace{1ex} 2]}$ is an
$N(N-3)/2 \times N(N-1)/2$ dimensional matrix, then
\begin{equation}\label{eq:sgamma}
\bm{S}_{\overline{\bm{\gamma}}'} = \left( \begin{array}{l} {\bm{S}}_{\overline{\bm{\gamma}}'}^{[N]} \\
{\bm{S}}_{\overline{\bm{\gamma}}'}^{[N-1, \hspace{1ex} 1]} \\
{\bm{S}}_{\overline{\bm{\gamma}}'}^{[N-2, \hspace{1ex} 2]}
\end{array} \right) =
\left( \begin{array}{l} W_{\overline{\bm{\gamma}}'}^{[N]} \, \overline{\bm{\gamma}}' \\
W_{\overline{\bm{\gamma}}'}^{[N-1, \hspace{1ex} 1]} \,
\overline{\bm{\gamma}}' \\
W_{\overline{\bm{\gamma}}'}^{[N-2, \hspace{1ex} 2]} \,
\overline{\bm{\gamma}}' \end{array} \right) \,,
\end{equation}
where
\begin{equation} \label{eq:Wsum2}
W_{\overline{\bm{\gamma}}'}^\alpha \, \overline{\bm{\gamma}}' =
\sum_{j=1}^N \sum_{i < j} \,\,
[W_{\overline{\bm{\gamma}}'}^\alpha]_{ij} \,
\overline{\gamma}'_{ij} \,.
\end{equation}
The symmetry coordinate column vector
${\bm{S}_{\overline{\bm{\gamma}}'}^{[N]}}$ is a one-element vector
and transforms under $D_{S_{\overline{\bm{\gamma}}'}}^{[N]}(R)$,
the one-dimensional irreducible matrix representation $[N]$ of
$S_N$, while the symmetry coordinate column vector
${\bm{S}}_{\overline{\bm{\gamma}}'}^{[N-1, \hspace{1ex} 1]}$ is an
$(N-1)$-element vector and transforms under
$D_{S_{\overline{\bm{\gamma}}'}}^{[N-1, \hspace{1ex} 1]}(R)$, the
$(N-1) \times (N-1)$-dimensional irreducible matrix representation
$[N-1, \hspace{1ex} 1]$ of $S_N$. The symmetry coordinate column
vector ${\bm{S}}_{\overline{\bm{\gamma}}'}^{[N-2, \hspace{1ex}
2]}$ is an $N(N-3)/2$-element vector and transforms under
$D_{S_{\overline{\bm{\gamma}}'}}^{[N-2, \hspace{1ex} 2]}(R)$, the
$\{N(N-3)/2\} \times \{N(N-3)/2\}$-dimensional irreducible matrix
representation $[N-2, \hspace{1ex} 2]$ of $S_N$.

\subsubsection{The Full Symmetry Coordinate Vector
${\mathbf{S}}$.} Symmetry coordinates
${\bm{S}}_{\bar{\bm{r}}'}^{\alpha}$ and
${\bm{S}}_{\overline{\bm{\gamma}}'}^{\beta}$ are said to belong to
the same species when $\alpha=\beta$. Now in what follows we will
find it useful to form a full symmetry coordinate vector as
follows:
\begin{equation}\label{eq:FSCV}
{\bm{S}} = P \left( \begin{array}{l} \bm{S}_{\bar{\bm{r}}'} \\
\bm{S}_{\overline{\bm{\gamma}}'} \end{array} \right) = \left( \begin{array}{l} {\bm{S}}_{\bar{\bm{r}}'}^{[N]} \\
{\bm{S}}_{\overline{\bm{\gamma}}'}^{[N]}  \\ \hline {\bm{S}}_{\bar{\bm{r}}'}^{[N-1, \hspace{1ex} 1]} \\
{\bm{S}}_{\overline{\bm{\gamma}}'}^{[N-1, \hspace{1ex} 1]} \\
\hline
{\bm{S}}_{\overline{\bm{\gamma}}'}^{[N-2, \hspace{1ex} 2]} \end{array} \right) = \left( \begin{array}{l} {\bm{S}}^{[N]} \\
{\bm{S}}^{[N-1, \hspace{1ex} 1]} \\
{\bm{S}}^{[N-2, \hspace{1ex} 2]} \end{array} \right) \,,
\end{equation}
where the orthogonal matrix
\begin{equation}\label{eq:p}
P = \left(
\begin{array}{ccccc}
  1 & 0 & 0 & 0 & 0 \\
  0 & 0 & 1 & 0 & 0 \\
  0 & 1 & 0 & 0 & 0 \\
  0 & 0 & 0 & 1 & 0 \\
  0 & 0 & 0 & 0 & 1
\end{array} \right) \,,
\end{equation}
and
\begin{equation}\label{eq:srep}
\begin{array}{cc@{\mbox{\hspace{2ex}and\hspace{2ex}}}c}
{\bm{S}}^{[N]} = \left( \begin{array}{l} {\bm{S}}_{\bar{\bm{r}}'}^{[N]} \\
{\bm{S}}_{\overline{\bm{\gamma}}'}^{[N]}
\end{array} \right) \,, &
{\bm{S}}^{[N-1, \hspace{1ex} 1]} = \left( \begin{array}{l} {\bm{S}}_{\bar{\bm{r}}'}^{[N-1, \hspace{1ex} 1]} \\
{\bm{S}}_{\overline{\bm{\gamma}}'}^{[N-1, \hspace{1ex} 1]}
\end{array} \right) &
\end{array}
{\bm{S}}^{[N-2, \hspace{1ex} 2]} =
{\bm{S}}_{\overline{\bm{\gamma}}'}^{[N-2, \hspace{1ex} 2]} \,.
\end{equation}
In ${\bm{S}}$, symmetry coordinates of the same species are
grouped together.

\subsection{Normal Coordinates} \label{subsec:normcond}
So how do symmetry coordinates simplify the solution of
Eqs.~(\ref{eq:qyt}), (\ref{eq:FGit}), (\ref{eq:character}),
(\ref{eq:normit}) and (\ref{eq:omega_p}) for the frequencies and
normal modes? Symmetry coordinates have maximum utility when
\begin{enumerate}
\item\label{it:teq0} all transformation matrices $t$, which relate
the equivalent irreducible representations of the symmetry group
under which symmetry coordinates of the same species transform
(see Eq.~(\ref{eq:DralphaDgammaalpha})), are equal to the identity
matrix. The irreducible representations are not just equivalent,
they are identical.
\item \label{it:ort} the symmetry coordinates transform under
unitary representations of the symmetry group (here $S_N$). In our
case we wish to have real symmetry coordinates and normal mode
coordinates, and so we require that the symmetry coordinates
transform under orthogonal representations of $S_N$.
\end{enumerate}
Now consider applying a transformation $W$ to Eqs.~(\ref{eq:qyt}),
(\ref{eq:FGit}), (\ref{eq:character}) and (\ref{eq:normit}), where
\begin{equation}\label{eq:W}
W= P \left( \begin{array}{c|c} W_{\bar{\bm{r}}'} & \bm{0}
\\ \hline \bm{0} & W_{\overline{\bm{\gamma}}'}
\end{array} \right) =
\left( \begin{array}{cc} W_{\bar{\bm{r}}'}^{[N]} & \bm{0} \\
\bm{0} & W_{\overline{\bm{\gamma}}'}^{[N]} \\
\hline
W_{\bar{\bm{r}}'}^{[N-1, \hspace{1ex} 1]} & \bm{0} \\
\bm{0} & W_{\overline{\bm{\gamma}}'}^{[N-1, \hspace{1ex} 1]} \\
\hline \bm{0} & W_{\overline{\bm{\gamma}}'}^{[N-2, \hspace{1ex}
2]}
\end{array} \right) \,,
\end{equation}
where
\begin{equation}\label{eq:SWy}
{\bm{S}} = W \, {\bar{\bm{y}}'} \,.
\end{equation}

\subsubsection{The Essential Equation and the Orthogonality of ${\mathbf{W}}$\,.}
In Appendix~\ref{app:proofOrthW} we show that
condition~\ref{it:ort} above implies that $W$ has to be an
orthogonal matrix, i.e.\ $W$ satisfies
\begin{equation}\label{eq:Worth}
W W^T = W^T W = I \,.
\end{equation}
In particular we show that $W$ is an orthogonal matrix when
\begin{equation}\label{eq:WaWaI}
W_{\bm{X}'}^\alpha [W_{\bm{X}'}^\alpha]^T = I_\alpha \,,
\end{equation}
where $\bm{X}'$ is $\bar{\bm{r}}'$ or $\overline{\bm{\gamma}}'$
and $I_\alpha$ is the unit matrix. \vspace{2em}

Thus Eq.~(\ref{eq:WaWaI}) is the {\em essential equation} for
$W_{\bm{X}'}^\alpha$ to satisfy.  \vspace{2em}

From Eq.~(\ref{eq:W})
\begin{equation}
W^T W = \left( \begin{array}{cc} \sum_{\alpha=[N]}^{[N-1,
\hspace{1ex} 1]} \, [W_{\bar{\bm{r}}'}^\alpha]^T W_{\bar{\bm{r}}'}^\alpha & \bm{0} \\
\bm{0} & \sum_{\alpha=[N]}^{[N-2, \hspace{1ex} 2]} \,
[W_{\overline{\bm{\gamma}}'}^\alpha]^T
W_{\overline{\bm{\gamma}}'}^\alpha
\end{array} \right)
\end{equation}
which with Eq.~(\ref{eq:Worth}) gives us
\begin{equation}\label{eq:IXdecomp}
\sum_{\alpha} \, [W_{\bm{X}'}^\alpha]^T W_{\bm{X}'}^\alpha =
I_{\bm{X}'} \,,
\end{equation}
where $I_{\bm{X}'}$ is the $N\times N$-dimensional unit matrix
when $\bm{X}'=\bar{\bm{r}}'$ or the $N(N-1)/2 \times
N(N-1)/2$-dimensional unit matrix when
$\bm{X}'=\overline{\bm{\gamma}}'$\,. From Eq.~(\ref{eq:WaWaI}) and
the automatically satisfied Eq.~(\ref{eq:wawapt}) we find
\begin{equation}
[W_{\bm{X}'}^{\alpha'}]^T W_{\bm{X}'}^{\alpha'} \,\,
[W_{\bm{X}'}^\alpha]^T W_{\bm{X}'}^\alpha =
\delta_{\alpha,\alpha'} \, [W_{\bm{X}'}^\alpha]^T
W_{\bm{X}'}^\alpha \,,
\end{equation}
and so Eq.~(\ref{eq:IXdecomp}) is a decomposition of the identity
into primitive idempotent projection operators which annul each
other on both sides. The $[W_{\bm{X}'}^\alpha]^T
W_{\bm{X}'}^\alpha$ project out orthogonal irreducible subspaces
while the $W_{\bm{X}'}^\alpha$ are the Clebsch-Gordon coefficients
of this decomposition.

\subsubsection{The Motion Associated with the Symmetry Coordinates
about the Lewis Structure Configuration.}
Equation~(\ref{eq:IXdecomp}) can be written as
\begin{eqnarray}\label{eq:IXdecompi}
I_{\bm{X}'} = \sum_{\alpha} \sum_\xi \,
[(W_{\bm{X}'}^\alpha)_\xi]^T \, (W_{\bm{X}'}^\alpha)_\xi \,, &&
\mbox{where}
\begin{array}[t]{l} \mbox{$\xi=1$ when $\alpha=[N]$\,,} \\
\mbox{$1 \leq \xi \leq N-1$ when $\alpha=[N-1, \hspace{1ex}
1]$\,,} \\
\mbox{$1 \leq \xi \leq N(N-3)/2$ when $\alpha=[N-2, \hspace{1ex}
2]$\,,} \end{array}
\end{eqnarray}
and the symbol $\xi$ denotes the $\xi^{\rm th}$ row of
$W_{\bm{X}'}^\alpha$\,. Thus we can write
\begin{equation}\label{eq:Xdecompi}
{\bm{X}}' = I_{\bm{X}'} {\bm{X}}' = \sum_{\alpha} \,
[W_{\bm{X}'}^\alpha]^T (W_{\bm{X}'}^\alpha \, {\bm{.}} \,
{\bm{X}}') = \sum_{\alpha} \sum_\xi \, \left(
(W_{\bm{X}'}^\alpha)_\xi \, {\bm{.}} \, {\bm{X}}' \right) \,\,
[(W_{\bm{X}'}^\alpha)_\xi]^T \,,
\end{equation}
where ${\bm{X}}'$ is $\bar{\bm{r}}'$ or $\overline{\bm{\gamma}}'$
of Eqs.~(\ref{eq:bfrp}) or (\ref{eq:bfgammap}) respectively.
According to Eqs.~(\ref{eq:wawapt}) and (\ref{eq:WaWaI}) the
$[(W_{\bm{X}'}^\alpha)_\xi]^T$ are column vectors which form a
complete set of orthonormal basis functions where, from
Eqs.~(\ref{eq:wawapt}) (\ref{eq:WaWaI}), one has
\begin{equation}
(W_{\bm{X}'}^\alpha)_\xi \, ([W_{\bm{X}'}^{\alpha'}]^T)_{\xi'} =
\delta_{\alpha,\, \alpha'} \, \delta_{\xi,\, \xi'} \,.
\end{equation}
From Eqs.~(\ref{eq:sr}) and (\ref{eq:sgamma})
\begin{equation}
[{\bm{S}}_{\bm{X}'}^\alpha]_\xi = (W_{\bm{X}'}^\alpha)_\xi \,
{\bm{.}} \, {\bm{X}}'
\end{equation}
and so Eq.~(\ref{eq:Xdecompi}) is a decomposition of the
$\bar{\bm{r}}'$ and $\overline{\bm{\gamma}}'$ vectors in terms of
the symmetry coordinates, i.e.\
\begin{equation} \label{eq:XdecompiSW}
{\bm{X}}' = \sum_{\alpha} \sum_\xi \, {\bm{X}}^{\prime \alpha}_\xi
\,,
\end{equation}
where
\begin{equation} \label{eq:Xpalphai}
{\bm{X}}^{\prime \alpha}_\xi = [{\bm{S}}_{\bm{X}'}^\alpha]_\xi \,
[(W_{\bm{X}'}^\alpha)_\xi]^T
\end{equation}
expresses the motion associated with the symmetry coordinate
$[{\bm{S}}_{\bm{X}'}^\alpha]_\xi$ in the original internal
displacement coordinate, ${\bm{X}}'$ ($\bar{\bm{r}}'$ or
$\overline{\bm{\gamma}}'$). Equation~(\ref{eq:XdecompiSW}) can be
written concisely as
\begin{equation} \label{eq:XdecompSW}
{\bar{\bm{y}}'} = W^T \, {\bm{S}}\,,
\end{equation}
where ${\bar{\bm{y}}'}$ is defined in Eqs.~(\ref{eq:ytransposeP}),
(\ref{eq:bfrp}) and (\ref{eq:bfgammap}), ${\bm{S}}$ is given by
Eqs.~(\ref{eq:FSCV}), (\ref{eq:p}) and (\ref{eq:srep}) while $W$
is given by Eq.~(\ref{eq:W}). Equation~(\ref{eq:XdecompSW}) may be
directly derived from Eq.~(\ref{eq:SWy}) using the orthogonality
of the $W$ matrix (Eq.~(\ref{eq:Worth})).

Using Eqs.~(\ref{eq:XdecompiSW}) and (\ref{eq:Xpalphai}) in
Eqs.~(\ref{eq:kappascale}), (\ref{eq:taylor1}) and
(\ref{eq:taylor2}) we find that
\begin{equation} \label{eq:yS}
{\bm{y}} = \left( \begin{array}{c} {\bm{r}} \\
\bm{\gamma} \end{array} \right) = \left(
\begin{array}{l} D^2 \overline{a}_{ho} \left( {\displaystyle
\overline{r}'_\infty {\bm{1}}_{\bar{\bm{r}}'} + \frac{1}{\sqrt{D}}
\sum_{\alpha} \sum_\xi \, \bar{\bm{r}}^{\prime \alpha}_\xi } \right) \\
{\displaystyle \overline{\gamma}_\infty
{\bm{1}}_{\overline{\bm{\gamma}}'} + \frac{1}{\sqrt{D}}
\sum_{\alpha} \sum_\xi \, \overline{\bm{\gamma}}^{\prime
\alpha}_\xi }
\end{array} \right)
\end{equation}
where according to Eq.~(\ref{eq:Xpalphai})
\begin{eqnarray}
\bar{\bm{r}}^{\prime \alpha}_\xi & = &
[{\bm{S}}_{\bar{\bm{r}}'}^\alpha]_\xi \,
[(W_{\bar{\bm{r}}'}^\alpha)_\xi]^T \,, \label{eq:rpaxi} \\
\overline{\bm{\gamma}}^{\prime \alpha}_\xi & = &
[{\bm{S}}_{\overline{\bm{\gamma}}'}^\alpha]_\xi \,
[(W_{\overline{\bm{\gamma}}'}^\alpha)_\xi]^T \,, \label{eq:gpaxi}
\end{eqnarray}
while
\begin{eqnarray} \label{eq:bf1i}
[{\bm{1}}_{\bar{\bm{r}}'}]_i= 1 & \forall & 1 \leq i \leq N
\end{eqnarray}
and
\begin{eqnarray} \label{eq:1eq1}
[{\bm{1}}_{\overline{\bm{\gamma}}'}]_{ij} = 1  & \forall & 1 \leq
i,j \leq N \,.
\end{eqnarray}
Equations~(\ref{eq:yS}), (\ref{eq:rpaxi}), (\ref{eq:gpaxi}),
(\ref{eq:bf1i}) and (\ref{eq:1eq1}) express the motion associated
with the symmetry coordinate about the Lewis structure
configuration.

\subsection{The Reduction of the Eigensystem Equations in the Symmetry
Coordinate Basis\,.} \label{subsec:eigreduct}
\subsubsection{The Central Theorem}
Let us start out by defining ${\bf Q_W}$ to be the ${\bf Q}$
matrix in the symmetry coordinate basis, i.e.\
\begin{equation} \label{eq:WQWT}
{\bf Q_W} = W {\bf Q} W^T \,,
\end{equation}
where ${\bf Q}$ is $\bm{F}$, $\bm{G}$ or $\bm{FG}$, and ${\bf
Q_W}$ is ${\bf F_W}$, ${\bf G_W}$ or $(\bm{FG})_W$\,. It has been
shown elsewhere\cite{WDC} that since $\bm{F}$, $\bm{G}$ and
$\bm{FG}$ are invariant matrices (under $S_N$), when
items~\ref{it:teq0} and \ref{it:ort} above are satisfied then
\begin{equation} \label{eq:Qw}
{\bf Q_W} = \left( \begin{array}{ccc} \bm{\sigma_{[N]}^Q} \otimes
{\bf I_{[N]}} &
\bm{0} & \bm{0} \\
\bm{0} & \bm{\sigma_{[N-1, \hspace{1ex} 1]}^Q}
\otimes {\bf I_{[N-1, \hspace{1ex} 1]}} & \bm{0} \\
\bm{0} & \bm{0} & \bm{\sigma_{[N-2, \hspace{1ex} 2]}^Q} \otimes
{\bf I_{[N-2, \hspace{1ex} 2]}}
\end{array} \right) \,,
\end{equation}
where the symbol $\otimes$ indicates the direct product,
$\bm{\sigma_{[N]}^Q}$ and $\bm{\sigma_{[N-1, \hspace{1ex} 1]}^Q}$
are $2 \times 2$ dimensional matrices and $\bm{\sigma_{[N-2,
\hspace{1ex} 2]}^Q}$ is a single element (a number). The matrix
${\bf I_{[N]}}$ is the $[N]$-sector identity matrix, simply the
number $1$, ${\bf I_{[N-1, \hspace{1ex} 1]}}$ is the $[N-1,
\hspace{1ex} 1]$-sector identity matrix, the $(N-1) \times
(N-1)$-dimensional unit matrix and ${\bf I_{[N-2, \hspace{1ex}
2]}}$ is the $[N-2, \hspace{1ex} 2]$-sector identity matrix, the
$N(N-3)/2 \times N(N-3)/2$-dimensional unit matrix. The matrix
elements
\begin{equation}\label{eq:sigmaQ}
[\bm{\sigma_\alpha^Q}]_{\bm{X}'_1,\,\bm{X}'_2} =
(W_{\bm{X}'_1}^\alpha)_\xi \, {\bf Q}_{\bm{X}'_1 \bm{X}'_2} \,
[(W_{\bm{X}'_2}^\alpha)_\xi]^T \,,
\end{equation}
where $\bm{X}'_1$ and $\bm{X}'_2$ are $\bar{\bm{r}}'$ or
$\overline{\bm{\gamma}}'$ (only $\overline{\bm{\gamma}}'$ for the
$[N-2, \hspace{1ex} 2]$ sector). The ${\bf Q}$-matrix quadrant,
${\bf Q}_{\bm{X}'_1 \bm{X}'_2}$, is given by Eqs.~(\ref{eq:Q}),
(\ref{eq:Qrr}), (\ref{eq:Qrg}), (\ref{eq:Qgr}) and (\ref{eq:Qgg}).
Note that although we are {\em not} summing over the repeated
index $\xi$ in Eq.~(\ref{eq:sigmaQ}), the $[{\bf
\sigma_\alpha^Q}]_{\bm{X}'_1,\,\bm{X}'_2}$ are {\em independent}
of the $W_{\bm{X}'}^\alpha$ row label, $\xi$\,. If, when we
calculate $W_{\bm{X}'}^\alpha$\,, the matrix element
$[\bm{\sigma_\alpha^Q}]_{\bm{X}'_1,\,\bm{X}'_2}$ turns out to
depend on the
$W_{\bm{X}'}^\alpha$ row label, $\xi$\,, then we
know that we have made a mistake calculating
$W_{\bm{X}'}^\alpha$\,. This is a strong check on the correctness
of our calculations in Sect.~\ref{sec:DetS}.

\subsubsection{The Reduction of the Eigenvalue Equation in the
Symmetry Coordinate Basis\,.}
Transforming the basic eigenvalue
equation, Eq.~(\ref{eq:FGit}), to the symmetry coordinate basis we
have
\begin{equation} \label{eq:FwGwcb}
W {\bf F} W^T \, W \bm{G} W^T \, W {\bm{b}} = {\bf F_W} \, {\bf
G_W} \, {\bm{c}}^{(b)} = \lambda_b \, {\bm{c}}^{(b)} \,,
\end{equation}
where
\begin{eqnarray}\label{eq:FGcW}
{\bf F_W} = W {\bf F} W^T & {\bf G_W} = W \bm{G} W^T &
{\bm{c}}^{(b)} = W {\bm{b}} \,.
\end{eqnarray}
Under this transformation we can also write Eq.~(\ref{eq:qyt}) for
the $b^{\rm th}$ normal-mode coordinate as
\begin{equation} \label{eq:cTS}
[q']_b = {\bm{b}}^T W^T W {\bar{\bm{y}}'} = \left[
{{\bm{c}}^{(b)}} \right]^T {\bm{\cdot}} \,\, {\bm{S}} \,,
\end{equation}
where $[q']_b$ is now directly expressed in terms of the symmetry
coordinates.

From Eqs.~(\ref{eq:Qw}) and (\ref{eq:FwGwcb}) the
normal-coordinate coefficient vector, ${\bm{c}}^{(b)}$, has the
form
\begin{equation} \label{eq:cb}
{\bm{c}}^{(b)} = \left(
\begin{array}{r@{\hspace{0.5ex}}c@{\hspace{0.5ex}}l}
\delta_{\alpha,\,[N]} \,\, {\mathsf{c}}^{[N]} & \otimes & 1 \\
\delta_{\alpha,\,[N-1, \hspace{1ex} 1]} \,\, {\mathsf{c}}^{[N-1,
\hspace{1ex} 1]} & \otimes &
{\bm{1}}^{[N-1, \hspace{1ex} 1]}_\xi\\
\delta_{\alpha,\,[N-2, \hspace{1ex} 2]} \,\, {\mathsf{c}}^{[N-2,
\hspace{1ex} 2]} & \otimes & {\bm{1}}^{[N-2, \hspace{1ex} 2]}_\xi
\end{array} \right) \,,
\end{equation}
where the ${\mathsf{c}}^{\alpha}$ satisfy the eigenvalue equations
\begin{equation} \label{eq:sceig}
\sigma_{\alpha}^{\bm{FG}} {\mathsf{c}}^{\alpha} = \lambda_\alpha
{\mathsf{c}}^{\alpha} \,.
\end{equation}
Note that $\sigma_{[N]}^{\bm{FG}}$ and $\sigma_{[N-1, \hspace{1ex}
1]}^{\bm{FG}}$ are $2 \times 2$-dimensional matrices, while
$\sigma_{[N-2, \hspace{1ex} 2]}^{\bm{FG}}$ is a one-dimensional
matrix. Thus there are five solutions to Eq.~(\ref{eq:sceig})
which we denote as
$\{\lambda^\pm_{[N]},\,{\mathsf{c}}_\pm^{[N]}\}$\,,
$\{\lambda^\pm_{[N-1, \hspace{1ex} 1]},\,{\mathsf{c}}_\pm^{[N-1,
\hspace{1ex} 1]}\}$ and $\{\lambda_{[N-2, \hspace{1ex}
2]},\,{\mathsf{c}}^{[N-2, \hspace{1ex} 2]}\}$\,, where $\bm{X}'$
($\bar{\bm{r}}'$ or $\overline{\bm{\gamma}}'$\,, only
$\overline{\bm{\gamma}}'$ for the $[N-2, \hspace{1ex} 2]$ sector)
labels the rows of the elements of the column vector
${\mathsf{c}}^{\alpha}$. The normal-coordinate label, $b$, has
been replaced by the labels $\alpha$, $\xi$ and $\pm$ on the right
hand side of Eq.~(\ref{eq:cb}), while the elements of the column
vectors ${\bm{1}}^{\alpha}_\xi$ are
\begin{equation}
[{\bm{1}}^{\alpha}_\xi]_\eta = \delta_{\xi \eta} \,,
\end{equation}
with $1 \leq \xi, \hspace{0.5ex} \eta \leq N-1$ when $\alpha=[N-1,
\hspace{1ex} 1]$, or $1 \leq \xi, \hspace{0.5ex} \eta \leq
N(N-3)/2$ when $\alpha=[N-2, \hspace{1ex} 2]$\,. The
${\mathsf{c}}^{\alpha}_\pm$ for the $[N]$ and $[N-1, \hspace{1ex}
1]$ sectors determine the amount of angular-radial mixing between
the symmetry coordinates in a normal coordinate of a particular
$\alpha$ since from Eqs.~(\ref{eq:FSCV}), (\ref{eq:srep}) and
(\ref{eq:cTS})
\begin{equation}\label{eq:q12}
[q']_b = [{\mathsf{c}}^{\alpha}_\pm]_{\bar{r}'} \,
[{\bm{S}}_{\bar{\bm{r}}'}^{\alpha}]_\xi \, + \,
[{\mathsf{c}}^{\alpha}_\pm]_{\overline{\bm{\gamma}}'} \,
[{\bm{S}}_{\overline{\bm{\gamma}}'}^{\alpha}]_\xi \,.
\end{equation}
For the $[N-2, \hspace{1ex} 2]$ sector we have
\begin{equation}
[q']_b = [{\mathsf{c}}^{[N-2, \hspace{1ex}
2]}]_{\overline{\bm{\gamma}}'} \,
[{\bm{S}}_{\overline{\bm{\gamma}}'}^{[N-2, \hspace{1ex} 2]}]_\xi
\,,
\end{equation}
i.e.\ the $[N-2, \hspace{1ex} 2]$ sector symmetry coordinates are
the $[N-2, \hspace{1ex} 2]$ sector normal coordinates up to a
normalization constant, $[{\mathsf{c}}^{[N-2, \hspace{1ex}
2]}]_{\overline{\bm{\gamma}}'}$ (see Sec.~\ref{subsubsec:norm}).

If we write ${\mathsf{c}}^{\alpha}$ as
\begin{equation} \label{eq:sfceqcthatc}
{\mathsf{c}}^{\alpha} = c^{\alpha} \times
\widehat{{\mathsf{c}}^{\alpha}}
\end{equation}
where $\widehat{{\mathsf{c}}^{\alpha}}$ is a vector satisfying the
normalization condition
\begin{equation} \label{eq:hatcnorm}
[\widehat{{\mathsf{c}}^{\alpha}}]^T \,
\widehat{{\mathsf{c}}^{\alpha}} = 1
\end{equation}
and $c^{\alpha}$ is a  normalization factor ensuring that
Eq.~(\ref{eq:normit}) is satisfied (see Sec.~\ref{subsubsec:norm}
below), then the reduced eigenvalue equation,
Eq.~(\ref{eq:sceig}), determines $\widehat{{\mathsf{c}}^{\alpha}}$
alone.

With the $[N-2, \hspace{1ex} 2]$ sector, Eq.~(\ref{eq:hatcnorm})
yields
\begin{equation}  \label{eq:hatcnormeq1}
\widehat{{\mathsf{c}}^{[N-2, \hspace{1ex} 2]}} = 1
\end{equation}
and so Eq.~(\ref{eq:sceig}) yields directly
\begin{equation} \label{eq:lNm2eqsig}
\lambda_{[N-2, \hspace{1ex} 2]} = \sigma_{[N-2, \hspace{1ex}
2]}^{\bm{FG}} \,.
\end{equation}
As for the $[N]$ and $[N-1, \hspace{1ex} 1]$ sectors, if we write
\begin{equation} \label{eq:hatcpm}
\widehat{{\mathsf{c}}_\pm^{\alpha}} = \left(
\begin{array}{c} \cos{\theta^\alpha_\pm} \\
\sin{\theta^\alpha_\pm} \end{array} \right) \,,
\end{equation}
then from Eqs.~(\ref{eq:q12}), (\ref{eq:sfceqcthatc}) and
(\ref{eq:hatcpm})
\begin{equation}
[q']_b = c_\pm^{\alpha} \left( \cos{\theta^\alpha_\pm} \,
[{\bm{S}}_{\bar{\bm{r}}'}^{\alpha}]_\xi \, + \,
\sin{\theta^\alpha_\pm} \,
[{\bm{S}}_{\overline{\bm{\gamma}}'}^{\alpha}]_\xi \right) \,.
\end{equation}
Writing
\begin{equation} \label{eq:sigmamat}
\sigma_{\alpha}^{\bm{FG}} = \left( \begin{array}{cc}
\protect[\bm{\sigma_\alpha^{\bm{FG}}}\protect]_{\bar{\bm{r}}',\,\bar{\bm{r}}'}
& \protect[\bm{\sigma_\alpha^{\bm{FG}}}\protect]_{\bar{\bm{r}}',\,
\overline{\bm{\gamma}}'} \\
\protect[\bm{\sigma_\alpha^{\bm{FG}}}\protect]_{\overline{\bm{\gamma}}',\,\bar{\bm{r}}'}
&
\protect[\bm{\sigma_\alpha^{\bm{FG}}}\protect]_{\overline{\bm{\gamma}}',\,\overline{\bm{\gamma}}'}
\end{array} \right) \,,
\end{equation}
then Eq.~(\ref{eq:sceig}) may be written as
\begin{equation} \label{eq:sigma12}
\left( \begin{array}{cc}
(\protect[\bm{\sigma_\alpha^{\bm{FG}}}\protect]_{\bar{\bm{r}}',\,\bar{\bm{r}}'}
- \lambda^\pm_\alpha) &
\protect[\bm{\sigma_\alpha^{\bm{FG}}}\protect]_{\bar{\bm{r}}',\,
\overline{\bm{\gamma}}'} \\
\protect[\bm{\sigma_\alpha^{\bm{FG}}}\protect]_{\overline{\bm{\gamma}}',\,\bar{\bm{r}}'}
&
(\protect[\bm{\sigma_\alpha^{\bm{FG}}}\protect]_{\overline{\bm{\gamma}}',\,\overline{\bm{\gamma}}'}
- \lambda^\pm_\alpha)
\end{array} \right) \left( \begin{array}{c}
\cos{\theta^\alpha_\pm} \\ \sin{\theta^\alpha_\pm} \end{array}
\right) = 0 \,,
\end{equation}
from which we derive
\begin{equation} \label{eq:lambda12pm}
\lambda^\pm_\alpha =
\frac{(\protect[\bm{\sigma_\alpha^{\bm{FG}}}\protect]_{\bar{\bm{r}}',\,\bar{\bm{r}}'}
+
\protect[\bm{\sigma_\alpha^{\bm{FG}}}\protect]_{\overline{\bm{\gamma}}',\,\overline{\bm{\gamma}}'})
\pm
\sqrt{(\protect[\bm{\sigma_\alpha^{\bm{FG}}}\protect]_{\bar{\bm{r}}',\,\bar{\bm{r}}'}
-
\protect[\bm{\sigma_\alpha^{\bm{FG}}}\protect]_{\overline{\bm{\gamma}}',\,\overline{\bm{\gamma}}'})^2
+ 4
\protect[\bm{\sigma_\alpha^{\bm{FG}}}\protect]_{\bar{\bm{r}}',\,
\overline{\bm{\gamma}}'}
\protect[\bm{\sigma_\alpha^{\bm{FG}}}\protect]_{\overline{\bm{\gamma}}',\,\bar{\bm{r}}'}}}{2}
\end{equation}
Equations~(\ref{eq:sigma12}) and (\ref{eq:lambda12pm}) then give
us
\begin{equation} \label{eq:tanthetaalphapm}
\tan{\theta^\alpha_\pm} = \frac{(\lambda^\pm_\alpha -
\protect[\bm{\sigma_\alpha^{\bm{FG}}}\protect]_{\bar{\bm{r}}',\,\bar{\bm{r}}'})}{\protect[\bm{\sigma_\alpha^{\bm{FG}}}\protect]_{\bar{\bm{r}}',\,
\overline{\bm{\gamma}}'}} =
\frac{\protect[\bm{\sigma_\alpha^{\bm{FG}}}\protect]_{\overline{\bm{\gamma}}',\,\bar{\bm{r}}'}}{(\lambda^\pm_\alpha
-
\protect[\bm{\sigma_\alpha^{\bm{FG}}}\protect]_{\overline{\bm{\gamma}}',\,\overline{\bm{\gamma}}'})}
\end{equation}

\subsubsection{The Normalization Condition.}
\label{subsubsec:norm} From Eqs.~(\ref{eq:normit}),
(\ref{eq:FGcW}), (\ref{eq:Qw}), (\ref{eq:sigmaQ}) and
(\ref{eq:cb}) the ${\mathsf{c}}^{\alpha}$ also satisfy the
normalization condition
\begin{equation} \label{eq:cnormeq}
[{\mathsf{c}}^{\alpha}]^T \sigma_{\alpha}^G {\mathsf{c}}^{\alpha}
= 1 \,.
\end{equation}
Thus Eqs.~(\ref{eq:sfceqcthatc}) and (\ref{eq:cnormeq}) mean that
\begin{equation} \label{eq:cnorm}
c^{\alpha} = \frac{1}{\sqrt{[\widehat{{\mathsf{c}}^{\alpha}}]^T
\sigma_{\alpha}^G \widehat{{\mathsf{c}}^{\alpha}}}} \,.
\end{equation}
For the $[N-2, \hspace{1ex} 2]$ sector,
Eqs.~(\ref{eq:hatcnormeq1}) and (\ref{eq:cnorm}) mean that
\begin{equation} \label{eq:c2norm}
c^{[N-2, \hspace{1ex} 2]} = \frac{1}{\sqrt{\sigma_{[N-2,
\hspace{1ex} 2]}^G}} \,.
\end{equation}
As for the normalization constant, $c_\pm^\alpha$, of the $[N]$
and $[N-1, \hspace{1ex} 1]$ sectors, Eqs.~(\ref{eq:hatcpm}) and
(\ref{eq:cnorm}) give us
\begin{equation} \label{eq:calphapm}
c_\pm^\alpha = \frac{1}{\sqrt{\left( \begin{array}{c}
\cos{\theta^\alpha_\pm} \\ \sin{\theta^\alpha_\pm} \end{array}
\right)^T \sigma_{\alpha}^{\bm{G}} \left( \begin{array}{c}
\cos{\theta^\alpha_\pm} \\ \sin{\theta^\alpha_\pm} \end{array}
\right)}} \,\,.
\end{equation}

\subsubsection{The Normal Coordinates.}
Thus the total transformation matrix, $V$\,, from the internal
displacement coordinates to the normal coordinates is
\begin{equation}
V = C \, W \,,
\end{equation}
where $W$ is given by Eq.~(\ref{eq:W}) and
\begin{equation}
C = \left(
\begin{array}{c@{\hspace{2em}}c|c@{\hspace{2em}}c|c} c^{[N]}_+
\cos{\theta^{[N]}_+} &
c^{[N]}_+ \sin{\theta^{[N]}_+} & 0 & 0 & 0 \\
c^{[N]}_- \cos{\theta^{[N]}_-} & c^{[N]}_- \sin{\theta^{[N]}_-} &
0 & 0 & 0 \\ \hline 0 & 0 & c^{[N-1, \hspace{1ex} 1]}_+
\cos{\theta^{[N-1, \hspace{1ex} 1]}_+} &
c^{[N-1, \hspace{1ex} 1]}_+ \sin{\theta^{[N-1, \hspace{1ex} 1]}_+} & 0 \\
0 & 0 & c^{[N-1, \hspace{1ex} 1]}_- \cos{\theta^{[N-1,
\hspace{1ex} 1]}_-} & c^{[N-1, \hspace{1ex} 1]}_-
\sin{\theta^{[N-1, \hspace{1ex} 1]}_-} & 0 \\ \hline 0 & 0 & 0 & 0
& c^{[N-2, \hspace{1ex} 2]}
\end{array} \right) \,,
\end{equation}
and so the normal-coordinate vector, ${\bm{q}'}$\,, is given by
\renewcommand{\jot}{1em}
\begin{eqnarray}
{\bm{q}'} & = & \left( \begin{array}{l} {\bm{q}'}^{[N]} \\
{\bm{q}'}^{[N-1, \hspace{1ex} 1]} \\
{\bm{q}'}^{[N-2, \hspace{1ex} 2]} \end{array} \right) = \left( \begin{array}{l} {\bm{q}'}_+^{[N]} \\
{\bm{q}'}_-^{[N]}  \\ \hline {\bm{q}'}_+^{[N-1, \hspace{1ex} 1]} \\
{\bm{q}'}_-^{[N-1, \hspace{1ex} 1]} \\ \hline {\bm{q}'}^{[N-2,
\hspace{1ex} 2]} \end{array} \right) = V \, {\bar{\bm{y}}'} = C \,
{\bm{S}} = C \left( \begin{array}{l} {\bm{S}}_{\bar{\bm{r}}'}^{[N]} \\
{\bm{S}}_{\overline{\bm{\gamma}}'}^{[N]}  \\ \hline {\bm{S}}_{\bar{\bm{r}}'}^{[N-1, \hspace{1ex} 1]} \\
{\bm{S}}_{\overline{\bm{\gamma}}'}^{[N-1, \hspace{1ex} 1]} \\
\hline {\bm{S}}_{\overline{\bm{\gamma}}'}^{[N-2, \hspace{1ex} 2]}
\end{array} \right) = \nonumber \\ & = & \left( \begin{array}{c}
c^{[N]}_+ \cos{\theta^{[N]}_+} {\bm{S}}_{\bar{\bm{r}}'}^{[N]} +
c^{[N]}_+ \sin{\theta^{[N]}_+} {\bm{S}}_{\overline{\bm{\gamma}}'}^{[N]} \\
c^{[N]}_- \cos{\theta^{[N]}_-} {\bm{S}}_{\bar{\bm{r}}'}^{[N]} +
c^{[N]}_-
\sin{\theta^{[N]}_-} {\bm{S}}_{\overline{\bm{\gamma}}'}^{[N]}  \\
\hline c^{[N-1, \hspace{1ex} 1]}_+ \cos{\theta^{[N-1, \hspace{1ex}
1]}_+} {\bm{S}}_{\bar{\bm{r}}'}^{[N-1, \hspace{1ex} 1]} + c^{[N-1,
\hspace{1ex} 1]}_+ \sin{\theta^{[N-1, \hspace{1ex} 1]}_+}
{\bm{S}}_{\overline{\bm{\gamma}}'}^{[N-1, \hspace{1ex} 1]} \\
c^{[N-1, \hspace{1ex} 1]}_- \cos{\theta^{[N-1, \hspace{1ex} 1]}_-}
{\bm{S}}_{\bar{\bm{r}}'}^{[N-1, \hspace{1ex} 1]} + c^{[N-1,
\hspace{1ex} 1]}_- \sin{\theta^{[N-1, \hspace{1ex} 1]}_-}
{\bm{S}}_{\overline{\bm{\gamma}}'}^{[N-1, \hspace{1ex} 1]} \\
\hline c^{[N-2, \hspace{1ex} 2]}
{\bm{S}}_{\overline{\bm{\gamma}}'}^{[N-2, \hspace{1ex} 2]}
\end{array} \right) \,. \label{eq:qvector}
\end{eqnarray}
\renewcommand{\jot}{0em}

\subsubsection{The Motion Associated with the Normal Coordinates
about the Lewis Structure Configuration.}
Consider now the inverse
transformation to express the internal displacement coordinates in
terms of the normal coordinates. From Eq.~(\ref{eq:Worth})
\begin{equation}
I = W^T W = W^T C^{-1} C W
\end{equation}
so that
\begin{equation} \label{eq:WTCm1q}
{\bar{\bm{y}}'} = W^T C^{-1} \, (V {\bar{\bm{y}}'}) = W^T C^{-1}
\, {\bm{q}'}  \,,
\end{equation}
where
\newlength{\lengthaa}
\settowidth{\lengthaa}{$\frac{-\sin{\theta^{[N]}_-}}{c^{[N]}_+}$}
\newlength{\lengthab}
\settowidth{\lengthab}{$\frac{-\cos{\theta^{[N]}_+}}{c^{[N]}_-}$}
\newlength{\lengthcc}
\settowidth{\lengthcc}{$\frac{-\sin{\theta^{[N-1, \hspace{1ex}
1]}_-}}{c^{[N-1, \hspace{1ex} 1]}_+}$}
\newlength{\lengthcd}
\settowidth{\lengthcd}{$\frac{-\cos{\theta^{[N-1, \hspace{1ex}
1]}_+}}{c^{[N-1, \hspace{1ex} 1]}_-}$}
\renewcommand{\jot}{1em}
\begin{eqnarray}
\lefteqn{C^{-1} =} \nonumber \\ & & \left(
\renewcommand{\arraystretch}{1.5}
\begin{array}{r@{}c@{}l|r@{}c@{}l|c}
\frac{1}{s(\theta^{[N]})} \left(
\vphantom{\begin{array}{c@{\hspace{2em}}c}
\frac{-\sin{\theta^{[N]}_-}}{c^{[N]}_+} &
\frac{\sin{\theta^{[N]}_+}}{c^{[N]}_-} \\
\frac{\cos{\theta^{[N]}_-}}{c^{[N]}_+} &
\frac{-\cos{\theta^{[N]}_+}}{c^{[N]}_-}
\end{array}}
\right. & \begin{array}{c@{\hspace{2em}}c}
\frac{-\sin{\theta^{[N]}_-}}{c^{[N]}_+} &
\frac{\sin{\theta^{[N]}_+}}{c^{[N]}_-} \\
\frac{\cos{\theta^{[N]}_-}}{c^{[N]}_+} &
\frac{-\cos{\theta^{[N]}_+}}{c^{[N]}_-} \end{array} & \left.
\vphantom{\begin{array}{c@{\hspace{2em}}c}
\frac{-\sin{\theta^{[N]}_-}}{c^{[N]}_+} &
\frac{\sin{\theta^{[N]}_+}}{c^{[N]}_-} \\
\frac{\cos{\theta^{[N]}_-}}{c^{[N]}_+} &
\frac{-\cos{\theta^{[N]}_+}}{c^{[N]}_-}
\end{array}} \right) & & \begin{array}{c@{\hspace{2em}}c}
\makebox[\lengthcc]{$0$} & \makebox[\lengthcd]{$0$} \\ 0 & 0
\end{array} & &
\begin{array}{c} 0 \\ 0 \end{array} \\ &&&&&& \protect\vspace{-5ex} \\ \hline
 &&&&&& \protect\vspace{-5ex} \\
& \begin{array}{c@{\hspace{2em}}c} \makebox[\lengthaa]{$0$} &
\makebox[\lengthab]{$0$} \\ 0 & 0 \end{array} & &
\frac{1}{s(\theta^{[N-1, \hspace{1ex} 1]})} \left(
\vphantom{\begin{array}{c@{\hspace{2em}}c}
\frac{-\sin{\theta^{[N-1, \hspace{1ex} 1]}_-}}{c^{[N-1,
\hspace{1ex} 1]}_+} &
\frac{\sin{\theta^{[N-1, \hspace{1ex} 1]}_+}}{c^{[N-1, \hspace{1ex} 1]}_-} \\
\frac{\cos{\theta^{[N-1, \hspace{1ex} 1]}_-}}{c^{[N-1,
\hspace{1ex} 1]}_+} & \frac{-\cos{\theta^{[N-1, \hspace{1ex}
1]}_+}}{c^{[N-1, \hspace{1ex} 1]}_-}
\end{array}} \right. &
\begin{array}{c@{\hspace{2em}}c} \frac{-\sin{\theta^{[N-1, \hspace{1ex} 1]}_-}}{c^{[N-1, \hspace{1ex} 1]}_+} &
\frac{\sin{\theta^{[N-1, \hspace{1ex} 1]}_+}}{c^{[N-1, \hspace{1ex} 1]}_-} \\
\frac{\cos{\theta^{[N-1, \hspace{1ex} 1]}_-}}{c^{[N-1,
\hspace{1ex} 1]}_+} & \frac{-\cos{\theta^{[N-1, \hspace{1ex}
1]}_+}}{c^{[N-1, \hspace{1ex} 1]}_-}
\end{array} & \left. \vphantom{\begin{array}{c@{\hspace{2em}}c} \frac{-\sin{\theta^{[N-1, \hspace{1ex} 1]}_-}}{c^{[N-1, \hspace{1ex} 1]}_+} &
\frac{\sin{\theta^{[N-1, \hspace{1ex} 1]}_+}}{c^{[N-1, \hspace{1ex} 1]}_-} \\
\frac{\cos{\theta^{[N-1, \hspace{1ex} 1]}_-}}{c^{[N-1,
\hspace{1ex} 1]}_+} & \frac{-\cos{\theta^{[N-1, \hspace{1ex}
1]}_+}}{c^{[N-1, \hspace{1ex} 1]}_-}
\end{array}}
\right) & \begin{array}{c} 0 \\ 0 \end{array} \\  &&&&&&
\protect\vspace{-5ex} \\ \hline &
\begin{array}{c@{\hspace{2em}}c} \makebox[\lengthaa]{$0$} &
\makebox[\lengthab]{$0$} \end{array} & & &
\begin{array}{c@{\hspace{2em}}c} \makebox[\lengthcc]{$0$} & \makebox[\lengthcd]{$0$} \end{array} & & \frac{1}{c^{[N-2, \hspace{1ex}
2]}}
\end{array} \renewcommand{\arraystretch}{1.5} \right) \nonumber \\
& & \label{eq:Cm1}
\end{eqnarray}
\renewcommand{\jot}{0em}
and
\begin{equation} \label{eq:s}
s(\theta^{\alpha}) = \sin{(\theta^{\alpha}_+ - \theta^{\alpha}_-)}
\,.
\end{equation}
Thus from Eqs.~(\ref{eq:ytransposeP}), (\ref{eq:WTCm1q}),
(\ref{eq:Cm1}) and (\ref{eq:s})
\begin{equation} \label{eq:yqp}
{\bar{\bm{y}}'} = \left( \begin{array}{c} \bar{\bm{r}}' \\
\overline{\bm{\gamma}}' \end{array} \right) =
\renewcommand{\arraystretch}{1.5}
\begin{array}[t]{cr@{}l} & {\displaystyle \frac{1}{s(\theta^{[N]})}} & \left(
\begin{array}{c}
\protect[W_{\bar{\bm{r}}'}^{[N]}\protect]^T
\left(\frac{-\sin{\theta^{[N]}_-}}{c^{[N]}_+} \, {\bm{q}'}_+^{[N]}
+ \frac{\sin{\theta^{[N]}_+}}{c^{[N]}_-} \,
{\bm{q}'}_-^{[N]} \right) \\
\protect[W_{\overline{\bm{\gamma}}'}^{[N]}\protect]^T \left(
\frac{\cos{\theta^{[N]}_-}}{c^{[N]}_+} \, {\bm{q}'}_+^{[N]} +
\frac{-\cos{\theta^{[N]}_+}}{c^{[N]}_-} \, {\bm{q}'}_-^{[N]}
\right) \end{array} \right) + \vspace{1ex} \\ + & {\displaystyle
\frac{1}{s(\theta^{[N-1, \hspace{1ex} 1]})}} & \left(
\begin{array}{c} \protect[W_{\bar{\bm{r}}'}^{[N-1, \hspace{1ex} 1]}\protect]^T
\left(\frac{-\sin{\theta^{[N-1, \hspace{1ex} 1]}_-}}{c^{[N-1,
\hspace{1ex} 1]}_+} \, {\bm{q}'}_+^{[N-1, \hspace{1ex} 1]} +
\frac{\sin{\theta^{[N-1, \hspace{1ex} 1]}_+}}{c^{[N-1,
\hspace{1ex} 1]}_-} \,
{\bm{q}'}_-^{[N-1, \hspace{1ex} 1]} \right) \\
\protect[W_{\overline{\bm{\gamma}}'}^{[N-1, \hspace{1ex}
1]}\protect]^T \left( \frac{\cos{\theta^{[N-1, \hspace{1ex}
1]}_-}}{c^{[N-1, \hspace{1ex} 1]}_+} \, {\bm{q}'}_+^{[N-1,
\hspace{1ex} 1]} + \frac{-\cos{\theta^{[N-1, \hspace{1ex}
1]}_+}}{c^{[N-1, \hspace{1ex} 1]}_-} \, {\bm{q}'}_-^{[N-1,
\hspace{1ex} 1]}
\right) \end{array} \right) + \vspace{1ex} \\
 + &  & \left(
\begin{array}{c} \bm{0} \\
\frac{1}{c^{[N-2, \hspace{1ex} 2]}} \,
\protect[W_{\overline{\bm{\gamma}}'}^{[N-2, \hspace{1ex}
2]}\protect]^T \, {\bm{q}'}^{[N-2, \hspace{1ex} 2]}
\end{array} \right)\,.
\end{array}
\renewcommand{\arraystretch}{1}
\end{equation}

Thus from Eqs.~(\ref{eq:kappascale}), (\ref{eq:ytranspose}),
(\ref{eq:taylor1}), (\ref{eq:taylor2}) and (\ref{eq:yqp}) we find
that
\begin{equation} \label{eq:yq}
{\bm{y}} = \left( \begin{array}{c} {\bm{r}} \\
\bm{\gamma} \end{array} \right) = {\bm{y}}_\infty +
\frac{1}{\sqrt{D}} \left(
\begin{array}[t]{@{}l@{}} {\displaystyle \hspace{2ex} \sum} \\ {\scriptstyle \alpha= \left\{
\renewcommand{\arraystretch}{0.5}
\begin{array}{@{}c@{}} {\scriptstyle
\protect[N\protect]\,,} \\ {\scriptstyle \protect[N-1,
\hspace{1ex} 1\protect]}
\end{array} \renewcommand{\arraystretch}{1} \right\}
} \end{array} \sum_\xi \sum_{\tau=\pm} \, _{\tau}{\bm{y}}^{\prime
\, \alpha}_\xi + \sum_\xi \, {\bm{y}}^{\prime \, \protect[N-2,
\hspace{1ex} 2\protect]}_\xi \right)
\end{equation}
where
\begin{equation} \label{eq:yqinfty}
{\bm{y}}_\infty = \left( \begin{array}{c} D^2
\overline{a}_{ho} \, \overline{r}'_\infty {\bm{1}}_{\bar{\bm{r}}'} \\
\overline{\gamma}_\infty {\bm{1}}_{\overline{\bm{\gamma}}'}
\end{array} \right) \,,
\end{equation}
with ${\bm{1}}_{\bar{\bm{r}}'}$ and
${\bm{1}}_{\overline{\bm{\gamma}}'}$ given in Eqs.~(\ref{eq:bf1i})
and (\ref{eq:1eq1}) respectively. The vectors
$_{+}{\bm{y}}^{\prime \, \alpha}_\xi$, \,\,$_{-}{\bm{y}}^{\prime
\, \alpha}_\xi$ and ${\bm{y}}^{\prime \, \protect[N-2,
\hspace{1ex} 2\protect]}_\xi$ are
\begin{equation} \label{eq:pyqalphaxi}
_{+}{\bm{y}}^{\prime \, \alpha}_\xi = \frac{1}{s(\theta^{\alpha})}
\left(
\renewcommand{\arraystretch}{1.5}
\begin{array}{c}
{\displaystyle D^2 \overline{a}_{ho}
\frac{-\sin{\theta^{\alpha}_-}}{c^{\alpha}_+} \,
[{\bm{q}'}_+^{\alpha}]_\xi \, \protect[(W_{\bar{\bm{r}}'}^{\alpha})_\xi \protect]^T } \\
{\displaystyle \frac{\cos{\theta^{\alpha}_-}}{c^{\alpha}_+} \,
[{\bm{q}'}_+^{\alpha}]_\xi \,
\protect[(W_{\overline{\bm{\gamma}}'}^{\alpha})_\xi \protect]^T }
\end{array} \right) \,,
\renewcommand{\arraystretch}{1}
\end{equation}
\begin{equation} \label{eq:myqalphaxi}
_{-}{\bm{y}}^{\prime \, \alpha}_\xi =
\renewcommand{\arraystretch}{1.5}
\frac{1}{s(\theta^{\alpha})} \left(
\begin{array}{c}
{\displaystyle D^2 \overline{a}_{ho}
\frac{\sin{\theta^{\alpha}_+}}{c^{\alpha}_-} \,
[{\bm{q}'}_-^{\alpha}]_\xi \, \protect[(W_{\bar{\bm{r}}'}^{\alpha})_\xi \protect]^T } \\
{\displaystyle \frac{-\cos{\theta^{\alpha}_+}}{c^{\alpha}_-} \,
[{\bm{q}'}_-^{\alpha}]_\xi \,
\protect[(W_{\overline{\bm{\gamma}}'}^{\alpha})_\xi \protect]^T }
\end{array} \right)
\renewcommand{\arraystretch}{1}
\end{equation}
for $\alpha =[N]$ or $[N-1, \hspace{1ex} 1]$\,, and
\begin{equation} \label{eq:yqnm2xi}
{\bm{y}}^{\prime \, \protect[N-2, \hspace{1ex} 2\protect]}_\xi =
\left(
\renewcommand{\arraystretch}{1.5}
\begin{array}{c} \bm{0} \\
{\displaystyle \frac{1}{c^{[N-2, \hspace{1ex} 2]}} \,
[{\bm{q}'}^{[N-2, \hspace{1ex} 2]}]_\xi \,
\protect[(W_{\overline{\bm{\gamma}}'}^{[N-2, \hspace{1ex}
2]})_\xi]^T }
\end{array} \renewcommand{\arraystretch}{1} \right) \,.
\end{equation}
Equations~(\ref{eq:yq}), (\ref{eq:yqinfty}),
(\ref{eq:pyqalphaxi}), (\ref{eq:myqalphaxi}) and
(\ref{eq:yqnm2xi}) express, in terms of the internal coordinates
${\bm{r}}$ and $\bm{\gamma}$\,, the motion associated with the
normal coordinates, ${\bm{q}'}$\,, about the Lewis structure
configuration ${\bm{y}}_\infty$\,.

\section{Determining the Symmetry Coordinates} \label{sec:DetS}
\subsection{Primitive Irreducible Coordinates}
We determine the symmetry coordinates, and hence the $W_{r}$ and
$W_{\gamma}$ of Eqs.~(\ref{eq:swr}) and (\ref{eq:swgamma})
respectively, in a two-step process:
\newcounter{twostep}
\newcounter{twostepseca}
\newcounter{twostepsecb}
\begin{list}{\alph{twostep}).}{\usecounter{twostep}\setlength{\rightmargin}{\leftmargin}}
\item \setcounter{twostepseca}{\value{twostep}} Determine two sets
of linear combinations of the elements of coordinate vector
$\bar{\bm{r}}'$ which transform under particular non-orthogonal
$[N]$ and $[N-1, \hspace{1ex} 1]$ irreducible representations of
$S_N$\,. Using these two sets of coordinates we then determine two
sets of linear combinations of the elements of coordinate vector
$\overline{\bm{\gamma}}'$ which transform under exactly the same
non-orthogonal $[N]$ and $[N-1, \hspace{1ex} 1]$ irreducible
representations of $S_N$ as the coordinate sets in the
$\bar{\bm{r}}'$ sector. In this way we satisfy item~\ref{it:teq0}
of Subsec.~\ref{subsec:normcond} above. Then another set of linear
combinations of the elements of coordinate vector
$\overline{\bm{\gamma}}'$ which transforms under a particular
non-orthogonal $[N-2, \hspace{1ex} 2]$ irreducible representation
of $S_N$. These sets of coordinates, which we term primitive
irreducible coordinates for reasons discussed below, have $W$
matrices which automatically satisfy Eq.~(\ref{eq:wawapt}).
\item \setcounter{twostepsecb}{\value{twostep}} Appropriate linear
combinations within each coordinate set from
item~\alph{twostepseca}).\ above are taken so that the results
transform under orthogonal irreducible representations of $S_N$\,.
These are then the symmetry coordinates of
Section~\ref{sec:symnorm} above. Care is taken to ensure that the
transformation from the coordinates which transform under the
non-orthogonal irreducible representations of $S_N$ of
item~\alph{twostepseca}).\ above, to the symmetry coordinates
which transform from the under orthogonal irreducible
representations of $S_N$ preserve the identity of equivalent
representations in the $\bar{\bm{r}}'$ and
$\overline{\bm{\gamma}}'$ sectors to ensure that
item~\ref{it:teq0} of Subsec.~\ref{subsec:normcond} above
continues to be satisfied. In this way both items~\ref{it:teq0}
and \ref{it:ort} of Subsec.~\ref{subsec:normcond} above are
satisfied.
\end{list}
What are the advantages of this two-step process? In
step~\alph{twostepseca}).\ above we set out to find sets of
coordinates transforming irreducibly under $S_N$ which have the
simplest functional form possible. This is why we call them
primitive irreducible coordinates. In step~\alph{twostepsecb}).\
above, we find appropriate linear combinations of the primitive
irreducible coordinates so that they transform under orthogonal
representations of $S_N$. Thus the symmetry coordinates are
composed of the building blocks of the primitive irreducible
coordinates. Furthermore, we choose one of the symmetry
coordinates to be just a single primitive irreducible coordinate,
and so it describes the simplest motion possible under the
requirement that it transforms irreducibly under $S_N$. The
succeeding symmetry coordinate is then chosen to be composed of
two primitive invariant coordinates and so on. In this way the
complexity of the motions described by the symmetry coordinates is
kept to a minimum and only builds up slowly as more symmetry
coordinates of a given species are considered.

\subsection{The Primitive Irreducible Coordinate Vector, ${\mathbf{\overline{\bm{S}}_{\bar{\bm{r}}'}}}$\,.}
The primitive irreducible coordinates,
$\overline{\bm{S}}_{\bar{\bm{r}}'}$, of the $\bar{\bm{r}}'$ sector
which transform under irreducible, though non-orthogonal,
representations of the group $S_N$ are given by
\begin{equation}\label{eq:sbwbr}
\overline{\bm{S}}_{\bar{\bm{r}}'} = \overline{W}_{\bar{\bm{r}}'}
\bar{\bm{r}}' \,,
\end{equation}
where
\begin{equation}\label{eq:wbr}
\overline{W}_{\bar{\bm{r}}'} = \left( \begin{array}{l} \overline{W}_{\bar{\bm{r}}'}^{[N]} \\
\overline{W}_{\bar{\bm{r}}'}^{[N-1, \hspace{1ex} 1]}
\end{array} \right) \,,
\end{equation}
$\overline{W}_{\bar{\bm{r}}'}^{[N]}$ is a $1 \times N$ dimensional
matrix and $\overline{W}_{\bar{\bm{r}}'}^{[N-1, \hspace{1ex} 1]}$
is an $(N-1) \times N$ dimensional matrix. Hence we can write
\begin{equation}\label{eq:sbr}
\overline{\bm{S}}_{\bar{\bm{r}}'} = \left( \begin{array}{l} \overline{\bm{S}}_{\bar{\bm{r}}'}^{[N]} \\
\overline{\bm{S}}_{\bar{\bm{r}}'}^{[N-1, \hspace{1ex} 1]} \end{array} \right) = \left( \begin{array}{l} \overline{W}_{\bar{\bm{r}}'}^{[N]} \bar{\bm{r}}' \\
\overline{W}_{\bar{\bm{r}}'}^{[N-1, \hspace{1ex} 1]} \bar{\bm{r}}'
\end{array} \right) \,.
\end{equation}

Since $\overline{W}_{\bar{\bm{r}}'}$ is a matrix which effects the
reduction of the representation, $D_{\bar{\bm{r}}'}(R)$, under
which $\bar{\bm{r}}'$ transforms, to particular irreducible,
though non-orthogonal, representations of the group $S_N$ under
which $\overline{\bm{S}}_{\bar{\bm{r}}'}$ transforms, then
\begin{equation}
\overline{W}_{\bar{\bm{r}}'} \, D_{\bar{\bm{r}}'}(R)\,
\overline{W}_{\bar{\bm{r}}'}^{-1} \,\,
\overline{\bm{S}}_{\bar{\bm{r}}'}=
\overline{D}_{\overline{S}_{\bar{\bm{r}}'}}(R) \,\,
\overline{\bm{S}}_{\bar{\bm{r}}'} \,,
\end{equation}
where
\begin{equation}
\overline{D}_{\overline{S}_{\bar{\bm{r}}'}}(R) = \left(
\begin{array}{c|c} \overline{D}_{\overline{S}_{\bar{\bm{r}}'}}^{[N]}(R) & \bm{0} \\ \hline \bm{0} &
\overline{D}_{\overline{S}_{\bar{\bm{r}}'}}^{[N-1, \hspace{1ex}
1]}(R) \,,
\end{array} \right)\,.
\end{equation}
\noindent In the above,
$\overline{D}_{\overline{S}_{\bar{\bm{r}}'}}^{[N]}(R)$ is the
one-dimensional irreducible matrix representation $[N]$ of $S_N$
and $\overline{D}_{\overline{S}_{\bar{\bm{r}}'}}^{[N-1,
\hspace{1ex} 1]}(R)$ is an $(N-1) \times (N-1)$-dimensional,
non-orthogonal, irreducible matrix representation $[N-1,
\hspace{1ex} 1]$ of $S_N$.

Thus the primitive irreducible coordinate column vector
${\overline{\bm{S}}_{\bar{\bm{r}}'}^{[N]}}$ is a one-element
vector and transforms under
$\overline{D}_{\overline{S}_{\bar{\bm{r}}'}}^{[N]}(R)$, a
one-dimensional non-orthogonal irreducible matrix representation
$[N]$ of $S_N$. The symmetry coordinate column vector
$\overline{\bm{S}}_{\bar{\bm{r}}'}^{[N-1, \hspace{1ex} 1]}$ is a
$(N-1)$-element vector and transforms under
$\overline{D}_{\overline{S}_{\bar{\bm{r}}'}}^{[N-1, \hspace{1ex}
1]}(R)$, the aforementioned $(N-1) \times (N-1)$-dimensional,
non-orthogonal, irreducible matrix representation $[N-1,
\hspace{1ex} 1]$ of $S_N$.

\subsection{The Primitive Irreducible Coordinate Vector, ${\mathbf{\overline{\bm{S}}_{\overline{\bm{\gamma}}'}}}$\,.}
Likewise the primitive irreducible coordinates,
$\overline{\bm{S}}_{\overline{\bm{\gamma}}'}$, of the
$\overline{\bm{\gamma}}'$ sector which transform under
irreducible, though non-orthogonal, representations of the group
$S_N$ are given by
\begin{equation}\label{eq:sbwbgamma}
\overline{\bm{S}}_{\overline{\bm{\gamma}}'} =
\overline{W}_{\overline{\bm{\gamma}}'} \, \overline{\bm{\gamma}}'
\end{equation}
and so writing
\begin{equation}\label{eq:wbgamma}
\overline{W}_{\overline{\bm{\gamma}}'} = \left( \begin{array}{l} \overline{W}_{\overline{\bm{\gamma}}'}^{[N]} \\
\overline{W}_{\overline{\bm{\gamma}}'}^{[N-1, \hspace{1ex} 1]} \\
\overline{W}_{\overline{\bm{\gamma}}'}^{[N-2, \hspace{1ex} 2]}
\end{array} \right) \,,
\end{equation}
where $\overline{W}_{\overline{\bm{\gamma}}'}^{[N]}$ is a $1
\times N(N-1)/2$ dimensional matrix,
$\overline{W}_{\overline{\bm{\gamma}}'}^{[N-1, \hspace{1ex} 1]}$
is an $(N-1) \times N(N-1)/2$ dimensional matrix and
$\overline{W}_{\overline{\bm{\gamma}}'}^{[N-2, \hspace{1ex} 2]}$
is an $N(N-3)/2 \times N(N-1)/2$ dimensional matrix, then
\begin{equation}\label{eq:sbgamma}
\overline{\bm{S}}_{\overline{\bm{\gamma}}'} = \left( \begin{array}{l} \overline{\bm{S}}_{\overline{\bm{\gamma}}'}^{[N]} \\
\overline{\bm{S}}_{\overline{\bm{\gamma}}'}^{[N-1, \hspace{1ex} 1]} \\
\overline{\bm{S}}_{\overline{\bm{\gamma}}'}^{[N-2, \hspace{1ex}
2]}
\end{array} \right) =
\left( \begin{array}{l} \overline{W}_{\overline{\bm{\gamma}}'}^{[N]} \, \overline{\bm{\gamma}}' \\
\overline{W}_{\overline{\bm{\gamma}}'}^{[N-1, \hspace{1ex} 1]} \,
\overline{\bm{\gamma}}' \\
\overline{W}_{\overline{\bm{\gamma}}'}^{[N-2, \hspace{1ex} 2]} \,
\overline{\bm{\gamma}}'
\end{array} \right) \,,
\end{equation}
where
\begin{equation} \label{eq:Wbsum2}
\overline{W}_{\overline{\bm{\gamma}}'}^\alpha \,
\overline{\bm{\gamma}}' = \sum_{j=1}^N \sum_{i < j} \,\,
[\overline{W}_{\overline{\bm{\gamma}}'}^\alpha]_{ij} \,
\overline{\gamma}'_{ij} \,.
\end{equation}

Since $\overline{W}_{\overline{\bm{\gamma}}'}$ is a matrix which
effects the reduction of the representation,
$\overline{D}_{\overline{\bm{\gamma}}'}(R)$, under which the
$\overline{\bm{\gamma}}'$ vector transforms, to particular
irreducible, though non-orthogonal, representations of the group
$S_N$ under which $\overline{\bm{S}}_{\overline{\bm{\gamma}}'}$
transforms, then
\begin{equation}
\overline{W}_{\overline{\bm{\gamma}}'} \,
D_{\overline{\bm{\gamma}}'}(R)\,
\overline{W}_{\overline{\bm{\gamma}}'}^{-1} \,\,
\overline{\bm{S}}_{\overline{\bm{\gamma}}'} =
\overline{D}_{\overline{S}_{\overline{\bm{\gamma}}'}}(R) \,\,
\overline{\bm{S}}_{\overline{\bm{\gamma}}'} \,,
\end{equation}
where
\begin{equation}
\overline{D}_{\overline{S}_{\overline{\bm{\gamma}}'}}(R) = \left(
\begin{array}{c|c|c} \overline{D}_{\overline{S}_{\overline{\bm{\gamma}}'}}^{[N]}(R) & \bm{0} & \bm{0} \\ \hline \bm{0} &
\overline{D}_{\overline{S}_{\overline{\bm{\gamma}}'}}^{[N-1, \hspace{1ex} 1]}(R) & \bm{0} \\
\hline \bm{0} & \bm{0} &
\overline{D}_{\overline{S}_{\overline{\bm{\gamma}}'}}^{[N-2,
\hspace{1ex} 2]}(R)
\end{array} \right) \,.
\end{equation}
\noindent In the above,
$\overline{D}_{\overline{S}_{\overline{\bm{\gamma}}'}}^{[N]}(R)$
is the one-dimensional irreducible matrix representation $[N]$ of
$S_N$,
$\overline{D}_{\overline{S}_{\overline{\bm{\gamma}}'}}^{[N-1,
\hspace{1ex} 1]}(R)$ is an $(N-1) \times (N-1)$-dimensional,
non-orthogonal, irreducible matrix representation $[N-1,
\hspace{1ex} 1]$ of $S_N$ and
$\overline{D}_{\overline{S}_{\overline{\bm{\gamma}}'}}^{[N-2,
\hspace{1ex} 2]}(R)$ is an $\{N(N-3)/2\} \times
\{N(N-3)/2\}$-dimensional, non-orthogonal, irreducible matrix
representation $[N-2, \hspace{1ex} 2]$ of $S_N$.

Note that according to item~\alph{twostepseca}).\ above,
\begin{equation} \label{eq:DbreqDbg}
\overline{D}_{\overline{S}_{\bar{\bm{r}}'}}^{\alpha}(R) =
\overline{D}_{\overline{S}_{\overline{\bm{\gamma}}'}}^{\alpha}(R)
= \overline{D}^{\alpha}(R) \,,
\end{equation}
where $\alpha$ is the partition $[N]$ or $[N-1, \hspace{1ex}
1]$\,.

Thus primitive irreducible coordinate column vector
$\overline{\bm{S}}_{\overline{\bm{\gamma}}'}^{[N]}$ is a
one-element vector and transforms under $\overline{D}^{[N]}(R)$,
the one-dimensional irreducible matrix representation $[N]$ of
$S_N$, while the primitive irreducible coordinate column vector
$\overline{\bm{S}}_{\overline{\bm{\gamma}}'}^{[N-1, \hspace{1ex}
1]}$ is an $(N-1)$-element vector and transforms under
$\overline{D}^{[N-1, \hspace{1ex} 1]}(R)$, an $(N-1) \times
(N-1)$-dimensional, non-orthogonal, irreducible matrix
representation $[N-1, \hspace{1ex} 1]$ of $S_N$. The primitive
irreducible coordinate column vector
$\overline{\bm{S}}_{\overline{\bm{\gamma}}'}^{[N-2, \hspace{1ex}
2]}$ is an $N(N-3)/2$-element vector and transforms under
$\overline{D}^{[N-2, \hspace{1ex} 2]}(R)$, the $\{N(N-3)/2\}
\times \{N(N-3)/2\}$-dimensional, non-orthogonal, irreducible
matrix representation $[N-2, \hspace{1ex} 2]$ of $S_N$.

We also note that according to Eq.~(\ref{eq:wawapt})
\begin{eqnarray} \label{eq:wbawbapt}
\overline{W}_{\bm{X}'}^\alpha [\overline{W}_{\bm{X}'}^{\alpha'}]^T
= 0 \,, & \forall & \alpha \neq \alpha' \,.
\end{eqnarray}

\subsection{The Full Primitive Irreducible Coordinate Vector,
${\mathbf{\overline{\bm{S}}}}$\,.} Like for the discussion of the
symmetry coordinates in Subsection~\ref{subsec:symCoor} above, it
is useful to form a full primitive irreducible coordinate vector
that groups primitive irreducible coordinates of the same species
together as follows:
\begin{equation}\label{eq:FSbCV}
\overline{\bm{S}} = P \left( \begin{array}{l} \overline{\bm{S}}_{\bar{\bm{r}}'} \\
\overline{\bm{S}}_{\overline{\bm{\gamma}}'} \end{array} \right) = \left( \begin{array}{l} \overline{\bm{S}}_{\bar{\bm{r}}'}^{[N]} \\
\overline{\bm{S}}_{\overline{\bm{\gamma}}'}^{[N]}  \\ \hline \overline{\bm{S}}_{\bar{\bm{r}}'}^{[N-1, \hspace{1ex} 1]} \\
\overline{\bm{S}}_{\overline{\bm{\gamma}}'}^{[N-1, \hspace{1ex}
1]}
\\ \hline
\overline{\bm{S}}_{\overline{\bm{\gamma}}'}^{[N-2, \hspace{1ex} 2]} \end{array} \right) = \left( \begin{array}{l} \overline{\bm{S}}^{[N]} \\
\overline{\bm{S}}^{[N-1, \hspace{1ex} 1]} \\
\overline{\bm{S}}^{[N-2, \hspace{1ex} 2]} \end{array} \right) \,,
\end{equation}
where the orthogonal matrix $P$ is given by Eq.~(\ref{eq:p}) and
\begin{equation}\label{eq:sbrep}
\begin{array}{cc@{\mbox{\hspace{2ex}and\hspace{2ex}}}c}
\overline{\bm{S}}^{[N]} = \left( \begin{array}{l} \overline{\bm{S}}_{\bar{\bm{r}}'}^{[N]} \\
\overline{\bm{S}}_{\overline{\bm{\gamma}}'}^{[N]} \end{array}
\right) \,, &
\overline{\bm{S}}^{[N-1, \hspace{1ex} 1]} = \left( \begin{array}{l} \overline{\bm{S}}_{\bar{\bm{r}}'}^{[N-1, \hspace{1ex} 1]} \\
\overline{\bm{S}}_{\overline{\bm{\gamma}}'}^{[N-1, \hspace{1ex}
1]}
\end{array} \right) &
\end{array}
\overline{\bm{S}}^{[N-2, \hspace{1ex} 2]} =
\overline{\bm{S}}_{\overline{\bm{\gamma}}'}^{[N-2, \hspace{1ex}
2]} \,.
\end{equation}

Now consider applying a transformation $\overline{W}$ to
Eqs.~(\ref{eq:qyt}), (\ref{eq:FGit}), (\ref{eq:character}) and
(\ref{eq:normit}), where
\begin{equation}\label{eq:Wb}
\overline{W}= P \left( \begin{array}{c|c} \overline{W}_{\bar{\bm{r}}'} & \bm{0} \\
\hline \bm{0} & \overline{W}_{\overline{\bm{\gamma}}'}
\end{array} \right) =
\left( \begin{array}{cc} \overline{W}_{\bar{\bm{r}}'}^{[N]} & \bm{0} \\
\bm{0} & \overline{W}_{\overline{\bm{\gamma}}'}^{[N]} \\
\hline
\overline{W}_{\bar{\bm{r}}'}^{[N-1, \hspace{1ex} 1]} & \bm{0} \\
\bm{0} & \overline{W}_{\overline{\bm{\gamma}}'}^{[N-1, \hspace{1ex} 1]} \\
\hline \bm{0} & \overline{W}_{\overline{\bm{\gamma}}'}^{[N-2,
\hspace{1ex} 2]}
\end{array} \right) \,,
\end{equation}
where
\begin{equation}\label{eq:SbWby}
{\overline{\bm{S}}} = \overline{W} \, {\bar{\bm{y}}'} \,.
\end{equation}
Eq.~(\ref{eq:SbWby}) implies that the representation under which
the displacement coordinate vector ${\bm{S}}$ transforms is
\begin{equation}\label{eq:DbSbWbDby}
\overline{D}_{\overline{S}} = \overline{W} D_{\bar{\bm{y}}'}(R) \,
\overline{W}^{-1} \,,
\end{equation}
where $\overline{D}_{\bar{\bm{y}}'}(R) $ is given by
Eq.~(\ref{eq:Dyb}). The non-orthogonal matrix representation
$D_{\overline{S}}(R)$ is a block diagonal matrix of irreducible,
though non-orthogonal, representations of the form
\begin{eqnarray}
\overline{D}_{\overline{S}}(R) &=& P \left( \begin{array}{c|c} \overline{D}_{\overline{S}_{\bar{\bm{r}}'}}(R) & \bm{0} \\
\hline \bm{0} & \overline{D}_{\overline{S}_{\overline{\bm{\gamma}}'}}(R) \end{array} \right) P^T \nonumber \\
 &=& \left(
\begin{array}{cc|cc|c}
\overline{D}^{[N]}(R) & \bm{0} & \bm{0} & \bm{0} & \bm{0} \\
\bm{0} & \overline{D}^{[N]}(R) & \bm{0} & \bm{0} & \bm{0} \\
\hline \bm{0} & \bm{0} & \overline{D}^{[N-1, \hspace{1ex} 1]}(R) &
\bm{0} & \bm{0} \\ \bm{0} & \bm{0} &
\bm{0} & \overline{D}^{[N-1, \hspace{1ex} 1]}(R) & \bm{0} \\
\hline \bm{0} & \bm{0} & \bm{0} & \bm{0} & \overline{D}^{[N-2,
\hspace{1ex} 2]}(R)
\end{array} \right) \,,
\label{eq:DbSb}
\end{eqnarray}
where we have also used Eq.~(\ref{eq:DbreqDbg}) from above in the
last step.

\subsection{The Transformation, ${\mathbf{U}}$\,, from Primitive Irreducible
Coordinates to Symmetry Coordinates.} The symmetry coordinate
vector ${\bm{S}}$ will be related to the primitive irreducible
coordinate vector ${\overline{\bm{S}}}$ by a non-orthogonal linear
transformation $U$, i.e.\
\begin{equation} \label{eq:SUSb}
{\bm{S}} = U \, {\overline{\bm{S}}}
\end{equation}
and likewise $D_S(R)$ will be related to
$\overline{D}_{\overline{S}}(R)$ by
\begin{equation} \label{eq:DUDbUi}
D_S(R) = U \, \overline{D}_{\overline{S}}(R) \, U^{-1} \,.
\end{equation}
Thus from Eqs.~(\ref{eq:SWy}), (\ref{eq:SbWby}), (\ref{eq:DSWDy})
and (\ref{eq:DbSbWbDby}), $W$ will be related to $\overline{W}$ by
the same transformation, i.e.\
\begin{equation} \label{eq:WUWb}
W = U \, \overline{W} \,,
\end{equation}
where
\begin{equation}
U = \left( \begin{array}{cc|cc|c}
    U_{\bar{\bm{r}}'}^{[N]} & 0 & 0 & 0 & 0 \\
    0 & U_{\overline{\bm{\gamma}}'}^{[N]} & 0 & 0 & 0 \\ \hline
    0 & 0 & U_{\bar{\bm{r}}'}^{[N-1, \hspace{1ex} 1]} & 0 & 0 \\
    0 & 0 & 0 & U_{\overline{\bm{\gamma}}'}^{[N-1, \hspace{1ex} 1]} & 0 \\ \hline
    0 & 0 & 0 & 0 & U_{\overline{\bm{\gamma}}'}^{[N-2, \hspace{1ex} 2]}
    \end{array} \right) \,,
\end{equation}
so that
\begin{equation} \label{eq:WUWbm}
W = \left( \begin{array}{cc} U_{\bar{\bm{r}}'}^{[N]} \, \overline{W}_{\bar{\bm{r}}'}^{[N]} & \bm{0} \\
\bm{0} & U_{\overline{\bm{\gamma}}'}^{[N]} \, \overline{W}_{\overline{\bm{\gamma}}'}^{[N]} \\
\hline
U_{\bar{\bm{r}}'}^{[N-1, \hspace{1ex} 1]} \, \overline{W}_{\bar{\bm{r}}'}^{[N-1, \hspace{1ex} 1]} & \bm{0} \\
\bm{0} & U_{\overline{\bm{\gamma}}'}^{[N-1, \hspace{1ex} 1]} \,
\overline{W}_{\overline{\bm{\gamma}}'}^{[N-1, \hspace{1ex} 1]} \\
\hline \bm{0} & U_{\overline{\bm{\gamma}}'}^{[N-2, \hspace{1ex}
2]} \, \overline{W}_{\overline{\bm{\gamma}}'}^{[N-2, \hspace{1ex}
2]}
\end{array} \right) \,.
\end{equation}
Note that from Eqs.~(\ref{eq:DS}), (\ref{eq:DbSb}) and
(\ref{eq:DUDbUi}),
\begin{equation} \label{eq:UgcUr}
U^\alpha_{\overline{\bm{\gamma}}'} = A_U^\alpha \,
U^\alpha_{\bar{\bm{r}}'} \,,
\end{equation}
where $\alpha$ is $[N]$ or $[N-1, \hspace{1ex} 1]$ and
$A_U^\alpha$ is a number (see below).

Comparing Eq.~(\ref{eq:WUWbm}) with Eq.~(\ref{eq:W}) we see that
\begin{equation} \label{eq:WaXUaXWbaX}
W^\alpha_{\bm{X}'} = U^\alpha_{\bm{X}'} \,
\overline{W}^\alpha_{\bm{X}'}
\end{equation}
and according to Eq.~(\ref{eq:WaWaI}) $U^\alpha_{\bm{X}'}$ must
satisfy
\begin{equation}\label{eq:aaI}
U^\alpha_{\bm{X}'} \, \left\{\overline{W}^\alpha_{\bm{X}'}
[\overline{W}^\alpha_{\bm{X}'}]^T \right\} \,
[U^\alpha_{\bm{X}'}]^T = I_\alpha \,.
\end{equation}
Thus to determine $U^\alpha_{\bm{X}'}$ we need to know
$\overline{W}^\alpha_{\bm{X}'} [\overline{W}^\alpha_{\bm{X}'}]^T$
beforehand. Note that when we use Eq.~(\ref{eq:UgcUr}) in
Eq.~(\ref{eq:aaI}) we must have
\begin{equation} \label{eq:Wbg2eq1oa2Wbr2}
\overline{W}^\alpha_{\overline{\bm{\gamma}}'}
[\overline{W}^\alpha_{\overline{\bm{\gamma}}'}]^T =
\frac{1}{(A_U^\alpha)^2} \, \overline{W}^\alpha_{\bar{\bm{r}}'}
[\overline{W}^\alpha_{\bar{\bm{r}}'}]^T
\end{equation}
when $\alpha$ is $[N]$ or $[N-1, \hspace{1ex} 1]$ \,.

\subsection{Symmetry Coordinates Belonging to the $\mathbf{[N]}$ Species}

\subsubsection{The $\mathbf{\bar{\bm{r}}'}$ Sector}

\paragraph{The Primitive Irreducible Coordinate}
The primitive irreducible coordinate
$\overline{\bm{S}}_{\bar{\bm{r}}'}^{[N]}$ which is invariant under
$S_N$ is readily seen to be
\begin{equation} \label{eq:SbNr}
\overline{\bm{S}}_{\bar{\bm{r}}'}^{[N]} = \sum_{i=1}^N
\overline{r}'_i \,.
\end{equation}
From Eq.~(\ref{eq:sbr}) we can identify
$\overline{W}^{[N]}_{\bar{\bm{r}}'}$ as
\begin{equation} \label{eq:WbNr}
[\overline{W}^{[N]}_{\bar{\bm{r}}'}]_i =
[{\bm{1}}_{\bar{\bm{r}}'}]_i
\end{equation}
where ${\bm{1}}_{\bar{\bm{r}}'}$ is defined in
Eq.~(\ref{eq:bf1i}). We leave the calculation of
$\overline{\bm{S}}_{\bar{\bm{r}}'}^{[N-1, \hspace{1ex} 1]}$ and
$\overline{W}^{[N-1, \hspace{1ex} 1]}_{\bar{\bm{r}}'}$ to a
subsequent paper\cite{paperII}.

\paragraph{The Symmetry Coordinate}
One readily sees that
\begin{equation}
\overline{W}^{[N]}_{\bar{\bm{r}}'}
[\overline{W}^{[N]}_{\bar{\bm{r}}'}]^T = N
\end{equation}
and so from Eq.~(\ref{eq:aaI})
\begin{equation} \label{eq:UNr}
U^{[N]}_{\bar{\bm{r}}'} = \frac{1}{\sqrt{N}} \,.
\end{equation}
Thus from Eqs.~(\ref{eq:SUSb}), (\ref{eq:WaXUaXWbaX}),
(\ref{eq:SbNr}), (\ref{eq:WbNr}) and (\ref{eq:UNr})
\begin{equation} \label{eq:WNeqsqrtN1}
[W^{[N]}_{\bar{\bm{r}}'}]_i = \frac{1}{\sqrt{N}} \,
[{\bm{1}}_{\bar{\bm{r}}'}]_i
\end{equation}
and
\begin{equation} \label{eq:SNr}
{\bm{S}}_{\bar{\bm{r}}'}^{[N]} = \frac{1}{\sqrt{N}} \,
\sum_{i=1}^N \overline{r}'_i \,.
\end{equation}
We leave the calculation of $W^{[N-1, \hspace{1ex}
1]}_{\bar{\bm{r}}'}$ and ${\bm{S}}_{\bar{\bm{r}}'}^{[N-1,
\hspace{1ex} 1]}$ to a subsequent paper\cite{paperII}.

\paragraph{Motions Associated with Symmetry
Coordinate ${\bm{S}}_{\bar{\bm{r}}'}^{[N]}$\,.} According to
Eqs.~(\ref{eq:yS}), (\ref{eq:bf1i}), (\ref{eq:rpaxi}) and
(\ref{eq:WNeqsqrtN1}) the motions associated with symmetry
coordinate ${\bm{S}}_{\bar{\bm{r}}'}^{[N]}$ in the unscaled
internal displacement coordinates ${\bm{r}}$ about the unscaled
Lewis structure configuration
\begin{equation} \label{eq:rinftyup}
r_\infty = D^2 \overline{a}_{ho} \overline{r}'_\infty
{\bm{1}}_{\bar{\bm{r}}'}
\end{equation}
are given by
\begin{equation}
{\bm{r}}^{[N]} = \overline{a}_{ho} \, D^{3/2} \,
\bar{\bm{r}}^{\prime [N]} = \overline{a}_{ho} \, D^{3/2} \,
{\bm{S}}_{\bar{\bm{r}}'}^{[N]} \, [(W_{\bar{\bm{r}}'}^{[N]})]^T =
\overline{a}_{ho} \, \sqrt{\frac{D^3}{N}} \,
{\bm{S}}_{\bar{\bm{r}}'}^{[N]} \, {\bm{1}}_{\bar{\bm{r}}'} \,.
\end{equation}
Thus we see that the motions associated with symmetry coordinate
${\bm{S}}_{\bar{\bm{r}}'}^{[N]}$ involve symmetric-stretch motions
about the Lewis structure configuration, where all of the radii
expand and contract together.

We leave the calculation of ${\bm{r}}^{[N-1, \hspace{1ex} 1]}$ to
a subsequent paper\cite{paperII}.

\subsubsection{The $\mathbf{\overline{\bm{\gamma}}'}$ Sector}

\paragraph{The Primitive Irreducible Coordinate}
According to item~\alph{twostepseca}).\ above,
$\overline{\bm{S}}_{\overline{\bm{\gamma}}'}^{[N]}$ should
transform under exactly the same non-orthogonal irreducible $[N]$
representation of $S_N$ as
$\overline{\bm{S}}_{\bar{\bm{r}}'}^{[N]}$. Since
$\overline{\bm{S}}_{\bar{\bm{r}}'}^{[N]}$ is invariant under all
of the elements, we simply require that
$\overline{\bm{S}}_{\overline{\bm{\gamma}}'}^{[N]}$ be invariant
under $S_N$\,.

Now according to Eq.~(\ref{eq:int_coords})
\begin{equation} \label{eq:gij}
\gamma_{ij} = \widehat{{\bm{r}}_i} \, {\bm{.}} \,
\widehat{{\bm{r}}_j} \,,
\end{equation}
where $\widehat{{\bm{r}}_i}$ is the unit vector from the center of
the confining field to particle $i$. Now $\sum_{i=1}^N
\widehat{{\bm{r}}_i}$ is invariant under $S_N$ and so
$(\sum_{i=1}^N \widehat{{\bm{r}}_i}) \, {\bm{.}} \, (\sum_{j=1}^N
\widehat{{\bm{r}}_j})$ is invariant under $S_N$\,. The
contribution from the terms with $i=j$, i.e.\ $\sum_{i=1}^N
\widehat{{\bm{r}}_i} \, {\bm{.}} \, \widehat{{\bm{r}}_i} = N$\,,
is also invariant on its own. Since this is just equal to the
number $N$, we subtract this term from $(\sum_{i=1}^N
\widehat{{\bm{r}}_i}) \, {\bm{.}} \, (\sum_{j=1}^N
\widehat{{\bm{r}}_j})$ and consider the invariant quantity
$(\sum_{i=1}^N \widehat{{\bm{r}}_i}) \, {\bm{.}} \, (\sum_{j \neq
i} \widehat{{\bm{r}}_j})$\,. According to Eq.~(\ref{eq:gij}) this
equals $\sum_{i=1}^N \sum_{j \neq i} \gamma_{ij} = 2 \sum_{j=2}^N
\sum_{i < j} \gamma_{ij}$ and upon using Eq.~(\ref{eq:taylor2}) we
obtain $[N(N-1)/2] \, \overline{\gamma}_{\infty} + \delta^{1/2}
\sum_{j=2}^N \sum_{i < j} \overline{\gamma}'_{ij}$\,. Since the
first term is invariant under $S_N$, $\sum_{j=2}^N \sum_{i < j}
\overline{\gamma}'_{ij}$ is invariant under $S_N$, and so we
identify the primitive irreducible coordinate of the
$\overline{\bm{\gamma}}'$ sector which transforms under the $[N]$
representation as
\begin{equation} \label{eq:SbgN}
\overline{\bm{S}}_{\overline{\bm{\gamma}}'}^{[N]} = \sum_{j=2}^N
\sum_{i < j} \overline{\gamma}'_{ij} \,.
\end{equation}
From Eqs.~(\ref{eq:bfgammap}), (\ref{eq:sbgamma}),
(\ref{eq:Wbsum2}) and (\ref{eq:SbgN})
\begin{equation} \label{eq:Wbeq1}
[\overline{W}^{[N]}_{\overline{\bm{\gamma}}'}]_{ij} =
[{\bm{1}}_{\overline{\bm{\gamma}}'}]_{ij} \,,
\end{equation}
where ${\bm{1}}_{\overline{\bm{\gamma}}'}$ is defined in
Eq.~(\ref{eq:1eq1}). We leave the calculation of
$\overline{\bm{S}}_{\overline{\bm{\gamma}}'}^{[N-1, \hspace{1ex}
1]}$,\,\, $\overline{\bm{S}}_{\overline{\bm{\gamma}}'}^{[N-2,
\hspace{1ex} 2]}$,\,\, $\overline{W}^{[N-1, \hspace{1ex}
1]}_{\overline{\bm{\gamma}}'}$ and $\overline{W}^{[N-2,
\hspace{1ex} 2]}_{\overline{\bm{\gamma}}'}$ to a subsequent
paper\cite{paperII}.

\paragraph{The Symmetry Coordinate}
One readily sees that
\begin{equation}
\overline{W}^{[N]}_{\overline{\bm{\gamma}}'}
[\overline{W}^{[N]}_{\overline{\bm{\gamma}}'}]^T = \sum_{j=1}^N
\sum_{i < j} \,\,
[\overline{W}^{[N]}_{\overline{\bm{\gamma}}'}]_{ij} \,
[\overline{W}^{[N]}_{\overline{\bm{\gamma}}'}]_{ij} =
\frac{N(N-1)}{2} \,,
\end{equation}
where we have used Eqs.~(\ref{eq:Wbeq1}) and (\ref{eq:1eq1}) in
the last step, and so from Eq.~(\ref{eq:aaI})
\begin{equation} \label{eq:UNg}
U^{[N]}_{\overline{\bm{\gamma}}'} = \sqrt{\frac{2}{N(N-1)}} \,.
\end{equation}
Thus from Eqs.~(\ref{eq:SUSb}), (\ref{eq:WaXUaXWbaX}),
(\ref{eq:SbgN}), (\ref{eq:Wbeq1}) and (\ref{eq:UNg})
\begin{equation} \label{eq:WNgeqsqrt2ontnm11}
[W^{[N]}_{\overline{\bm{\gamma}}'}]_{ij} = \sqrt{\frac{2}{N(N-1)}}
\,\, [{\bm{1}}_{\overline{\bm{\gamma}}'}]_{ij}
\end{equation}
and
\begin{equation} \label{eq:SgN}
{\bm{S}}_{\overline{\bm{\gamma}}'}^{[N]} = \sqrt{\frac{2}{N(N-1)}}
\,\,\, \sum_{j=2}^N \sum_{i < j} \overline{\gamma}'_{ij} \,.
\end{equation}

We leave the calculation of $W^{[N-1, \hspace{1ex}
1]}_{\overline{\bm{\gamma}}'}$,\,\, $W^{[N-2, \hspace{1ex}
2]}_{\overline{\bm{\gamma}}'}$,\,\,
${\bm{S}}_{\overline{\bm{\gamma}}'}^{[N-1, \hspace{1ex} 1]}$ and
${\bm{S}}_{\overline{\bm{\gamma}}'}^{[N-2, \hspace{1ex} 2]}$to a
subsequent paper\cite{paperII}.

\paragraph{Motions Associated with Symmetry
Coordinate ${\bm{S}}_{\overline{\bm{\gamma}}'}^{[N]}$\,.}
According to Eqs.~(\ref{eq:yS}), (\ref{eq:gpaxi}), (\ref{eq:1eq1})
and (\ref{eq:WNgeqsqrt2ontnm11}) the motions associated with
symmetry coordinate ${\bm{S}}_{\overline{\bm{\gamma}}'}^{[N]}$ in
the unscaled internal displacement coordinates $\bm{\gamma}$ about
the Lewis structure configuration
\begin{equation} \label{eq:ginfty1}
\bm{\gamma}_\infty = \overline{\gamma}_\infty
{\bm{1}}_{\overline{\bm{\gamma}}'}
\end{equation}
are given by
\begin{equation} \label{eq:gammaN}
\bm{\gamma}^{[N]} = \frac{1}{\sqrt{D}} \, \bm{\gamma}^{\prime [N]}
= \frac{1}{\sqrt{D}} \, {\bm{S}}_{\overline{\bm{\gamma}}'}^{[N]}
\, [(W_{\overline{\bm{\gamma}}'}^{[N]})]^T =
\sqrt{\frac{2}{N(N-1)\, D}} \,
{\bm{S}}_{\overline{\bm{\gamma}}'}^{[N]} \,
{\bm{1}}_{\overline{\bm{\gamma}}'} \,.
\end{equation}
Thus we see that the motions associated with symmetry coordinate
${\bm{S}}_{\overline{\bm{\gamma}}'}^{[N]}$ involve symmetric-bend
motions about the Lewis structure configuration, where all of the
angles open out and contract together like a parasol, but with the
radii unchanged.

We leave the calculation of $\bm{\gamma}^{[N-1, \hspace{1ex} 1]}$
and $\bm{\gamma}^{[N-2, \hspace{1ex} 2]}$ to a subsequent
paper\cite{paperII}.

\section{The Frequencies and Normal-Mode Coordinates of the \\ $\mathbf{[N]}$
Species of the System.} \label{sec:FreqNorModN}
\subsection{The $\mathbf{G}$ and $\mathbf{FG}$ matrices in the
Symmetry-Coordinate Basis.} We can use the $W_{\bm{X}'}^{[N]}$
matrices of Eqs.~(\ref{eq:WNeqsqrtN1}) and
(\ref{eq:WNgeqsqrt2ontnm11}) to calculate the reduced $\bm{G}$ and
$\bm{FG}$ matrix elements,
$[\bm{\sigma_{[N]}^{\bm{G}}}]_{\bm{X}'_1,\,\bm{X}'_2}$ and
$[\bm{\sigma_{[N]}^{\bm{FG}}}]_{\bm{X}'_1,\,\bm{X}'_2}$\,, of
Eq.~(\ref{eq:sigmaQ}). An outline of this calculation is to be
found in Appendix~\ref{app:sigmacalc}, with the following results.

\subsubsection{The Matrix Elements $\mathbf{[
\sigma_{[N]}^{\bm{G}}]_{\bm{X}'_1,\,\bm{X}'_2}}$\,.} Using
Eqs.~(\ref{eq:Q}), (\ref{eq:Qrr}), (\ref{eq:Qrg}), (\ref{eq:Qgr}),
(\ref{eq:Qgg}), (\ref{eq:Gsub}), (\ref{eq:Gsym}),
(\ref{eq:Goneorzero}), (\ref{eq:sigmaQ}), (\ref{eq:WNeqsqrtN1})
and (\ref{eq:WNgeqsqrt2ontnm11}) we derive
\begin{equation} \label{eq:sigmaNG}
\bm{\sigma_{[N]}^{\bm{G}}} = \left(
\renewcommand{\arraystretch}{1.5}
\begin{array}{l@{\hspace{1.5em}}l} \protect[\bm{\sigma_{[N]}^{\bm{G}}}\protect]_{\bar{\bm{r}}',\,\bar{\bm{r}}'} =
\tilde{a}'
& {\displaystyle
\protect[\bm{\sigma_{[N]}^{\bm{G}}}\protect]_{\bar{\bm{r}}',\,
\overline{\bm{\gamma}}'} =
0} \\
{\displaystyle
\protect[\bm{\sigma_{[N]}^{\bm{G}}}\protect]_{\overline{\bm{\gamma}}',\,\bar{\bm{r}}'}
= 0} & {\displaystyle
\protect[\bm{\sigma_{[N]}^{\bm{G}}}\protect]_{\overline{\bm{\gamma}}',\,\overline{\bm{\gamma}}'}
= \left( \tilde{g}' + 2(N-1) \tilde{h}' \right)}
\end{array} \renewcommand{\arraystretch}{1} \right) \,.
\end{equation}
We leave the calculation of the matrix elements
$[\bm{\sigma_{[N-1, \hspace{1ex}
1]}^{\bm{G}}}]_{\bm{X}'_1,\,\bm{X}'_2}$ and $\bm{\sigma_{[N-2,
\hspace{1ex} 2]}^{\bm{G}}} = [\bm{\sigma_{[N-2, \hspace{1ex}
2]}^{\bm{FG}}}]_{\overline{\bm{\gamma}}',\,\overline{\bm{\gamma}}'}$
to a subsequent paper\cite{paperII}.

\subsubsection{The Matrix Elements $\mathbf{[
\sigma_{[N]}^{\bm{FG}}]_{\bm{X}'_1,\,\bm{X}'_2}}$\,.} Using
Eqs.~(\ref{eq:Q}), (\ref{eq:Qrr}), (\ref{eq:Qrg}), (\ref{eq:Qgr}),
(\ref{eq:Qgg}), (\ref{GFsub}), (\ref{GFsym}), (\ref{eq:sigmaQ}),
(\ref{eq:sigmamat}), (\ref{eq:WNeqsqrtN1}) and
(\ref{eq:WNgeqsqrt2ontnm11}) we derive
\begin{equation}  \label{eq:sigmaNFG}
\bm{\sigma_{[N]}^{\bm{FG}}} = \left(
\renewcommand{\arraystretch}{1.5}
\begin{array}{l@{\hspace{1.5em}}l} \protect[\bm{\sigma_{[N]}^{\bm{FG}}}\protect]_{\bar{\bm{r}}',\,\bar{\bm{r}}'} =
(\tilde{a}+\tilde{b}N) & {\displaystyle
\protect[\bm{\sigma_{[N]}^{\bm{FG}}}\protect]_{\bar{\bm{r}}',\,
\overline{\bm{\gamma}}'} =
\sqrt{2(N-1)} \,\, \left( \tilde{e}+\frac{N}{2}\tilde{f} \right)} \\
{\displaystyle
\protect[\bm{\sigma_{[N]}^{\bm{FG}}}\protect]_{\overline{\bm{\gamma}}',\,\bar{\bm{r}}'}
= \sqrt{2(N-1)} \,\, \left( \tilde{c}+\frac{N}{2}\tilde{d}
\right)} & {\displaystyle
\protect[\bm{\sigma_{[N]}^{\bm{FG}}}\protect]_{\overline{\bm{\gamma}}',\,\overline{\bm{\gamma}}'}
= \left(
\tilde{g}+2(N-1)\tilde{h}+\frac{N(N-1)}{2}\,\,\tilde{\iota}
\right)}
\end{array} \renewcommand{\arraystretch}{1} \right) \,.
\end{equation}
We leave the calculation of the matrix elements
$[\bm{\sigma_{[N-1, \hspace{1ex}
1]}^{\bm{FG}}}]_{\bm{X}'_1,\,\bm{X}'_2}$ and $\bm{\sigma_{[N-2,
\hspace{1ex} 2]}^{\bm{FG}}} = [\bm{\sigma_{[N-2, \hspace{1ex}
2]}^{\bm{FG}}}]_{\overline{\bm{\gamma}}',\,\overline{\bm{\gamma}}'}$
to a subsequent paper\cite{paperII}.

\subsection{The Frequencies and Normal Modes.}
Using Eqs.~(\ref{eq:sigmaNFG}) in Eq.~(\ref{eq:lambda12pm}) we
obtain $\lambda^\pm_{[N]}$\,. The frequencies are then determined
from Eq.~(\ref{eq:omega_p}).
The $\bar{\bm{r}}'$-$\overline{\bm{\gamma}}'$ mixing angles,
$\theta^{[N]}_\pm$ for the $[N]$ species are determined from
Eq.~(\ref{eq:tanthetaalphapm}). The normalization constant
$c^{[N]}$ of the reduced normal-coordinate coefficient vector,
${\mathsf{c}}^{[N]}$, of Eqs.~(\ref{eq:cb}) and
(\ref{eq:sfceqcthatc}) are determined from Eqs.~(\ref{eq:c2norm})
and (\ref{eq:calphapm}). One then determines the normal mode
vector, ${\bm{q}'}$\,, through Eqs.~(\ref{eq:qvector}),
(\ref{eq:SNr}) and (\ref{eq:SgN}). Thus we arrive at
\renewcommand{\jot}{0.5em}
\begin{eqnarray}
q_+^{\prime \, [N]} & = & c^{[N]}_+ \cos{\theta^{[N]}_+}
\frac{1}{\sqrt{N}} \, \sum_{i=1}^N \overline{r}'_i + c^{[N]}_+
\sin{\theta^{[N]}_+} \sqrt{\frac{2}{N(N-1)}} \,\,\,
\sum_{j=2}^N \sum_{i < j} \overline{\gamma}'_{ij} \\
q_-^{\prime \, [N]} & = & c^{[N]}_- \cos{\theta^{[N]}_-}
\frac{1}{\sqrt{N}} \, \sum_{i=1}^N \overline{r}'_i + c^{[N]}_-
\sin{\theta^{[N]}_-} \sqrt{\frac{2}{N(N-1)}} \,\,\, \sum_{j=2}^N
\sum_{i < j} \overline{\gamma}'_{ij}
\end{eqnarray}
\renewcommand{\jot}{0em}
We leave the calculation of ${\bm{q}'}_+^{[N-1, \hspace{1ex} 1]}$,
${\bm{q}'}_-^{[N-1, \hspace{1ex} 1]}$ and ${\bm{q}'}^{[N-2,
\hspace{1ex} 2]}$ to a subsequent paper\cite{paperII}.

\subsection{The Motions Associated with the Normal Modes.}
From Eqs.~(\ref{eq:yq}), (\ref{eq:yqinfty}),
(\ref{eq:pyqalphaxi}), (\ref{eq:myqalphaxi}), (\ref{eq:yqnm2xi}),
(\ref{eq:WNeqsqrtN1}) and (\ref{eq:WNgeqsqrt2ontnm11})
\begin{equation} \label{eq:yqf}
{\bm{y}} = \left( \begin{array}{c} {\bm{r}} \\
\bm{\gamma} \end{array} \right) = \,\, {\bm{y}}_\infty \,\, +
\begin{array}[t]{@{}l@{}} {\displaystyle \hspace{2ex} \sum} \\ {\scriptstyle \alpha= \left\{
\renewcommand{\arraystretch}{0.5}
\begin{array}{@{}c@{}} {\scriptstyle
\protect[N\protect]\,,} \\ {\scriptstyle \protect[N-1,
\hspace{1ex} 1\protect]}
\end{array} \renewcommand{\arraystretch}{1} \right\}
} \end{array} \sum_\xi \sum_{\tau=\pm} \, \left( \begin{array}{c} _{\tau}{\bm{r}}^{\alpha}_\xi \\
_{\tau}\bm{\gamma}^{\alpha}_\xi
\end{array} \right) \,\,\, + \,\,\, \sum_\xi \,
\left( \begin{array}{c} {\bm{0}} \\
\bm{\gamma}^{\protect[N-2, \hspace{1ex} 2\protect]}_\xi
\end{array} \right) \,,
\end{equation}
where ${\bm{y}}_\infty$ is given by Eq.~(\ref{eq:yqinfty}). The
$\xi$ sum for the $[N]$ species only includes one term and so
\begin{eqnarray}
_+{\bm{r}}^{[N]} & = & - \, \overline{a}_{ho} \,
\sqrt{\frac{D^3}{N}} \,
\frac{\sin{\theta^{[N]}_-}}{s(\theta^{[N]}) \, c^{[N]}_+} \,
\, q_+^{\prime \, [N]} {\bm{1}}_{\bar{\bm{r}}'} \,, \\
_+\bm{\gamma}^{[N]} & = & \sqrt{\frac{2}{N(N-1)\, D}} \,
\frac{\cos{\theta^{[N]}_-}}{s(\theta^{[N]}) \, c^{[N]}_+} \,\,
q_+^{\prime \, [N]} {\bm{1}}_{\overline{\bm{\gamma}}'} \,,
\end{eqnarray}
\begin{eqnarray}
_-{\bm{r}}^{[N]} & = & \overline{a}_{ho} \, \sqrt{\frac{D^3}{N}}
\, \frac{\sin{\theta^{[N]}_+}}{s(\theta^{[N]}) \, c^{[N]}_-} \,
\, q_-^{\prime \, [N]} {\bm{1}}_{\bar{\bm{r}}'} \,, \\
_-\bm{\gamma}^{[N]} & = & - \, \sqrt{\frac{2}{N(N-1)\, D}} \,
\frac{\cos{\theta^{[N]}_+}}{s(\theta^{[N]}) \, c^{[N]}_-} \,\,
q_-^{\prime \, [N]} {\bm{1}}_{\overline{\bm{\gamma}}'} \,.
\end{eqnarray}
We leave the calculation of $_{\tau}{\bm{r}}^{[N-1, \hspace{1ex}
1]}_\xi$\,,\,\, $_{\tau}\bm{\gamma}^{[N-1, \hspace{1ex} 1]}_\xi$
and $\bm{\gamma}^{[N-2, \hspace{1ex} 2]}_\xi$ to a subsequent
paper\cite{paperII}.

\section{Summary and Conclusions} \label{sec:SumConc}
In earlier papers\cite{FGpaper,energy,loeser} we have developed the method of
dimensional perturbation theory\cite{copen92} at low orders to examine
the energies of quantum-confined systems such as atoms, quantum
dots and Bose-Einstein condensates, both the ground and excited
states. Dimensional perturbation theory (DPT) has many advantages.
These include the fact that the number of particles, $N$\,, enters
into the theory as a parameter, and so it is easy to calculate
results for an arbitrary number of particles. Also the theory is a
beyond-mean-field method, directly accounting for each
interaction, rather than some average representation of the
interactions. This means that it is appropriate for confined
systems of both weakly interacting and strongly interacting
particles. In the case of a trapped gaseous atomic Bose-Einstein
condensate (BEC) the system is typically a weakly interacting
system for which a mean-field approximation is valid. However, if
$N$ is increased, or Feshbach resonances are exploited to increase
the effective scattering length, or the system is squeezed in one
or more directions, it will transition into a strongly-interacting
regime where the mean-field approach breaks down. Such systems
have been created in the laboratory. Dimensional perturbation
theory, however, is equally applicable to both regimes and may be
used to study the transition from weak to strongly interacting
regimes. Even at lowest order DPT includes beyond-mean-field
effects.

In this paper we have begun to address the problem of calculating
the zeroth-order DPT wave function for quantum-confined systems,
and it is a considerably expanded discussion of work briefly
outlined in a previous letter\cite{PRL}. At large dimensions the
Jacobian-weighted wave function of the system becomes harmonic,
with the system oscillating about a configuration termed the Lewis
structure. Notwithstanding the relatively simple form of the
large-dimension, zeroth-order wave function, it includes
beyond-mean-field effects.

With the wave function many more properties of the system become
accessible. The normal mode coordinates of the zeroth-order wave
function reveals the nature of the excitations of the system and
the lowest-order wave function yields expectation values and
transition matrix elements to zeroth order in DPT. For macroscopic
quantum-confined systems, such as the BEC, the wave function is
uncloaked in an explicit way as the density profile may be viewed
in a direct fashion. This is readily calculable at zeroth order in
DPT from the zeroth-order wave function. Also, calculating
energies and wave functions to higher orders in $1/D$ requires as
input the lowest-order wave function.

In this paper we limit ourselves to setting up the general theory
of the zeroth order DPT wave function for a quantum-confined
system. We illustrate this by applying it to the breathing and
center of mass modes of the $[N]$ species. In the case of the BEC
the ground state energies and excitation frequencies of these
modes have been calculated using DPT in a previous paper. The
modes of the $[N-1, \hspace{1ex} 1]$ and $[N-2, \hspace{1ex} 2]$
species are derived in a subsequent paper. The presence of these
additional normal-mode coordinates in the theory indicates the
existence of modes of excitation of the BEC beyond those which are
typically considered. The zeroth-order density profile of the
ground state condensate is to be considered in another paper, and
is relatively simple to calculate once the zeroth-order wave
function is known.

This paper restricts itself to consideration of quantum-confined
systems with a confining potential of spherical symmetry. The
extension of this method to systems with cylindrical symmetry is
relatively straightforward and is to be discussed in subsequent
papers. This is particularly important for the BEC as almost all current
laboratory traps have cylindrical symmetry.

As we have discussed in an earlier paper\cite{energy} considering
energies and excitation frequencies of the BEC, low-order DPT
calculations of confined interacting particles, and this includes
strongly interacting systems, are accurate out to $N$ equals a few
thousand particles. To obtain more accurate energies, and to
extend the calculation beyond a few thousand particles we need to
go beyond these low-order calculations of the energies and
excitation frequencies. These calculations are facilitated by the
present paper since, as we have already mentioned, it requires the
zeroth-order DPT wave function. This will be pursued in later
papers.

\section{Acknowledgments}
We would like to thank the Army Research Office for ongoing
support. This work was also supported in part by the Office of
Naval Research.

\appendix
\renewcommand{\theequation}{A\arabic{equation}}
\setcounter{equation}{0}
\section{The Wilson FG matrix method}\label{app:wilson}
In this appendix, we derive the Wilson FG matrix method\cite{dcw}
which is at the heart of our obtaining the normal mode
frequencies, the first-order energy correction, the normal
coordinates and the wave function at lowest order. The derivation
involves a transformation to the set of coordinates called
normal-coordinates in which both the differential term and the
potential term of Eq.~(\ref{Gham}) are diagonal. We begin by
defining a symmetric transformation, ${\bf A}$, that transforms
from the vector $\bar{\bm{y}}'$, defined by
Eq.~(\ref{eq:ytransposeP}), to $\bar{\bm{z}}'$
($\bar{\bm{z}}'={\bf A} \bar{\bm{y}}'$). ${\bf A}$ is an active
transformation and it has the property that it diagonalizes the
symmetric $\bm{G}$ matrix to unity. That is, ${\bf A}$ satisfies
\begin{eqnarray}
&& {\bf A}^{T}\bm{G}{\bf A}={\bf I},
\end{eqnarray}
which can also be written as
\begin{eqnarray}
\label{A2G} && \bm{G}={\bf A}^{-1}({\bf A}^{-1})^T,
\end{eqnarray}
where we have used the property that ${\bf A}$ is symmetric (${\bf
A}={\bf A}^{T}$) in the derivation of Eq.~(\ref{A2G}).
$\widehat{H}_1$ then becomes
\begin{equation}
\label{eq:H1} \widehat{H}_1 \to
-\frac{1}{2}\partial_{\bar{\bm{z}}'}^{T}\partial_{\bar{\bm{z}}'}
+\frac{1}{2}{\bar{\bm{z}}}^{\prime T} ({\bf A}^{-1})^{T}{\bf F}
{\bf A}^{-1} \bar{\bm{z}}'.
\end{equation}

Next we focus our attention on the potential term, the term
involving the matrix ${\bf F}$. We introduce another
transformation, ${\bf U}$ (${\bm{q}'}={\bf U} \bar{\bm{z}}'$),
that diagonalizes the potential term while simultaneously leaving
the differential term unchanged. This orthogonal transformation
(${\bf U}^T{\bf U}={\bf I}$, where ${\bf I}$ is the identity
matrix) leaves the differential term in the same form as in
Eq.~(\ref{eq:H1}), and the potential term becomes
\begin{equation}\label{tempeig}
{\bf U} ({\bf A}^{-1})^T {\bf F} {\bf A}^{-1} {\bf U}^T = {\bf
\Lambda},
\end{equation}
where $\Lambda$ is a diagonal matrix.  That is,
\begin{equation}\label{eq:appH1}
\widehat{H}_1 \to
-\frac{1}{2}\partial_{{\bm{q}'}}^{T}\partial_{{\bm{q}'}}
+\frac{1}{2}{\bm{q}'}^{T} {\bf \Lambda} {\bm{q}'}.
\end{equation}
From Eq.~(\ref{eq:appH1}) it is clear that ${\bf \Lambda}$ is the
diagonal matrix of the squares of the frequencies of the normal
modes (see Eq.~(\ref{eq:omega_p})).

Now if we introduce the projection operators
\begin{equation}
P_b = {\bm{b}}'' {\bm{b}}''^T \,,
\end{equation}
where the $c^{th}$ element of the column vector ${\bm{b}}''$ is
\begin{equation} \label{eq:ippdik}
({\bm{b}}'')_c = \delta_{bc}
\end{equation}
with $b$ and $c$ running from 1 to $(N+1)/2$, i.e.\ all of the
elements of the column vector are zero except the $b^{th}$ element
which is equal to unity, then
\begin{equation} \label{eq:Liij}
{\bf \Lambda} \, {\bm{b}}'' {\bm{b}}''^T \, {\bm{q}'} = [{\bf
\Lambda}]_{bb} \, {\bm{b}}'' [{\bm q'}]_b = \lambda_b \,
{\bm{b}}'' [{\bm q'}]_b \,,
\end{equation}
where $[{\bm q'}]_b$\,, the $b^{\rm th}$ normal-mode coordinate,
is given by
\begin{equation} \label{eq:qp}
[{\bm q'}]_b = {\bm{b}}''^T \, {\bm{q}'}
\end{equation}
Since $[{\bm q'}]_b$ is just a multiplier on all sides of
Eq.~(\ref{eq:Liij}), Eq.~(\ref{eq:Liij}) reduces to
\begin{equation} \label{eq:Liik}
{\bf \Lambda} \, {\bm{b}}'' = \lambda_b \, {\bm{b}}'' \,,
\end{equation}
Using Eq.~(\ref{tempeig}), Eq.~(\ref{eq:Liik}) reads
\begin{equation} \label{eq:Lii}
{\bf U} ({\bf A}^{-1})^T {\bf F} {\bf A}^{-1} {\bf U}^T \,
{\bm{b}}'' = \lambda_b \, {\bm{b}}'' \,.
\end{equation}

There are two important ways we can proceed from here.
\begin{enumerate}
\item \label{it:metric} If we now define
\begin{equation} \label{eq:ppup}
{\bm{b}}'' =  {\bf U} {\bf A} {\mathcal{B}}
\end{equation}
then Eq.~(\ref{eq:qp}), the equation for the $b^{\rm th}$
normal-mode coordinate, reads
\begin{equation} \label{eq:qGm1y}
[{\bm q'}]_b = {\mathcal{B}}^T \, {\bf A}^2 {\bar{\bm{y}}'} =
{\mathcal{B}}^T \, \bm{G}^{-1} {\bar{\bm{y}}'} \,,
\end{equation}
where we have used Eqs.~(\ref{A2G}) and (\ref{tempeig}), and from
Eq.~(\ref{eq:Lii}) we find that ${\mathcal{B}}$ satisfies the
eigenvalue equation
\begin{equation} \label{eq:GFi}
\bm{G} \, {\bf F} \, {\mathcal{B}}= \lambda_b \, {\mathcal{B}} \,.
\end{equation}
The normalization of ${\mathcal{B}}$ is determined by the
normalization condition for ${\bm{b}}''$,
\begin{equation} \label{eq:ippnorm}
{\bm{b}}''^T \, {\bm{b}}'' = 1
\end{equation}
(see Eq.~(\ref{eq:ippdik})). Using Eqs.~(\ref{A2G}) and
(\ref{tempeig}) in Eq.~(\ref{eq:ippnorm}) we obtain
\begin{equation} \label{eq:normI}
{\mathcal{B}}^T \bm{G}^{-1} {\mathcal{B}} = 1.
\end{equation}
Notice how the  constant matrix $\bm{G}^{-1}$ is the contravariant
metric tensor for the space. This is to be expected from the form
of Eq.~(\ref{Gham}).
\item All of the work applying DPT to large-$N$ quantum-confined
systems up till now has been focussed on calculating energies and
adopts the procedure outlined in item~\ref{it:metric}. The
frequency squared, $\lambda_b$, is calculated from
Eq.~(\ref{eq:GFi}) by solving
\begin{equation}
\label{GF} \det(\bm{G}{\bf F}-\lambda{\bf B})=0,
\end{equation}
Since we are also interested in large-$D$ wave functions and
normal coordinates, our life would be easier if ${\mathcal{B}}$
were the coefficient vector of the displacement coordinates
contributing to the normal coordinate $[{\bm q'}]_b$\,. Looking at
Eq.~(\ref{eq:qGm1y}) we see that this is not so as there is a
$\bm{G}^{-1}$ operation involved. Thus instead of following the
procedure outlined in item~\ref{it:metric} above, in this paper we
solve directly for the coefficient vector, ${\bm{b}}$, of the
displacement coordinates contributing to the normal coordinate
$[{\bm q'}]_b$\,,
\begin{equation} \label{eq:iI}
{\bm{b}} = \bm{G}^{-1} {\mathcal{B}} \,.
\end{equation}
Thus in terms of ${\bm{b}}$, Eq.~(\ref{eq:ppup}) for the normal
mode coordinate takes on the desired form
\begin{equation} \label{eq:qy}
[{\bm q'}]_b = {\bm{b}}^T {\bar{\bm{y}}'} \,.
\end{equation}
Using Eq.~(\ref{eq:iI}) in Eq.~(\ref{eq:GFi}) we find that the
coefficient vector satisfies the eigenvalue equation
\begin{equation} \label{eq:FGi}
{\bf F} \, \bm{G} \, {\bm{b}}= \lambda_b \, {\bm{b}}
\end{equation}
with the resultant secular equation
\begin{equation} \label{eq:detFG}
\det({\bf F}\bm{G}-\lambda{\bf I})=0.
\end{equation}
From Eq.~(\ref{eq:normI}) we see that the coefficient vector
satisfies the normalization condition
\begin{equation} \label{eq:normi}
{\bm{b}}^T \bm{G} \, {{\bm{b}}} = 1.
\end{equation}
\end{enumerate}

\renewcommand{\theequation}{B\arabic{equation}}
\setcounter{equation}{0}
\section{Gramian Determinants}\label{app:gram}
The Gramian determinant\cite{gantmacher} is defined as:
\begin{equation}
\Gamma\equiv\det(\gamma_{ij}),
\end{equation}
where $\gamma_{ij}={\bf r}_{i}\cdot{\bf r}_{j}/r_i r_j$, the angle
cosines between the particle radii ${\bf r}_{i}$ and ${\bf
r}_{j}$, represents the elements of an $N \times N$ matrix.  A
related quantity used in the main text is the principle minor of
the Gramian, $\Gamma^{(\alpha)}$, defined as the determinant of
the $\gamma_{ij}$ matrix with the $\alpha^{th}$ row and column
removed.

A most challenging part of calculating the large-$D$ minimum and
the ${\bf F}$ matrix elements in the systems discussed in this
paper is handling the Gramian determinants and their derivatives.
What makes these calculations feasible is the very high symmetry
of the infinite-dimension, symmetric minimum.  In this appendix we
discuss the Gramian determinant and show how to obtain its
derivatives at the infinite-dimension, symmetric minimum. We will
now demonstrate how this is done and summarize the results.

In general $\gamma_{ii}=1$ and at the infinite-$D$ symmetric
minimum all of the remaining direction cosines are equal,
$\gamma_{ij}=\overline{\gamma}_{\infty}$.  Hence, the Gramian
determinant is an $(N \times N)$ matrix of the form
\begin{equation}
\Gamma \Big|_{\infty} = \det \left(
\begin{array}{cccc}
1 & \overline{\gamma}_{\infty}        & \cdots & \overline{\gamma}_{\infty}      \\
\overline{\gamma}_{\infty}        & 1 &  \ddots        & \vdots   \\
\vdots  &  \ddots          & \ddots &  \overline{\gamma}_{\infty}      \\
\overline{\gamma}_{\infty}        & \cdots  &   \overline{\gamma}_{\infty}     & 1 \\
\end{array}
\right),
\end{equation}
in Appendices B and D of Ref.~\cite{FGpaper} we show that
(Eq.~(D4))
\begin{equation} \label{eq:gammainf}
\Gamma\Big|_{\infty} =
[1+(N-1)\overline{\gamma}_{\infty}](1-\overline{\gamma}_{\infty})^{N-1}.
\end{equation}
The principle minor evaluated at the infinite-$D$ symmetric
minimum $\Gamma^{(\alpha)}\Big|_{\infty}$ is simply related to the
corresponding Gramian determinant (\ref{eq:gammainf}) by $N \to
N-1$.

To calculate the infinite-$D$ symmetric minimum of the Gramian
derivatives, we expand $\Gamma$ in terms of its cofactors.  The
cofactor, denoted by $C_{ij}$, of the element $\gamma_{ij}$ in
$\Gamma$ is $(-1)^{i+j}$ multiplied by the determinant of the
matrix obtained by deleting the $i^{th}$ row and $j^{th}$ column
of $\Gamma$.  We may then write $\Gamma$ as
\begin{equation}
\Gamma=\sum_{j=1}^N \gamma_{ij} C_{ij}.
\end{equation}
Then the partial derivative of $\Gamma$ in terms of the cofactor
is
\begin{equation}
\frac{\partial \Gamma}{\partial \gamma_{ij}}=2C_{ij}.
\end{equation}
From this equation, the partial derivative of $\Gamma$ evaluated
at the infinite-$D$ symmetric minimum is
\begin{equation}
\left. \frac{\partial \Gamma}{\partial
\gamma_{ij}}\right|_{\infty}=-2 C_{\infty}^{(N-1)},
\end{equation}
where we have defined the following determinant of an $(N-1)
\times (N-1)$ matrix:
\begin{equation}
C_{\infty}^{(N-1)}=\det \left(
\begin{array}{ccccc}
\overline{\gamma}_{\infty}& \overline{\gamma}_{\infty}&\overline{\gamma}_{\infty}& \cdots & \overline{\gamma}_{\infty}\\
\overline{\gamma}_{\infty} & 1 &  \overline{\gamma}_{\infty}&  & \vdots \\
\overline{\gamma}_{\infty}&  \overline{\gamma}_{\infty}& 1 & \ddots & \vdots    \\
\vdots & \ddots & \ddots & \ddots & \overline{\gamma}_{\infty} \\
\overline{\gamma}_{\infty} & \cdots    & \cdots   &   \overline{\gamma}_{\infty}     & 1 \\
\end{array}
\right),
\end{equation}
and the superscript $(N-1)$ simply indicates the size of the
matrix. From this matrix one can show that the following recursion
relation holds:
\begin{equation}
C_{\infty}^{(N-1)}=\overline{\gamma}_{\infty}\left(
\Gamma\Big|^{(N-2)}_{\infty} - (N-2) C_{\infty}^{(N-2)} \right),
\end{equation}
or equivalently
\begin{equation}\label{eq:recursion}
C_{\infty}^{(N)}=\overline{\gamma}_{\infty}\left(
\Gamma\Big|^{(N-1)}_{\infty} - (N-1) C_{\infty}^{(N-1)} \right),
\end{equation}
where the $(N)$ superscript in the notation
$\Gamma\Big|^{(N)}_{\infty}$ again refers to the size of the
matrix $\Gamma\Big|_{\infty}$. From the recursion relation
(\ref{eq:recursion}), one can easily prove by induction the
conjecture that
$C_{\infty}^{(N)}=\overline{\gamma}_{\infty}(1-\overline{\gamma}_{\infty})^{N-2}$
and so,
\begin{equation}\label{eq:gamderiv}
\left. \frac{\partial \Gamma}{\partial
\gamma_{ij}}\right|_{\infty}=-2
\overline{\gamma}_{\infty}(1-\overline{\gamma}_{\infty})^{N-2}.
\end{equation}
The derivative of the principle minor evaluated at the
infinite-$D$ symmetric minimum is simply related to the
corresponding Gramian determinant derivative (\ref{eq:gamderiv})
by $N \to N-1$.

To summarize the above results, the following expressions are
needed when calculating the minimum of the effective potential
(\ref{minimum1} and \ref{minimum2}):
\begin{equation}\label{firstgammas}
\begin{array}{ll}
\left. \frac{\partial{\Gamma}}{\partial
\gamma_{ij}}\right|_{\infty}=-2\overline{\gamma}_{\infty}(1-\overline{\gamma}_{\infty})^{N-2}
& \left. \frac{\partial{\Gamma^{(\alpha)}}}{\partial
\gamma_{ij}}\right|_{\infty}=-2\overline{\gamma}_{\infty}(1-\overline{\gamma}_{\infty})^{N-3}
\\
\left.
\Gamma\right|_{\infty}=[1+(N-1)\overline{\gamma}_{\infty}](1-\overline{\gamma}_{\infty})^{N-1}
& \left.
\Gamma^{(\alpha)}\right|_{\infty}=[1+(N-2)\overline{\gamma}_{\infty}](1-\overline{\gamma}_{\infty})^{N-2}.
\end{array}
\end{equation}

And when evaluating the ${\bf F}$ matrix elements at the
infinite-$D$ symmetric minimum, the following six second-order
derivatives of the Gramian determinants are needed:
\begin{equation}\label{secondgammas}
\begin{array}{lll}
\left. \frac{\partial^2{\Gamma}}{\partial \gamma_{ij}\partial
\gamma_{kl}}\right|_{\infty}=0 & \left.
\frac{\partial^2{\Gamma}}{\partial \gamma_{ij}^2}\right|_{\infty}
=
-2(1-\overline{\gamma}_{\infty})^{N-3}(1+(N-3)\overline{\gamma}_{\infty})
& \left. \frac{\partial^2{\Gamma}}{\partial \gamma_{ij}\partial
\gamma_{jk}}\right|_{\infty}=2\overline{\gamma}_{\infty}(1-\overline{\gamma}_{\infty})^{N-3} \\ \\
\left. \frac{\partial^2{\Gamma^{(\alpha)}}}{\partial
\gamma_{ij}\partial \gamma_{kl}}\right|_{\infty}=0 & \left.
\frac{\partial^2{\Gamma^{(\alpha)}}}{\partial
\gamma_{ij}^2}\right|_{\infty} =
-2(1-\overline{\gamma}_{\infty})^{N-4}(1+(N-4)\overline{\gamma}_{\infty})
& \left. \frac{\partial^2{\Gamma^{(\alpha)}}}{\partial
\gamma_{ij}\partial \gamma_{jk}}\right|_{\infty} =
2\overline{\gamma}_{\infty}(1-\overline{\gamma}_{\infty})^{N-4}.
\end{array}
\end{equation}

\renewcommand{\theequation}{C\arabic{equation}}
\setcounter{equation}{0}
\section{The Symmetric Group and the Theory of Group Characters}\label{app:Char}
A group of transformations is a set of transformations which
satisfy the composition law $ab=c$ where $a$ and $b$ are any two
elements of the group and $c$ also belongs to the group. A group
also contains the identity element $I$ such that $aI=Ia=a$ and for
every element in the group its inverse is also to be found in the
group.

The symmetric group $S_N$ is the group of all permutations of $N$
objects and, as such, has $N!$ elements. A permutation may be
written as
\begin{equation}\label{eq:permutation}
\left(
\begin{array}{c@{\hspace{1.0ex}}c@{\hspace{1.0ex}}c@{\hspace{1.0ex}}c@{\hspace{1.0ex}}
c@{\hspace{1.0ex}}c@{\hspace{1.0ex}}c@{\hspace{1.0ex}}c}
1&2&3&4&5&6&7&8 \\
2&3&1&5&4&7&6&8 \end{array} \right) \end{equation}
This denotes the following transformation
\begin{center}
\begin{tabular}{|ccc|} \hline
\begin{tabular}{l}Object before \\ transformation \end{tabular} &
$\longrightarrow$ & \begin{tabular}{l} is transformed \\ to object
\end{tabular} \\ \hline
1 & $\longrightarrow$ & 2 \\
2 & $\longrightarrow$ & 3 \\
3 & $\longrightarrow$ & 1 \\
4 & $\longrightarrow$ & 5 \\
5 & $\longrightarrow$ & 4 \\
6 & $\longrightarrow$ & 7 \\
7 & $\longrightarrow$ & 6 \\
8 & $\longrightarrow$ & 8 \\ \hline \end{tabular}
\end{center}
A cycle is a particular kind of permutation where the object
labels are permuted into each other cyclically. For example the
cycle $(abc)$ means that object $a$ is transformed into object
$b$, object $b$ is transformed into object $c$ and object $c$ is
transformed into object $a$. The cycle $(abc)$ is termed a 3-cycle
since it cycles three objects. Like wise $(3479)$ is a 4-cycle and
$(5)$ is a 1-cycle, the latter transforms object five into itself.
All $N!$ permutations of the group $S_N$ may be decomposed into
cycles. For example, the permutation of Eq.~(\ref{eq:permutation})
may be decomposed into cycles as
\begin{equation}\label{eq:decomposition}
\left(
\begin{array}{c@{\hspace{1.0ex}}c@{\hspace{1.0ex}}c@{\hspace{1.0ex}}
c@{\hspace{1.0ex}}c@{\hspace{1.0ex}}c@{\hspace{1.0ex}}c@{\hspace{1.0ex}}c}
1&2&3&4&5&6&7&8 \\
2&3&1&5&4&7&6&8 \end{array} \right) = (123)(45)(67)(8)
\end{equation}
This consists of one 1-cycle, two 2-cycles and one 3-cycle. We can
denote the cycle structure of a permutation by the symbol
$(1^{\nu_1},2^{\nu_2},3^{\nu_3},\ldots,N^{\nu_N})$,\cite{hamermesh}
where the notation $j^{\nu_j}$ means a cycle of length $j$ and
$\nu_j$ equals the number of cycles of length $j$ in that
permutation. In the case of the permutation of
Eq.~(\ref{eq:decomposition}), its cycle structure is
$(1^1,2^2,3^1)$.

A matrix representation of a group is a set of nonsingular
matrices, including the unit matrix, which has the same
composition law as the elements of the group. The character of an
element of an matrix representation of a group is the trace of the
matrix. The character, as the trace of a matrix, is invariant
under a similarity transformation and thus elements of equivalent
representations have the same character. The set of all the
distinct characters of the elements of an irreducible
representation of the group uniquely specify the irreducible
representation up to an equivalence transformation. The characters
of irreducible representations of a group are termed {\em simple}
characters. All elements of a group which are related by a
similarity transformation are said to belong to the same class.
The character of the elements of a group belonging to a particular
class all have the same character. Thus there are as many distinct
characters for a group as there are classes.

For the group $S_N$ all elements with the same cycle structure
belong to the same class and so all elements of a matrix
representation of $S_N$ with the same cycle structure have the
same character.

A reducible matrix representation of a group may be bought to
block diagonal form by a similarity transformation, where the
individual blocks are irreducible matrix representations of the
same group with lower dimensionality. Thus the characters,
$\chi(R)$, of a reducible group are the sums of the characters of
the irreducible matrix representations into which it can be
decomposed, i.e.\
\begin{equation}\label{eq:chi_p}
\chi(R) = \sum_{p} \chi_p(R) \,,
\end{equation}
where $R$ denotes the element of the group, $p$ labels all of the
irreducible blocks into which the reducible matrix representation
of the group may be decomposed and $\chi_p(R)$ is the character of
the irreducible representation of the block labelled by $p$. Now
in a particular reducible representation a given irreducible
representation may be repeated along the diagonal $a_\alpha$ times
and so Eq.~(\ref{eq:chi_p}) may be rewritten as
\begin{equation}\label{eq:chi_a_alpha}
\chi(R) = \sum_{\alpha} a_\alpha \chi_\alpha(R) \,.
\end{equation}
The decomposition of $\chi(R)$ of Eq.~(\ref{eq:chi_a_alpha}) into
simple characters $\chi_\alpha(R)$ is unique, i.e.\ there is not
another decomposition of the form
\begin{equation}\label{eq:chi_b_alpha}
\chi(R) = \sum_{\alpha} b_\alpha \chi_\alpha(R) \,,
\end{equation}
where at least one of the $b_\alpha$ is different from the
corresponding $a_\alpha$. This follows from the fact that quite
generally
\begin{equation}\label{eq:num_rep}
a_\alpha = \frac{1}{h} \sum_R \chi^*_\alpha(R) \, \chi(R)
\end{equation}
and that the simple characters satisfy the orthogonality condition
\begin{equation}\label{eq:chi_orth}
\sum_R \chi^*_\alpha(R) \, \chi_\beta(R) = h \, \delta_{\alpha
\beta} \,,
\end{equation}
where $h$ is the number of elements in the group. For
Eqs.~(\ref{eq:chi_b_alpha}), (\ref{eq:num_rep}) and
(\ref{eq:chi_orth}) to be consistent we must have
\begin{equation}\label{eq:beqa}
b_\alpha = a_\alpha \,.
\end{equation}

The irreducible matrix representations of $S_N$ may be labelled by
a Young diagram ( = Young pattern = Young shape). A Young diagram
is a is a set of $N$ adjacent squares such that the row below a
given row of squares is equal to or shorter in length. The set of
all Young diagrams that can be formed from $N$ boxes of all
possible irreducible matrix representations of $S_N$.

A given Young diagram may be denoted by a partition. A partition,
$[\lambda_1, \hspace{1ex} \lambda_2, \hspace{1ex} \lambda_3,
\hspace{1ex} \ldots, \hspace{1ex} \lambda_N]$ is a series of $N$
numbers $\lambda_i$ such that $\lambda_1 \geq \lambda_2 \geq
\lambda_3 \geq \ldots \geq \lambda_N$ such that $\lambda_1 +
\lambda_2 + \lambda_3 + \cdots + \lambda_N = N$. The number
$\lambda_i$ is the number of boxes in row $i$ of the Young
diagram. Thus the set of all possible partitions of length $N$
labels all of the possible irreducible matrix representations of
$S_N$ and so the irreducible representation labels $\alpha$ and
$\beta$ in Eqs.~(\ref{eq:chi_a_alpha}), (\ref{eq:chi_b_alpha}),
(\ref{eq:num_rep}), (\ref{eq:chi_orth}) and (\ref{eq:beqa}) above,
for the group $S_N$ may be taken to run over the set of all
possible partitions.

A Young diagram can have up to $N$ rows. For an $N$-row
(one-column) Young diagram all of the $\lambda_i$s are non zero.
However only one Young diagram will have $N$ row; all of the rest
will have less than $N$ rows. Thus in all but one case the last
few $\lambda_i$s will be zero. It is standard practice to drop the
zeros and use a partition with less than $N$ numbers. Thus the
partition $[3,0,0]$ labelling an irreducible representation of
$S_3$ is usually abbreviated to just $[3]$.

\renewcommand{\theequation}{D\arabic{equation}}
\setcounter{equation}{0}
\section{The Inequivalence of Subspaces ${\mathbf{\alpha}}$
and ${\bm{\alpha'}}$ under ${\bm{S_N}}$ when ${\bm{\alpha \neq
\alpha'}}$ -- A Useful Equation.}\label{app:Ineqsusp} Here we wish
to note something that proves useful in Sec.~\ref{subsec:normcond}
above. Consider the quantity
\begin{equation}\label{eq:wwd}
W_{\bm{X}'}^\alpha \Upsilon [W_{\bm{X}'}^{\alpha'}]^T \,
D_{S_{\bm{X}'}}^{\alpha'}(R) \,,
\end{equation}
where $\bm{X}'$ is $\bar{\bm{r}}'$ or $\overline{\bm{\gamma}}'$,
$\alpha$ and $\alpha'$ denote the partitions $[N]$, $[N-1,
\hspace{1ex} 1]$ or $[N-2, \hspace{1ex} 2]$\,, and $\Upsilon$ is
an unspecified matrix. The matrix $W_{\bm{X}'}^\alpha
 \Upsilon [W_{\bm{X}'}^{\alpha'}]^T$ effects a similarity mapping from the
irreducible $\alpha'$ and $S_{\bm{X}'}$ space onto an irreducible
space denoted by $\alpha$. Thus one can construct an irreducible
matrix representation $D_c^{\alpha}(R)$ belonging to the set of
equivalent representations denoted by $\alpha$ such that
\begin{equation}\label{eq:dwwd}
D_c^\alpha(R) \,\, W_{\bm{X}'}^\alpha \Upsilon
[W_{\bm{X}'}^{\alpha'}]^T = W_{\bm{X}'}^\alpha \Upsilon
[W_{\bm{X}'}^{\alpha'}]^T \, D_{S_{\bm{X}'}}^{\alpha'}(R)
\end{equation}
for all elements $R$ of the group. According to Schur's Lemma
however, any such similarity mapping, $W_{\bm{X}'}^\alpha \Upsilon
[W_{\bm{X}'}^{\alpha'}]^T$, that satisfies Eq.~(\ref{eq:dwwd}) is
identically zero unless $\alpha=\alpha'$, i.e.\
\begin{eqnarray} \label{eq:wauwapt}
W_{\bm{X}'}^\alpha \Upsilon [W_{\bm{X}'}^{\alpha'}]^T = 0 \,, &
\forall & \alpha \neq \alpha' \,.
\end{eqnarray}
Equation~(\ref{eq:wauwapt}) is a statement of the inequivalence of
the irreducible spaces labelled by $\alpha$ and $\alpha'$ under
$S_N$ when $\alpha \neq \alpha'$. In particular we have
\begin{eqnarray} \label{eq:wawapt}
W_{\bm{X}'}^\alpha [W_{\bm{X}'}^{\alpha'}]^T = 0 \,, & \forall &
\alpha \neq \alpha' \,.
\end{eqnarray}
Schur's Lemma also implies $D_c^{\alpha'}(R)\neq
D_{S_{\bm{X}'}}^{\alpha'}(R)$ unless $W_{\bm{X}'}^{\alpha'}
[W_{\bm{X}'}^{\alpha'}]^T \propto I_{\alpha'}$, where
$I_{\alpha'}$ is the unit matrix.

\renewcommand{\theequation}{E\arabic{equation}}
\setcounter{equation}{0}
\section{Proof of the Orthogonality of ${\mathbf{W}}$\,.}
\label{app:proofOrthW} We shall show that condition~\ref{it:ort}
of Sec~\ref{subsec:normcond} above implies that $W$ is an
orthogonal matrix. This can be seen as follows. The displacement
vector ${\bar{\bm{y}}'}$ of Eq.~(\ref{eq:ytransposeP}) transforms
under an orthogonal, though reducible, representation of $S_N$
since $\bar{\bm{y}}^{\prime T} \, {\bar{\bm{y}}'}$ is invariant
under $S_N$. Writing this orthogonal representation as
\begin{equation} \label{eq:Dyb}
D_{\bar{\bm{y}}'}(R) = \left( \begin{array}{c|c}
D_{\bar{\bm{r}}'}(R) & \bm{0} \\ \hline \bm{0} &
D_{\overline{\bm{\gamma}}'}(R)
\end{array} \right) \,,
\end{equation}
Eq.~(\ref{eq:SWy}) implies that the representation under which the
displacement coordinate vector ${\bm{S}}$ transforms is
\begin{equation}\label{eq:DSWDy}
D_S = W D_{\bar{\bm{y}}'}(R) \, W^{-1} \,.
\end{equation}
Since $D_{\bar{\bm{y}}'}(R)$ is an orthogonal matrix
representations of $S_N$ then from Eq.~(\ref{eq:DSWDy}) we have
\begin{equation}\label{eq:DyDy}
D_{\bar{\bm{y}}'}^T(R) D_{\bar{\bm{y}}'}(R) = W^T D_S^T(R)
[W^{-1}]^T W^{-1} D_S(R) W = 1 \,.
\end{equation}
From Eq.~(\ref{eq:DyDy}) we can write
\begin{equation}
D_S^T(R) [W W^T]^{-1} D_S(R) = [W W^T]^{-1} = D_S^T(R) D_S(R) [W
W^T]^{-1} \,,
\end{equation}
where we have used the orthogonality of $D_S(R)$ in the last step.
Left multiplying by $[D_S^T]^{-1}(R)$ we obtain
\begin{equation}
[W W^T]^{-1} D_S(R) = D_S(R) [W W^T]^{-1} \,,
\end{equation}
from which we derive
\begin{equation}\label{eq:CWWTD}
W W^T \, D_S(R) = D_S(R) \, W W^T \,.
\end{equation}
We notice that
\begin{equation}\label{eq:WWTeqI}
W W^T = I
\end{equation}
satisfies Eq.~(\ref{eq:CWWTD}). We will now show that it is the
only solution. We note that $D_S(R)$ is a block diagonal matrix of
irreducible representations of the form
\begin{eqnarray}
D_S(R) &=& P \left( \begin{array}{c|c} D_{S_{\bar{\bm{r}}'}}(R) & \bm{0} \\
\hline \bm{0} & D_{S_{\overline{\bm{\gamma}}'}}(R) \end{array} \right) P^T \nonumber \\
 &=& \left(
\begin{array}{cc|cc|c}
D^{[N]}(R) & \bm{0} & \bm{0} & \bm{0} & \bm{0} \\
\bm{0} & D^{[N]}(R) & \bm{0} & \bm{0} & \bm{0} \\
\hline \bm{0} & \bm{0} & D^{[N-1, \hspace{1ex} 1]}(R) & \bm{0} &
\bm{0} \\ \bm{0} & \bm{0} &
\bm{0} & D^{[N-1, \hspace{1ex} 1]}(R) & \bm{0} \\
\hline \bm{0} & \bm{0} & \bm{0} & \bm{0} & D^{[N-2, \hspace{1ex}
2]}(R)
\end{array} \right) \,,
\label{eq:DS}
\end{eqnarray}
where we have also used condition~\ref{it:teq0} from
Sec.~\ref{subsec:normcond} above in the last step. Now from
Eq.~(\ref{eq:W})
\begin{equation}\label{eq:WtW}
\begin{array}{@{}l@{}} W W^T = \\
\left(
\begin{array}{@{}c@{\hspace{-1.5ex}}c@{\hspace{-1.5ex}}c@{\hspace{-1.5ex}}cc@{}}
W_{\bar{\bm{r}}'}^{[N]} [W_{\bar{\bm{r}}'}^{[N]}]^T & \bm{0} &
W_{\bar{\bm{r}}'}^{[N]} [W_{\bar{\bm{r}}'}^{[N-1,
\hspace{1ex} 1]}]^T & \bm{0} & \bm{0} \\
\bm{0} & W_{\overline{\bm{\gamma}}'}^{[N]}
[W_{\overline{\bm{\gamma}}'}^{[N]}]^T & \bm{0} &
W_{\overline{\bm{\gamma}}'}^{[N]}
[W_{\overline{\bm{\gamma}}'}^{[N-1, \hspace{1ex} 1]}]^T &
W_{\overline{\bm{\gamma}}'}^{[N]} [W_{\overline{\bm{\gamma}}'}^{[N-2, \hspace{1ex} 2]}]^T \\
W_{\bar{\bm{r}}'}^{[N-1, \hspace{1ex} 1]}
[W_{\bar{\bm{r}}'}^{[N]}]^T & \bm{0} & W_{\bar{\bm{r}}'}^{[N-1,
\hspace{1ex} 1]} [W_{\bar{\bm{r}}'}^{[N-1, \hspace{1ex} 1]}]^T &
\bm{0} & \bm{0} \\
\bm{0} & W_{\overline{\bm{\gamma}}'}^{[N-1, \hspace{1ex} 1]}
[W_{\overline{\bm{\gamma}}'}^{[N]}]^T & \bm{0} &
W_{\overline{\bm{\gamma}}'}^{[N-1, \hspace{1ex} 1]}
[W_{\overline{\bm{\gamma}}'}^{[N-1, \hspace{1ex} 1]}]^T &
W_{\overline{\bm{\gamma}}'}^{[N-1, \hspace{1ex} 1]}
[W_{\overline{\bm{\gamma}}'}^{[N-2, \hspace{1ex} 2]}]^T \\
\bm{0} & W_{\overline{\bm{\gamma}}'}^{[N-2, \hspace{1ex} 2]}
[W_{\overline{\bm{\gamma}}'}^{[N]}]^T & \bm{0} &
W_{\overline{\bm{\gamma}}'}^{[N-2, \hspace{1ex} 2]}
[W_{\overline{\bm{\gamma}}'}^{[N-1, \hspace{1ex} 1]}]^T &
W_{\overline{\bm{\gamma}}'}^{[N-2, \hspace{1ex} 2]}
[W_{\overline{\bm{\gamma}}'}^{[N-2, \hspace{1ex} 2]}]^T
\end{array} \right) \end{array}
\end{equation}
Substituting Eqs.~(\ref{eq:DS}) and (\ref{eq:WtW}) in
Eq.~(\ref{eq:CWWTD}) we obtain
\begin{equation}\label{eq:wwtDc}
W_{\bm{X}'}^\alpha [W_{\bm{X}'}^{\alpha'}]^T \, D^{\alpha'}(R) =
D^\alpha(R) \,\, W_{\bm{X}'}^\alpha [W_{\bm{X}'}^{\alpha'}]^T \,,
\end{equation}
where $\bm{X}'$ is $\bar{\bm{r}}'$ or $\overline{\bm{\gamma}}'$,
and $\alpha$ and $\alpha'$ are the partitions $[N]$, $[N-1,
\hspace{1ex} 1]$ or $[N-2, \hspace{1ex} 2]$. As we note in
Appendix~\ref{app:Ineqsusp}, Eq.~(\ref{eq:wawapt}),
$W_{\bm{X}'}^\alpha [W_{\bm{X}'}^{\alpha'}]^T = 0$ for $\alpha
\neq \alpha'$, and so Eq.~(\ref{eq:wwtDc}) is automatically
satisfied when $\alpha \neq \alpha'$. It does not give us a new
condition on the $W_{\bm{X}'}^\alpha$. When $\alpha = \alpha'$
however, we have
\begin{equation}\label{eq:wwtaaDc}
W_{\bm{X}'}^\alpha [W_{\bm{X}'}^\alpha]^T \, D^\alpha(R) =
D^\alpha(R) \,\, W_{\bm{X}'}^\alpha [W_{\bm{X}'}^\alpha]^T \,.
\end{equation}
The second part of Schur's Lemma\cite{SchurII} then requires that
\begin{equation}\label{eq:WaWaIapp}
W_{\bm{X}'}^\alpha [W_{\bm{X}'}^\alpha]^T = I_\alpha \,,
\end{equation}
where $I_\alpha$ is the unit matrix.

Thus Eq.~(\ref{eq:WaWaIapp}) is the essential equation to satisfy.

When Eq.~(\ref{eq:WaWaIapp}) is satisfied,
Eq.~(\ref{eq:WaWaIapp}), together with the automatically satisfied
Eq.~(\ref{eq:wawapt}), ensures that Eq.~(\ref{eq:WWTeqI}) is
satisfied. Multiplying Eq.~(\ref{eq:WWTeqI}) on the right by $W^T$
and on the left by $[W^T]^{-1}$ we obtain $W^T W = I$ and so $W$
is an orthogonal transformation satisfying
\begin{equation}
W W^T = W^T W = I \,.
\end{equation}

\renewcommand{\theequation}{F\arabic{equation}}
\setcounter{equation}{0}
\section{Calculation of $\mathbf{[
\sigma_{[N]}^{\bm{G}}]_{\bm{X}'_1,\,\bm{X}'_2}}$ and $\mathbf{[
\sigma_{[N]}^{\bm{FG}}]_{\bm{X}'_1,\,\bm{X}'_2}\,}$\,, the reduced
${\bm G}$ and ${\bm {FG}}$ matrix elements, in the
symmetry-coordinate basis.}\label{app:sigmacalc} In this appendix
we use the $W_{\bm{X}'}^{[N]}$ matrices ($\bm{X}'= \bar{\bm{r}}'$
or $\overline{\bm{\gamma}}'$) to calculate the reduced $\bm{G}$
and $\bm{FG}$ matrix elements,
$[\bm{\sigma_{[N]}^{\bm{G}}}]_{\bm{X}'_1,\,\bm{X}'_2}$ and
$[\bm{\sigma_{[N]}^{\bm{FG}}}]_{\bm{X}'_1,\,\bm{X}'_2}$\,, using
Eq.~(\ref{eq:sigmaQ}):
\begin{equation}
[\bm{\sigma_{[N]}^Q}]_{\bm{X}'_1,\,\bm{X}'_2} =
(W_{\bm{X}'_1}^{[N]})_\xi \, {\bf Q}_{\bm{X}'_1 \bm{X}'_2} \,
[(W_{\bm{X}'_2}^{[N]})_\xi]^T \,,
\end{equation}
where $\bm{Q} = \bf G$ or $\bf FG$ and $\xi$ is a row label.

Using Eq.~(\ref{eq:WNeqsqrtN1})
\begin{equation}
[W^{[N]}_{\bar{\bm{r}}'}]_i = \frac{1}{\sqrt{N}} \,
[{\bm{1}}_{\bar{\bm{r}}'}]_i
\end{equation}
and Eq.~(\ref{eq:WNgeqsqrt2ontnm11})
\begin{equation}
[W^{[N]}_{\overline{\bm{\gamma}}'}]_{ij} = \sqrt{\frac{2}{N(N-1)}}
\,\, [{\bm{1}}_{\overline{\bm{\gamma}}'}]_{ij}
\end{equation}
with Eq.~(\ref{eq:Gsub})
\begin{equation}
\bm{G} = \left( \begin{array}{cc}
\tilde{a}' {\bf I}_N & \bm{0} \\
\bm{0} & \tilde{g}' {\bf I}_M + \tilde{h}' {\bf R}^T {\bf R}
\end{array} \right) \,,
\end{equation}
where ${\bf I}_N$\,, ${\bf I}_M$\,, ${\bf R}$ are defined in
Sect.~\ref{subsec:Qsubm} after Eq.~(56) and $ \tilde{g}'$ and
$\tilde{h}'$ are defined by Eq.~(65), we find:
\renewcommand{\jot}{1em}
\begin{eqnarray}
[\bm{\sigma_{[N]}^G}]_{\bar{\bm{r}}',\,\bar{\bm{r}}'} & = &
\sum_{i,j=1}^N [W_{\bar{\bm{r}}'}^{[N]}]_{i} \, [{\bm
G}_{\bar{\bm{r}}' \bar{\bm{r}}'}]_{ij}
\, [(W_{\bar{\bm{r}}'}^{[N]})^{T}]_{j} \,  \nonumber \\
& = & \frac{\tilde{a}'}{N} \sum_{i,j=1}^N [{\bm{1}}_{\bar{\bm{r}}'}]_{i} \,
\delta_{ij} [{\bm{1}}_{\bar{\bm{r}}'}]_{j}  \nonumber \\
& = & \frac{\tilde{a}'}{N} \sum_{i}^N 1  \nonumber \\
& = & \tilde{a}' \,,
\end{eqnarray}
\renewcommand{\jot}{0em}
\renewcommand{\jot}{1em}
\begin{eqnarray}
[\bm{\sigma_{[N]}^G}]_{\bar{\bm{r}}',\,\overline{\bm{\gamma}}'} &
= & \sum_{i=1}^N \sum_{k=2}^N \sum_{j=1}^{k-1}
[W_{\bar{\bm{r}}'}^{[N]}]_{i} \, [{\bm G}_{\bar{\bm{r}}'
\overline{\bm{\gamma}}'}]_{i,\,jk}
\, [(W_{\overline{\bm{\gamma}}'}^{[N]})^{T}]_{jk} \,  \nonumber \\
& = & 0 \,,
\end{eqnarray}
\renewcommand{\jot}{0em}
\renewcommand{\jot}{1em}
\begin{eqnarray}
[\bm{\sigma_{[N]}^G}]_{\overline{\bm{\gamma}}',\, \bar{\bm{r}}'} &
= & \sum_{j=2}^N \sum_{i=1}^{j-1} \sum_{k=1}^{N}
[W_{\overline{\bm{\gamma}}'}^{[N]}]_{ij} \, [{\bm
G}_{\overline{\bm{\gamma}}' \bar{\bm{r}}'}]_{ij,\,k}
\, [(W_{\bar{\bm{r}}'}^{[N]})^{T}]_{k} \,  \nonumber \\
& = & 0
\end{eqnarray}
\renewcommand{\jot}{0em}
and
\renewcommand{\jot}{1em}
\begin{eqnarray}
[\bm{\sigma_{[N]}^G}]_{\overline{\bm{\gamma}}',\,
\overline{\bm{\gamma}}'} & = & \sum_{j=2}^N \sum_{i=1}^{j-1}
\sum_{l=2}^{N} \sum_{k=1}^{l-1}
[W_{\overline{\bm{\gamma}}'}^{[N]}]_{ij} \, [{\bm
G}_{\overline{\bm{\gamma}}' \overline{\bm{\gamma}}'}]_{ij,\,kl}
\, [(W_{\overline{\bm{\gamma}}'}^{[N]})^{T}]_{kl} \,  \nonumber \\
& = & \frac{2}{N(N-1)} \sum_{j=2}^N \sum_{i= 1}^{j-1}
\sum_{l=2}^{N} \sum_{k=1}^{l-1}
    [{\bm{1}}_{\overline{\bm{\gamma}}'}]_{ij}
\bigl(\tilde{g'} (\delta_{ik} \delta_{jl} + \delta_{il}
\delta_{jk}) + \tilde{h'} (\delta_{ik} + \delta_{il} + \delta_{jk}
+ \delta_{jl}))
[{\bm{1}}_{\overline{\bm{\gamma}}'}]_{kl}   \nonumber \\
& = & \frac{2}{4N(N-1)} \Bigl(
\renewcommand{\arraystretch}{1.5} \begin{array}[t]{l}
{\displaystyle \sum_{i,j,k,l=1}^N \bigl[ \tilde{g'} (\delta_{ik}
\delta_{jl} + \delta_{il} \delta_{jk})
+ \tilde{h'} (\delta_{ik} + \delta_{il} + \delta_{jk} + \delta_{jl}) \bigr] } \\
{\displaystyle - \sum_{i,j,k = 1}^N \bigl[ \tilde{g'} (\delta_{ik}
\delta_{jk} + \delta_{ik} \delta_{jk})
+ \tilde{h'} (\delta_{ik} + \delta_{ik} + \delta_{jk} + \delta_{jk}) \big] } \\
{\displaystyle - \sum_{i,k,l = 1}^N \bigl[ \tilde{g'} (\delta_{ik}
\delta_{il} + \delta_{il} \delta_{ik})
+ \tilde{h'} (\delta_
{ik} + \delta_{il} + \delta_{ik} + \delta_{il}) \bigr] } \\
{\displaystyle + \sum_{i,k = 1}^N \bigl[ \tilde{g'} (\delta_{ik}
\delta_{ik} + \delta_{ik} \delta_{ik}) + \tilde{h'} (\delta_{ik} +
\delta_{ik} + \delta_{ik} + \delta_{ik}) \bigr]
\Bigr)  } \end{array} \renewcommand{\arraystretch}{1} \nonumber \\
& = & \frac{1}{2N(N-1)} \bigl[(2 \tilde{g'} N^2 - 4 \tilde{h'}
N^3) -2 (2 \tilde{g'} N + 4 \tilde{h'} N^2) + (2 \tilde{g'} N + 4
\tilde{h'} N) \bigr]  \nonumber \\
& = & \tilde{g'} + 2 \tilde{h'} (N-1) \,.
\end{eqnarray}
\renewcommand{\jot}{0em}
Thus we obtain Eq.~(\ref{eq:sigmaNG}):
\begin{equation}
\bm{\sigma_{[N]}^{\bm{G}}} = \left(
\renewcommand{\arraystretch}{1.5}
\begin{array}{l@{\hspace{1.5em}}l} \protect[\bm{\sigma_{[N]}^{\bm{G}}}\protect]_{\bar{\bm{r}}',\,\bar{\bm{r}}'} =
\tilde{a}' & {\displaystyle
\protect[\bm{\sigma_{[N]}^{\bm{G}}}\protect]_{\bar{\bm{r}}',\,
\overline{\bm{\gamma}}'} =
0} \\
{\displaystyle
\protect[\bm{\sigma_{[N]}^{\bm{G}}}\protect]_{\overline{\bm{\gamma}}',\,\bar{\bm{r}}'}
= 0} & {\displaystyle
\protect[\bm{\sigma_{[N]}^{\bm{G}}}\protect]_{\overline{\bm{\gamma}}',\,\overline{\bm{\gamma}}'}
= \left( \tilde{g}' + 2(N-1) \tilde{h}' \right)}
\end{array} \renewcommand{\arraystretch}{1} \right) \,.
\end{equation}

Similarly letting ${\bf Q} = \bm{FG}$ and using Eq.~(62):
\begin{equation}
\bm{FG}= \left(
\begin{array}{cc}
\tilde{a} {\bf I}_N + \tilde{b} {\bf J}_N & \tilde{e} {\bf R}
+ \tilde{f} {\bf J}_{NM} \\
\tilde{c} {\bf R}^T + \tilde{d} {\bf J}_{MN} & \tilde{g} {\bf I}_M
+ \tilde{h} {\bf R}^T {\bf R} + \tilde{\iota} {\bf J}_M
\end{array}\right) \,,
\end{equation}
we can derive:
\renewcommand{\jot}{1em}
\begin{eqnarray}
[\bm{\sigma_{[N]}^{\bm{FG}}}]_{\bar{\bm{r}}',\,\bar{\bm{r}}'} & =
& \sum_{i,j=1}^N [W_{\bar{\bm{r}}'}^{[N]}]_{i} \, [{\bm
FG}_{\bar{\bm{r}}' \bar{\bm{r}}'}]_{ij} \,
[(W_{\bar{\bm{r}}'}^{[N]})^{T}]_{j} \,  \nonumber \\
& = & \frac{1}{N} \sum_{i,j=1}^N [{\bm{1}}_{\bar{\bm{r}}'}]_{i} (
\tilde{a}
\delta_{ij} + \tilde{b} 1_{ij}) [{\bm{1}}_{\bar{\bm{r}}'}]_{j}  \nonumber \\
& = & \frac{1}{N} [ N \tilde{a} + \tilde{b} N^2 ]  \nonumber \\
& = & \tilde{a} + \tilde{b} N \,,
\end{eqnarray}
\renewcommand{\jot}{0em}
\renewcommand{\jot}{1em}
\begin{eqnarray}
[\bm{\sigma_{[N]}^{\bm{FG}}}]_{\bar{\bm{r}}',\,\overline{\bm{\gamma}}'}
& = & \sum_{i=1}^N \sum_{k=2}^N \sum_{j=1}^{k-1}
[W_{\bar{\bm{r}}'}^{[N]}]_{i} \, [{\bm FG}_{\bar{\bm{r}}'
\overline{\bm{\gamma}}'}]_{i,\,jk}
\, [(W_{\overline{\bm{\gamma}}'}^{[N]})^{T}]_{jk} \,  \nonumber \\
& = & \frac{1}{N} \sqrt{\frac{2}{N-1}} \sum_{i=1}^N \sum_{k=2}^N
\sum_{j=1}^{k-1} [{\bm{1}}_{\bar{\bm{r}}'}]_{i} \Bigl(\tilde{e}
(\delta_{ij} + \delta_{ik})
+ \tilde{f}  \Bigr) [{\bm{1}}_{\overline{\bm{\gamma}}'}]_{jk}  \nonumber \\
& = & \frac{1}{N \sqrt{2(N-1)}} \,\, \Bigl( \sum_{i,j,k=1}^N
\tilde{e} ( \delta_{ij} + \delta_{ik}) + \tilde{f} - \sum_{i,j =
1}^N \tilde{e}
( \delta_{ij} + \delta_{ij}) + \tilde{f} \Bigr)  \nonumber \\
& = & \frac{1}{N \sqrt{2 (N-1)}} \,\, ( 2 \tilde{e} N^2 +
\tilde{f} N^3
  - 2 \tilde{e} N - \tilde{f} N^2)  \nonumber \\
& = & \sqrt{ 2 (N-1)} \, \left( \tilde{e} + \frac{N}{2} \tilde{f}
\right) \,,
\end{eqnarray}
\renewcommand{\jot}{0em}
\renewcommand{\jot}{1em}
\begin{eqnarray}
[\bm{\sigma_{[N]}^{\bm{FG}}}]_{\overline{\bm{\gamma}}',\,
\bar{\bm{r}}'} & = & \sum_{j=2}^N \sum_{i=1}^{j-1} \sum_{k=1}^{N}
[W_{\overline{\bm{\gamma}}'}^{[N]}]_{ij} \, [{\bm
FG}_{\overline{\bm{\gamma}}' \bar{\bm{r}}'}]_{ij,\,k} \,
[(W_{\bar{\bm{r}}'}^{[N]})^{T}]_{k} \,  \nonumber \\
& = & \frac{1}{N} \sqrt{\frac{2}{N-1}} \sum_{j=2}^N
\sum_{i=1}^{j-1} \sum_{k=1}^N
[{\bm{1}}_{\overline{\bm{\gamma}}'}]_{ij} \Bigl[\tilde{c}
(\delta_{ki} + \delta_{kj})
+ \tilde{d}  \Bigr] [{\bm{1}}_{\bar{\bm{r}}'}]_{k}  \nonumber \\
& = & \frac{1}{N \sqrt{2(N-1)}} \,\, \Bigl( \sum_{i,j,k = 1}^N
\bigl[ \tilde{c} ( \delta_{ki} + \delta_{kj}) + \tilde{d} \bigr] -
\sum_{ik = 1}^N \bigl[ \tilde{c}
( \delta_{ki} + \delta_{ki}) + \tilde{d} \bigr] \Bigr)  \nonumber \\
& = & \frac{1}{N \sqrt{2 (N-1)}} \,\, ( 2 \tilde{c} N^2 +
\tilde{d} N^3
  - 2 \tilde{c} N - \tilde{d} N^2)  \nonumber \\
& = & \sqrt{2(N-1)} \, \left(  \tilde{c} + \frac{N}{2} \,
\tilde{d} \right)
\end{eqnarray}
\renewcommand{\jot}{0em}
and
\renewcommand{\jot}{1em}
\begin{eqnarray}
[\bm{\sigma_{[N]}^{\bm{FG}}}]_{\overline{\bm{\gamma}}',\,
\overline{\bm{\gamma}}'} & = & \sum_{j=2}^N \sum_{i=1}^{j-1}
\sum_{l=2}^{N} \sum_{k=1}^{l-1}
[W_{\overline{\bm{\gamma}}'}^{[N]}]_{ij} \, [{\bm
FG}_{\overline{\bm{\gamma}}' \overline{\bm{\gamma}}'}]_{ij,\,kl}
\, [(W_{\overline{\bm{\gamma}}'}^{[N]})^{T}]_{kl} \,  \nonumber \\
& = & \frac{2}{N(N-1)} \Bigl( \sum_{j=2}^N \sum_{i= 1}^{j-1}
\sum_{l=2}^{N} \sum_{k=1}^{l-1}
[{\bm{1}}_{\overline{\bm{\gamma}}'}]_{ij} \bigl[ \tilde{g}
(\delta_{ik} \delta_{jl} + \delta_{il} \delta_{jk}) + \tilde{h}
(\delta_{ik} + \delta_{il} + \delta_{jk} + \delta_{jl})
+ \tilde{i} \bigr] [{\bm{1}}_{\overline{\bm{\gamma}}'}]_{kl} \Bigr)  \nonumber \\
& = & \frac{1}{2N(N-1)} \Bigl(
\renewcommand{\arraystretch}{1.5} \begin{array}[t]{l} {\displaystyle \sum_{i,j,k,l = 1}^N
 \bigl[ \tilde{g}
(\delta_{ik} \delta_{jl} + \delta_{il} \delta_{jk}) + \tilde{h}
(\delta_{ik} + \delta_{il} + \delta_{jk} + \delta_{jl})
+ \tilde{i} \bigr]  } \\
{\displaystyle - \sum_{i,j,k = 1}^N \bigl( \tilde{g} (\delta_{ik}
\delta_{jk} + \delta_{ik} \delta_{jk}) + \tilde{h} (\delta_{ik} +
\delta_{ik} + \delta_{jk} + \delta_{jk})
+ \tilde{i} \bigr) } \\
{\displaystyle - \sum_{i,k,l = 1}^N \bigl( \tilde{g} (\delta_{ik}
\delta_{il} + \delta_{il} \delta_{ik}) + \tilde{h} (\delta_{ik} +
\delta_{il} + \delta_{ik} + \delta_{il})
+ \tilde{i} \bigr)  } \\
{\displaystyle + \sum_{i,k = 1}^N \bigl( \tilde{g} (\delta_{ik}
\delta_{ik} + \delta_{ik} \delta_{ik}) + \tilde{h} (\delta_{ik} +
\delta_{ik} + \delta_{ik} + \delta_{ik}) + \tilde{i} \bigr)
\Bigr)  } \end{array} \renewcommand{\arraystretch}{1} \nonumber \\
& = & \frac{1}{2N(N-1)} \bigl(  2N(N-1) \tilde{g} + 4N( N^2-2N +
1) \tilde{h}
+ N^2 (N^2 - 2N + 1) \tilde{i} \bigr)  \nonumber \\
& = & \tilde{g} + 2 (N-1) \tilde{h} + \frac{N (N-1)}{2} \,\,
\tilde{i} \,.
\end{eqnarray}
\renewcommand{\jot}{0em}
Thus we obtain  Eq.~(\ref{eq:sigmaNFG}):
\begin{equation}
\bm{\sigma_{[N]}^{\bm{FG}}} = \left(
\renewcommand{\arraystretch}{1.5}
\begin{array}{l@{\hspace{1.5em}}l} \protect[\bm{\sigma_{[N]}^{\bm{FG}}}\protect]_{\bar{\bm{r}}',\,\bar{\bm{r}}'} =
(\tilde{a}+\tilde{b}N) & {\displaystyle
\protect[\bm{\sigma_{[N]}^{\bm{FG}}}\protect]_{\bar{\bm{r}}',\,
\overline{\bm{\gamma}}'} =
\sqrt{2(N-1)} \,\, \left( \tilde{e}+\frac{N}{2}\tilde{f} \right)} \\
{\displaystyle
\protect[\bm{\sigma_{[N]}^{\bm{FG}}}\protect]_{\overline{\bm{\gamma}}',\,\bar{\bm{r}}'}
= \sqrt{2(N-1)} \,\, \left( \tilde{c}+\frac{N}{2}\tilde{d}
\right)} & {\displaystyle
\protect[\bm{\sigma_{[N]}^{\bm{FG}}}\protect]_{\overline{\bm{\gamma}}',\,\overline{\bm{\gamma}}'}
= \left(
\tilde{g}+2(N-1)\tilde{h}+\frac{N(N-1)}{2}\,\,\tilde{\iota}
\right)}
\end{array} \renewcommand{\arraystretch}{1} \right) \,.
\end{equation}


\begin{thebibliography}{99}
\bibitem{different} In Ref.~\cite{loeser}, Eq.~(\ref{eq:E1}) would read
%
\begin{eqnarray}
\overline{E} &=& \overline{E}_{\infty} + \delta \overline{E}_o + O(\delta^2) \nonumber \\
&=&V_{\mathtt{eff}}(\bar{r}_{\infty},\overline{\gamma}_{\infty}) +
\hspace{0.50em} \delta \left\{  \hspace{-0.5em}
\sum_{\renewcommand{\arraystretch}{0}
\begin{array}[t]{r@{}l@{}c@{}l@{}l} \scriptstyle \mu = \{
  & \scriptstyle \bm{0}^\pm,\hspace{0.5ex}
  & \scriptstyle \bm{1}^\pm & , & \\
  & & \scriptstyle \bm{2} & & \scriptstyle  \}
            \end{array}
            \renewcommand{\arraystretch}{1} }
\hspace{-1.5em} ( n_{\mu}+\frac{d_\mu}{2} ) \, \bar{\omega}_{\mu}
+v_o \right\} + O(\delta^2) \,,
\end{eqnarray}
%
where $n_{\mu}$ is the total number of quanta in all the normal
modes with the same frequency $\bar{\omega}_{\mu}$, i.e.\
%
\begin{equation}
n_\mu = \sum_{\mathsf{n}_{\mu}=0}^\infty
           {\mathsf{n}}_{\mu} \, d_{\mu,\mathsf{n}_{\mu}} \,.
\end{equation}
\end{thebibliography}
\end{document}